\documentclass[reprint,bibnotes,amsmath,amssymb,aps,prb,showpacs,floatfix,superscriptaddress,longbibliography]{revtex4-1}

\usepackage{amsmath}
\usepackage{amsfonts}
\usepackage{amssymb}
\usepackage{amsxtra}
\usepackage{xcolor}
\usepackage{graphicx}
\usepackage{subfigure}
\usepackage{dcolumn}
\usepackage{float}
\usepackage{bm}
\usepackage[breaklinks=true,colorlinks,citecolor=blue,linkcolor=blue,urlcolor=blue]{hyperref}

\newcommand{\nn}{\nonumber}

\newcommand{\bsigma}{\boldsymbol{\sigma}}
\DeclareMathAlphabet{\bi}{OML}{cmm}{b}{it}
\def\be{\begin{equation}}
\def\ee{\end{equation}}
\def\bearr{\begin{eqnarray}}
\def\eearr{\end{eqnarray}}

\def\bs{\boldsymbol}

\begin{document}
\title{Giant optical activity and Kerr effect in type-I and type-II Weyl semimetals}
\author{Kabyashree Sonowal}
\affiliation{Department of Physics, Indian Institute of Technology Kanpur, Kanpur - 208016, India}
\author{Ashutosh Singh}
\affiliation{Department of Physics, Indian Institute of Technology Kanpur, Kanpur - 208016, India}
\affiliation{Department of Physics and astronomy, McMaster University, Hamilton, Ontario L8S 4M1, Canada}
\author{Amit  Agarwal}
\email{amitag@iitk.ac.in}
\affiliation{Department of Physics, Indian Institute of Technology Kanpur, Kanpur - 208016, India}

\date{\today}

\begin{abstract}
We explore optical activity in thin films and bulk of type-I and type-II Weyl semimetals (WSM), and demonstrate the existence of a giant Kerr effect in both. In time-reversal symmetry broken WSM thin films, the polarization rotation is caused by the optical Hall conductivity including the anomalous Hall term. The Kerr angle is found to be $\propto Q/\omega$, with $Q$ and $\omega$ being the Weyl node separation and the optical frequency, respectively. In contrast, the optical activity in the bulk WSM is dominated by axion 
electrodynamics, which persists even in the Pauli blocked regime of no optical transitions. In bulk WSM, $Q$ acts analogous to the magnetization in magnetic materials, leading to large `polar Kerr effect' (linear in $Q$) when light is incident on WSM surface without Fermi arc states, and the `Voigt effect' (quadratic in $Q$), 
when light is incident on surface with Fermi arc states. 
%
%
\end{abstract}

%
\maketitle

\section{Introduction}

Weyl semimetals host even pairs of linearly dispersing massless quasi-particles of opposite chirality with novel Fermi-arc surface states \cite{Nielsen83,PhysRevB.83.205101,alexey,Jia,Bansil16,Yan,RevModPhys.90.015001}. Their existence has been recently demonstrated in several crystalline materials with broken time reversal or space-inversion symmetry. Some examples include, topological insulator material~\cite{PhysRevLett.107.127205}, pyrochlore iridates~\cite{PhysRevB.83.205101,PhysRevLett.120.096801}, WTe$_{2}$~\cite{PhysRevB.94.241119, Li2017},
MoTe$_{2}$~\cite{PhysRevB.92.161107,Deng,Lunan,Wang,Tamai,Jiang}, Mo$_{x}$W$_{1-x}$Te$_{2}$~\cite{PhysRevB.94.085127,Chang}, TaAs~\cite{Lv,Lv1,Yang}, and TaP~\cite{Xu,Xue1501092, Sun}, among others. 
In contrast to their high energy counterparts, Weyl semimetals can also have Lorentz symmetry violating tilted energy spectrum. This allows their classification into a type-I (partially tilted) WSM with vanishing DOS at the Weyl point, and a type-II (overtilted) phase with a finite density of states at the Fermi energy and an electron and a hole pocket on either side of the Weyl point \cite{alexey}. The type-II Weyl state has been experimentally demonstrated in noncentrosymmetric TaIrTe$_4$\cite{Haubold}, MoTe$_2$\cite{Deng}, WTe$_2$\cite{Jiang}, and LaAlGe\cite{Xue1603266}, among other materials. 


The Weyl nodes in a WSM act as a source or sink of the Berry curvature, which acts as a magnetic field in the momentum space \cite{PhysRevB.96.115202, PhysRevLett.108.046602, Shou,Kamal2018}. This finite Berry phase induces several very interesting electronic transport and optical phenomena, including the the anomalous quantum Hall (AQH) effect \cite{Zyuzin_Burkov12,Son12,Burkov14,Kim14,PhysRevLett.119.036601,PhysRevLett.115.117403,PhysRevB.92.161110,PhysRevB.91.115135,PhysRevB.91.081106,Burkov17,PhysRevB.96.115202,PhysRevB.97.035403,Wu2016,Ma2017,PhysRevLett.117.217204,Mehdi,Kamal2018,Yang1,Chen2018,Klaus2018,Kotov2,2019arXiv190303072S,2019arXiv190301205D}. Another intriguing aspect of optical activity in WSM is its connection to axion electrodynamics \cite{Frank,Wu1124}, which modifies the Maxwell's equations. The combined effect of these has led to the prediction of several interesting effects in WSM interacting with light, such as giant polarisation rotation\cite{Mehdi}, tunable perfect absorption\cite{Klaus2018} and creation of novel waveguide modes\cite{Kotov2}, among others. Contact free optical techniques such as polarization rotation are also regularly used to explore time reversal symmetry breaking states in ferromagnets\cite{Tesarova}, multiferroics\cite{Cheong2018}, superconductors\cite{Superconductor}, and the AQH state 
in graphene\cite{PhysRevLett.107.097402} and topological insulators\cite{PhysRevLett.105.057401}.
Here, we demonstrate the existence of giant polarization rotation in the electrodynamic response in tilted WSM with broken time reversal symmetry. Our results generalize the study of Ref.~[\onlinecite{Mehdi}] to include the impact of the tilt in type-I and type-II WSM. 

In WSM thin films, the giant Kerr rotation (GKR), polarization rotation in the reflected beam, originates from from the optical Hall conductivity including the AQH effect. 
We analytically calculate the full optical conductivity matrix $\sigma_{ij}(\omega)$ for tilted type-I and type-II WSM and find that all the diagonal components  $\sigma_{ii} \propto \omega$, while the off-diagonal Hall term $\sigma_{xy} \propto Q$, where $Q$ denotes the Weyl node separation (in the $z$-direction) and $\sigma_{xz} = \sigma_{yz} = 0$.  
For light incident on the WSM thin film surface without the Fermi arc states (perpendicular to the node separation), the Kerr angle  $\Theta_{\rm Kerr}\propto \sigma_{xy}/\sigma_{xx} \propto Q/\omega$. For realistic material parameters, the Kerr angle is found to be of the order of $10^{-1}$ radian  for optical frequencies below $10^{14}$ Hz. 
This is `giant' compared to the usually observed values of $10^{-6} - 10^{-4}$ radians in topological insulators, magnetic and other materials\cite{Mehdi,Tesarova,PhysRevLett.107.097402,Superconductor}.  For light incident on the surface with Fermi arc states (parallel to the node separation), we find that  $\Theta_{\rm Kerr} \propto \sigma_{yz}/\sigma_{zz} = 0$.

In bulk WSM large optical activity is predominantly caused by the axion electrodynamics induced changes in the Maxwell's equations. The effective dielectric constant of the modified Maxwell's equation can be expressed as $\epsilon_{ij}' = \epsilon_{ij} + \varepsilon_{ijk} Q_k \times 2i \alpha_F c/(\pi \omega)$, where the second term establishes $Q/\omega$ to be the effective `gyrotropy' constant, analogous to magnetization in magnetic systems.  This axion induced gyrotropy is what leads to optical activity in bulk WSM, even in the Pauli blocked regime with forbidden optical transitions. For light incident parallel to the Weyl node spearation (on surface without Fermi arc states), only circular eigenmodes are allowed in the WSM, leading to large circular bifriengence and circular dichroism, along with that of giant Kerr effect which is odd in $Q$. 
For light incident on surface perpendicular to the Weyl node separation (on surface with Fermi arc states), there is large 
linear bifriengence and dichrosim, along with a polarization angle dependent giant Kerr effect which is an even function of $Q$.

This paper is organized as follows: In Sec.~\ref{Model} we discuss the formalism for calculating optical conductivity for a generic two band
model, based on the optical Bloch equation. In Sec.~\ref{conductivity}, we obtain optical conductivity matrix for tilted type-I and type-II WSM, including the QAH contribution. In Sec.~\ref{application1} we discuss the electromagnetic response in thin films of tilted WSM, and demonstrate the existence of giant polarisation rotation. This is followed by a discussion of axion electrodynamics induced large optical activity in bulk WSM in Sec.~\ref{Application2}. Finally, we summarize our findings in Sec.~ \ref{summary}. 
%
\section{Light-matter interaction in WSM}\label{Model}
The low energy properties of a generic two band system can be described using the following $2 \times 2$ Hamiltonian,
\be\label{H_gen} 
 \hat H_0 = \sum_{\bs k}{\bf h}_{\bs k} \cdot \bsigma~.
\ee%
Here, $ {\bf h}_{\bs k} = (h_{0{\bs k}},h_{1{\bs k}}, h_{2{\bs k}}, h_{3{\bs k}})$ is a vector composed of real scalar functions of $\bs k$, and
$ \bsigma = (\openone_2, \sigma_x, \sigma_y, \sigma_z)$ is a vector composed of the identity and the three $2 \times 2$ Pauli matrices. 
The interaction with an  electro-magnetic field is modelled using the dipole approximation, $\hat H = \hat H_0 + e{\bf E}\cdot{\hat{\bf r}}$. 
Here $e$ is the electronic charge, and ${\bf E}$ is the electric field. 

Now, physical observables can be calculated using the optical Bloch equation for the dynamics of the density matrix \cite{Singh,Singh2018}. 
The corresponding inter-band contribution to the current (for a given $\bs k$) in the {\it linear response regime} is given by,
\be\label{current_fourier} 
{\bf J_k}(\omega) = -\frac{in_{\bs k}^{\rm eq}}{\hbar\omega_{\bs k}}
\left[\frac{\left({\bf E}\cdot{\bf M}^{vc}_{\bs k}\right){\bf M}^{cv}_{\bs k}}{\omega + \omega_{\bs k} + i\gamma} + 
\frac{\left({\bf E}\cdot{\bf M}^{cv}_{\bs k}\right){\bf M}^{vc}_{\bs k}}{\omega - \omega_{\bs k} + i\gamma}\right].
\ee
Here, $n_{\bs k}^{\rm eq} =  f(\epsilon^c_{\bf k},\mu) - f(\epsilon^v_{\bf k},\mu) $ is the equilibrium population difference between the conduction and the valance band and 
${\bf M}^{vc}_{\bs k} \equiv \langle \psi^{v}_{\bs k}| e\nabla_{\bs k}\hat H_0/\hbar |\psi^{c}_{\bs k}\rangle = ({\bf M}^{cv}_{\bs k})^*$ 
is the optical matrix element responsible for vertical transition between valence and conduction band. 
The transition energy is denoted by 
$\hbar\omega_{\bs k} = \varepsilon^c_{\bs k}-\varepsilon^v_{\bs k}$ and $\gamma$ is the phenomenological damping term for the interband coherence (off-diagonal elements of the density matrix).
The inter-band optical conductivity obtained from Eq. \eqref{current_fourier} in the limiting case of $\gamma \to 0$, is equivalent to the Kubo formula for 
a two band system. It is given by %
\bearr\nn\label{Kubo}
\sigma_{\alpha\beta}(\omega) &=& -\lim_{\gamma\to0}\sum_{{\bs k}} \frac{in_{\bs k}^{\rm eq}}{\hbar\omega_{\bs k}} \\
& & \times 
\left[\frac{\left({\bf M}^{vc}_{\bs k} \otimes {\bf M}^{cv}_{\bs k}\right)_{\alpha \beta}}{\omega + \omega_{\bs k} + i \gamma}
+ \frac{\left({\bf M}^{cv}_{\bs k} \otimes{\bf M}^{vc}_{\bs k}\right)_{\alpha \beta}}{\omega - \omega_{\bs k} + i \gamma}\right]
\eearr
with ${\otimes}$ denoting the outer product of the optical matrix element vectors.
 
Now let us consider a simple continuum model of a WSM with a pair of oppositely tilted Weyl nodes,  with chirality $\xi = \pm 1$ located in the Brilluoin zone  at ${\bs k} = \{0,0,\mp Q\}$. 
The low energy Hamiltonian of a Weyl node\cite{Zyuzin2016,PhysRevB.96.085114,PhysRevB.97.035144,PhysRevB.97.045150} of chirality $\xi$ 
can be expressed in the form of Eq.~\eqref{H_gen}, with the following mapping: 
$h_{0\bf k} = \hbar \xi v_t k_z^{\xi}$, $h_{1\bf k} = \xi\hbar v_F k_x$, $h_{2\bf k} = \xi\hbar v_F k_y$
and $h_{3\bf k} = \xi\hbar v_F k_z^{\xi}$, to yield  
\be\label{H_weyl}
\hat H_{\xi} =  \hbar \xi v_t k_z^{\xi} ~\openone + \xi \hbar v_F \left[k_x \sigma_x + k_y \sigma_y + k_z^{\xi} \sigma_z \right].
\ee
where $k_z^{\xi} = k_z + \xi Q$. Here, $v_t$ is the tilt velocity of the $\xi = +1$, Weyl node and $v_F$ is the Fermi velocity. The degree of tilt of the Weyl nodes is characterized by $\alpha_t = v_t/v_F$, with 
$|\alpha_t| < 1$ being a type-I Weyl node, and $|\alpha_t| > 1$ being a type-II Weyl node. 
Eigenvalues of Eq.~\eqref{H_weyl} are given by 
$\varepsilon^{\lambda}_{\bs k} = \hbar \xi v_t k_z^{\xi} + \lambda \hbar v_F k_{\xi},
$ where $\lambda =  1$ ($-1$) denotes the conduction (valance) band. Here, we have defined $k_\xi = [k_x^2 + k_y^2 + (k_z^{\xi})^2]^{1/2}$.  


The dimensionless optical matrix element corresponding to Eq.~\eqref{H_weyl} is given by $\tilde{\bf M}_{\bs k}^{vc} \equiv {\bf M}_{\bs k}^{vc}/{ev_F} $
\bearr \nn
&& = \left(i\sin\phi_{\bs k}-\xi \tilde{k}_z \cos\phi_{\bs k} , -i\cos\phi_{\bs k}-\xi \tilde{k}_z \sin\phi_{\bs k}, \xi\frac{k_{\perp}}{k_{\xi}}\right). \\
\eearr
Here, we have defined $\tilde{k}_z = k_z^{\xi} /{k_{\xi}}$,  $\phi_{\bs k} = \tan^{-1}(k_y/k_x)$ and $k_{\perp} = (k_x^2 + k_y^2)^{1/2}$. 
Note that $\tilde{\bf M}_{\bs k}^{vc}$ does not depend on the tilt velocity $v_t$ at all.

We now calculate the optical conductivity matrix of the tilted type-I and type-II WSM, in the next section. 
These will be used to calculate the polarization rotation (Kerr angle and ellipticity) of a linearly polarized optical beam reflected from WSM thin films, 
and from bulk WSM. In both cases, the polarization rotation ($\Phi$) and the ellipticity angle ($\Psi$) can be expressed in terms of a complex dimensionless quantity ($\chi$) via the relation \cite{Yoshino:13}, 
\be \label{Kerr_defn}
\tan(2 \Phi) = \frac{2 {\rm Re}[\chi]}{1 - |\chi|^2}~,~~{\rm and}~~~\sin(2 \Psi) = \frac{2 {\rm Im}[\chi]}{1 +|\chi|^2}~.
\ee
In the limiting case of $|\chi| \ll 1$, Eq.~\eqref{Kerr_defn}, can be simplified as $\Phi \approx {\rm Re}[\chi]$, and $\Psi \approx {\rm Im}[\chi]$. However unlike the case of magnetic materials\cite{MLD} where $|\chi| \ll 1$, 
for the case of WSM we find that in general $|\chi| \sim 1$, and the exact Eq.~\eqref{Kerr_defn} has to be used.


\section{Optical conductivity matrix of type-I and type-II WSM}\label{conductivity}

Calculating the optical conductivity of the tilted type-I and type-II WSM described by Eq.~\eqref{H_weyl}, we find that 
$\sigma_{xx} = \sigma_{yy}$, $\sigma_{xy} = -\sigma_{yx}$ 
and the remaining off-diagonal elements of the conductivity matrix are zero,  
$\sigma_{xz} = \sigma_{yz} = \sigma_{zx} = \sigma_{zy} = 0$.  
The vertical transitions responsible for the optical conductivities of type-I and type-II WSM, are shown in Fig.~\ref{fig1}(a), and 
(b), respectively. %
For simplicity, we will work in the zero temperature regime where the Fermi function can be replaced by the corresponding Heaviside step function.

\subsection{${\sigma}_{xx}(\Omega)$}

\begin{figure}[t!]
\includegraphics[width = \linewidth]{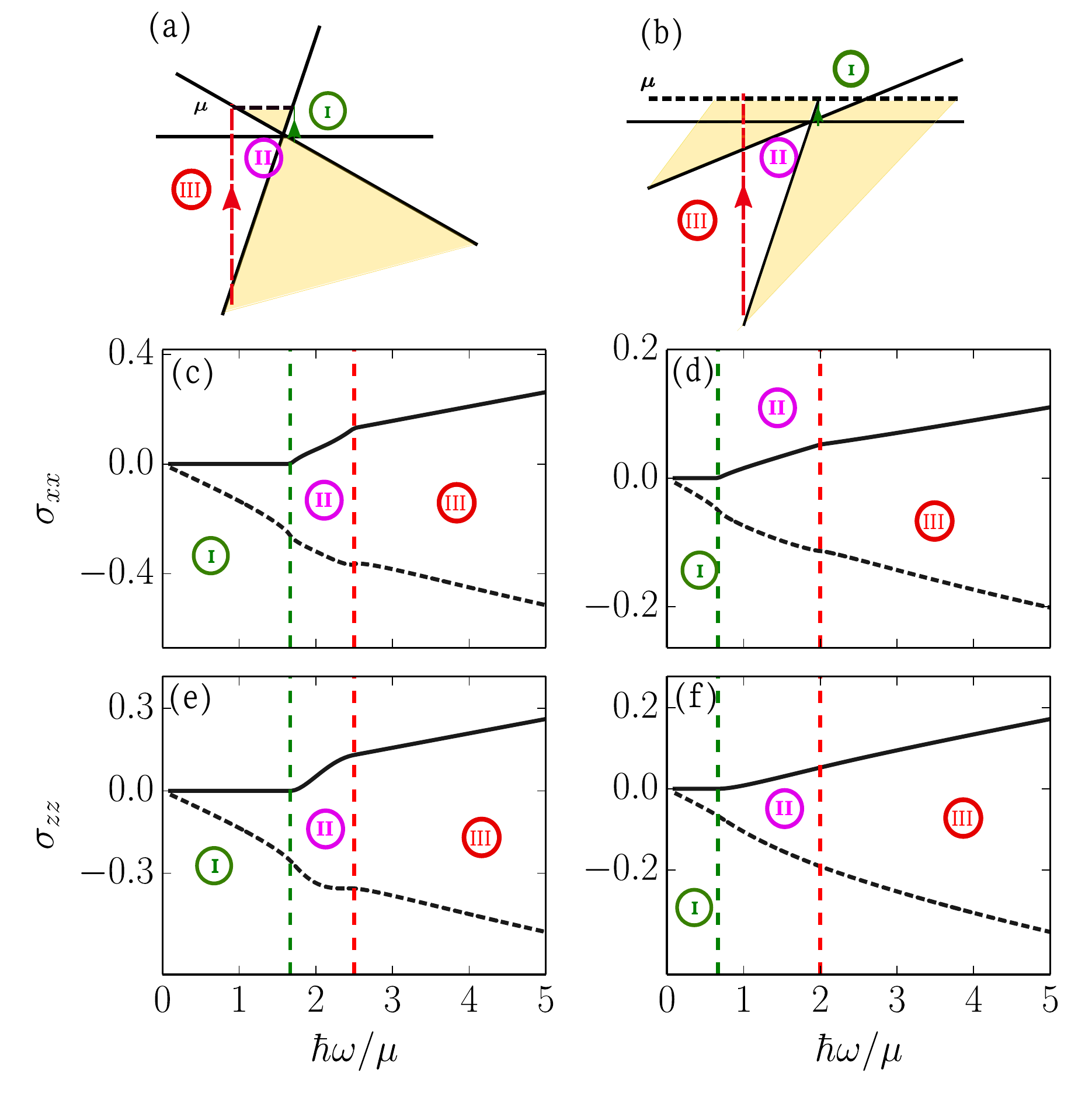}
\caption{(a)-(b) Schematic of the Pauli-blocking and allowed vertical optical transitions in a type-I and a type-II WSM, respectively. (c),(d) show  the real (solid line) and the imaginary part (dashed line) of the $\sigma_{xx}$, in a type-I ($\alpha_t = 0.2$) and a type-II ($\alpha_t = 2.0$) WSM, respectively. 
(e),(f) show  the real (solid line) and the imaginary part (dashed line) of $\sigma_{zz}$, in a type-I ($\alpha_t = 0.2$) and a type-II ($\alpha_t = 2.0$) WSM, respectively. The vertical lines mark the boundary values of transition energy, $\hbar \omega_l = 2 \mu/(1+|\alpha_t|)$(green), and $\hbar\omega_u ={2\mu}/(1-|\alpha_t|)$(red) or $\hbar\omega'_u ={2\mu}/(|\alpha_t| -1)$ (red) in a type-II WSM. 
Here, the conductivities are scaled in units of $\sigma_Q = e^2 Q/(\pi h)$, and we have chosen $\mu/(\hbar v_FQ)$ = 0.1.
}
\label{fig1}
\end{figure}

The real part of the longitudinal conductivity $\sigma_{xx}$ has been calculated in Ref.~[\onlinecite{PhysRevB.94.165111}], and our calculations (detailed in appendix~\ref{LOC_cal}) based on Eq.~\eqref{Kubo} reproduce their results. 
To start with, we calculate the real and imaginary part separately by using the Dirac identity on ~Eq.~\eqref{Kubo}. For a type-I WSM hosting a pair of oppositely tilted Weyl nodes with $|\alpha_t| < 1$, we obtain  
\be \label{Resigmaxx_tl1}
    {\rm Re}[{\sigma}_{xx}(\omega)] = \left\{\begin{array}{lr}
        0, & {\rm I}:~~ \omega < \omega_{l}\\
        \sigma_\omega(1/2 - \eta_1), & {\rm II}:~~ \omega_{l} < \omega < \omega_{u}\\
        \sigma_\omega & {\rm III}:~~ \omega > \omega_{u}~.
        \end{array}~\right.
\ee
Here we have defined a conductivity scale set by the optical frequency: $\sigma_\omega = e^2\omega/(6 h v_F)$. 
The photon energy bounds are $\hbar \omega_{l} = {2{\mu}}/(1 + |\alpha_t|)$, $\hbar \omega_{u} = {2{\mu}}/(1 - |\alpha_t|)$, and 
\be
\eta_1 = \frac{3}{8|\alpha_t|}\left(\frac{2{\mu}}{\hbar\omega}-1\right)\left[ 1 + \frac{1}{3\alpha_t^2}\left(\frac{2{\mu}}{\hbar \omega}-1\right)^2\right].
\ee
%
Basically, in region I for $\omega < \omega_l$, the phase space for vertical transitions is Pauli blocked, as shown in Fig.~\ref{fig1}. For region II, $\omega \in (\omega_l,\omega_u$), the Pauli blocking gets removed with increasing energy resulting in finite vertical transitions and finite ${\rm Re}[\sigma_{xx}(\omega)]$. In the $\alpha_t \to 0$ limit we have $\omega_l \to \omega_u = 2 \mu/\hbar$, and we have ${\rm Re}[{\sigma}_{xx}(\Omega)] = \sigma_\omega$ for $\hbar \omega \ge 2 \mu$. In the other limit $\alpha_t \to 1$, we have 
$\omega_u \to \infty$ with  the region II extending to higher energies. 

For the type-II Weyl node with $|\alpha_t| > 1$, we obtain 
\be \label{Resigmaxx_tl2}
    {\rm Re}[{\sigma}_{xx}(\omega)] = \left\{\begin{array}{lr}
        0,& {\rm I}:~~\omega < \omega_{l}\\
        \sigma_\omega(1/2 - \eta_1), & {\rm II}:~~ \omega_{l} < \omega < \omega'_{u}\\
        \sigma_\omega\eta_2, & {\rm III}:~~ \omega > \omega'_{u}~.
        \end{array}\right.
\ee
Here we have defined  $\hbar \omega'_{u} = {2{\mu}}/(|\alpha_t| - 1)$, and 
\be
\eta_2 = \frac{3}{4|\alpha_t|}\left[ 1 + \frac{1}{3\alpha_t^2} + \left(\frac{2{\mu}}{{\alpha_t}\hbar \omega}\right)^2\right].
\ee

The imaginary part of the longitudinal optical conductivity ${\sigma}_{xx}(\omega)$, for a type-I WSM, for $|\alpha_t|<1$ is calculated to be 
\begin{widetext}
\bearr \label{ImSxx}
\frac{{\rm Im}[\sigma_{xx}]}{\sigma_{\omega}} & = & \frac{-1}{4 \pi}\bigg( \tau(\alpha_t) \log\bigg[\frac{|\omega_u^2 - \omega^2|}{|\omega_l^2 -\omega^2|}\bigg] + \frac{8}{\alpha_t^2}\bigg(\frac{\mu}{\hbar\omega}\bigg)^2 - \bigg(\frac{\mu}{\hbar\omega}\bigg)^3\Pi(\omega,\alpha_t,\mu)\log\bigg[\frac{|\omega_u-\omega|(\omega_l+\omega)}{|\omega_l-\omega|(\omega_u+\omega)}\bigg] \nn \\
& + &   \frac{6}{|\alpha_t|^3}\bigg(\frac{\mu}{\hbar\omega}\bigg)^2 \log\bigg[\frac{|\omega_u^2-\omega^2|\omega_l^2}{|\omega_l^2-\omega^2|\omega_u^2}\bigg] + 4 \log\bigg[\frac{|\omega_c^2-\omega^2|}{|\omega_u^2-\omega^2|}\bigg]\bigg)~.
\eearr
Here $\omega_c \equiv  v_F k_c$ is the ultra-violet cutoff frequency (see appendix \ref{LOC_cal} for details). For a type-II WSM, with $|\alpha_t|>1$, we obtain 
\bearr \label{ImSxx1}
\frac{{\rm Im}[\sigma_{xx}]}{\sigma_{\omega}} & = & \frac{-1}{4 \pi}\bigg( \tau(\alpha_t) \log\bigg[\frac{|\omega_u'^2 - \omega^2|}{|\omega_l^2 -\omega^2|}\bigg] + \frac{8}{\alpha_t^3}\bigg(\frac{\mu}{\hbar\omega}\bigg)^2 - \bigg(\frac{\mu}{\hbar\omega}\bigg)^3\Pi(\omega,\alpha_t,\mu)\log\bigg[\frac{|\omega_u'-\omega|(\omega_l+\omega)}{|\omega_l-\omega|(\omega_u'+\omega)}\bigg]  \\
& + & \frac{6}{|\alpha_t|^3}\bigg(\frac{\mu}{\hbar\omega}\bigg)^2 \log\bigg[\frac{|\omega_u'^2-\omega^2|\omega_l^2}{|\omega_l^2-\omega^2|\omega_u'^2}\bigg] + \bigg(\frac{3}{|\alpha_t|} + \frac{1}{|\alpha_t|^3}\bigg) \log\bigg[\frac{|\omega_c^2-\omega^2|}{|\omega_u'^2-\omega^2|}\bigg]\bigg) + \frac{12}{|\alpha_t|^3}\bigg(\frac{\mu}{\hbar \omega}\bigg)^2\log\bigg[\frac{|\omega_c^2-\omega^2|\omega_u'^2}{|\omega_u'^2-\omega^2|\omega_c^2}\bigg]\bigg)~. \nn
\eearr
\end{widetext}
In Eqs.~\eqref{ImSxx}-\eqref{ImSxx1}, we have defined the following functions:
\be
\tau(\alpha_t) = \frac{1}{2}\bigg(4 + \frac{1}{|\alpha_t|^3} + \frac{3}{|\alpha_t|}\bigg)~,
\ee
\be
\Pi({\mu},\omega,\alpha_t) = \frac{4}{|\alpha_t|^3} + 3\bigg(\frac{\hbar\omega}{\mu}\bigg)^2\bigg(\frac{1}{|\alpha_t|^3} + \frac{1}{|\alpha_t|}\bigg)~.
\ee
In the limiting case of $\alpha_t \to 0$, Eq.~\eqref{ImSxx} reduces to 
\be\label{w_0_im}
{\rm Im}[{\sigma}_{xx}(\omega)] = -\frac{\sigma_\omega}{\pi}\ln \bigg|\frac{\omega_c^2-\omega^2}{\omega^2-4{(\mu/\hbar)}^2}\bigg|~.
\ee
In the intrinsic limit of $\mu \to 0$, we obtain
\be
   \lim_{{\mu}\to 0} {\rm Im}[{\sigma}_{xx}(\omega)] =  \frac{-\sigma_\omega}{\pi} \ln \bigg|\frac{\omega_c^2-\omega^2}{\omega^2}\bigg| \left\{\begin{array}{lr}
       1, & {\rm Type~I~~~}\\
      \frac{1+3\alpha_t^2}{2 \alpha_t^3} & {\rm Type~II~.}\\
        \end{array}\right.
\ee
Of these, the type-I result is independent of the tilt, and has been derived earlier in Refs.~[\onlinecite{Mehdi,Chen2018}].

The real, and the imaginary part of ${\sigma}_{xx}(\omega)$ is shown in Fig.~\ref{fig1}(c) and (d). The tilt forces only selective region of the momentum space to be available for vertical transitions in region II, leading to a discontinuity in the optical response at the onset of region III.

\subsection{${\sigma}_{zz}(\omega)$}
We calculate calculate ${\sigma}_{zz}(\omega)$ in a similar manner (see appendix \ref{LOC_cal} for details of the calculation). For the case of type-I WSM with $|\alpha_t|<1$, the real part is given by 
\be\label{Resigmazz_tl1}
   {\rm Re}[{\sigma}_{zz}(\omega)] = \left\{\begin{array}{lr}
        0,& {\rm I:}~~~~ \omega < \omega_{l}\\
       \sigma_\omega\eta_4, & {\rm II:}~~~~ \omega_{l} < \omega < \omega_{u}\\
        \sigma_\omega, & {\rm III:}~~~~ \omega > \omega_{u}~.
        \end{array}\right.
\ee
Here, we have defined 
\be
\eta_4 = \frac{1}{2} + \frac{(2{\mu}-\hbar\omega)^3}{4\hbar^3\omega^3 |\alpha_t|^3} + \frac{1}{|\alpha_t|}\bigg(\frac{3}{4} - \frac{3{\mu}}{2\hbar\omega}\bigg)~.
\ee
For the type-II WSM, with $|\alpha_t| > 1$, we obtain\cite{PhysRevB.94.165111},
\be \label{Resigmazz_tl2}
   {\rm Re}[{\sigma}_{zz}(\Omega)] = \left\{\begin{array}{lr}
        0,& {\rm \rm I:}~~~~ \omega < \omega_{l}\\
         \sigma_\omega\eta_4, & {\rm II:}~~~~ \omega_{l} < \omega < \omega'_{u}\\
         \sigma_\omega\eta_5, & {\rm III:}~~~~\omega > \omega'_{u}~.
        \end{array}\right.\,
\ee
Here, we have defined 
\be
\eta_5 = \frac{-1}{2|\alpha_t|^3}\left[ 1 - {3\alpha_t^2} + \frac{12{\mu}^2}{\hbar^2\omega^2}\right].
\ee

The imaginary part of $\sigma_{zz}$, for a type-I WSM is calculated to be 
\begin{widetext}
\bearr \nn
\frac{{\rm Im}[\sigma_{zz}]}{\sigma_{\omega}} & = & \frac{-1}{2\pi}\bigg( \tau'(\alpha_t) \log\bigg[\frac{|\omega_u^2 - \omega^2|}{|\omega_l^2 -\omega^2|}\bigg] - \frac{8}{\alpha_t^2}\bigg(\frac{\mu}{\hbar\omega}\bigg)^2 + \bigg(\frac{\mu}{\hbar\omega}\bigg)^3\Pi^{\prime}({\mu},\omega,\alpha_t)\log\bigg[\frac{|\omega_u-\omega|(\omega_l+\omega)}{|\omega_l-\omega|(\omega_u +\omega)}\bigg] \\ 
& - & \frac{6}{|\alpha_t|^3}\bigg(\frac{\mu}{\hbar\omega}\bigg)^2 \log\bigg[\frac{|\omega_u^2-\omega^2|\omega_l^2}{|\omega_l^2-\omega^2|\omega_u^2}\bigg]  + 2 \log\bigg[\frac{|\omega_c^2-\omega^2|}{|\omega_u^2-\omega^2|}\bigg] \bigg)~.
\eearr
For a type-II WSM with $|\alpha_t|>1$, we obtain 
\bearr
\frac{{\rm Im}[\sigma_{zz}]}{\sigma_{\omega}} & = & \frac{-1}{2\pi}\bigg( \tau'(\alpha_t) \log\bigg[\frac{|\omega_u'^2 - \omega^2|}{|\omega_l^2 -\omega^2|}\bigg] - \frac{8}{\alpha_t^3}\bigg(\frac{\mu}{\hbar\omega}\bigg)^2 + \bigg(\frac{\mu}{\hbar\omega}\bigg)^3\Pi^{\prime}({\mu},\omega,\alpha_t)\log\bigg[\frac{|\omega_u'-\omega|(\omega_l+\omega)}{|\omega_l-\omega|(\omega_u' +\omega)}\bigg] \\ 
& -&   \frac{6}{|\alpha_t|^3}\bigg(\frac{\mu}{\hbar\omega}\bigg)^2 \log\bigg[\frac{|\omega_u'^2-\omega^2|\omega_l^2}{|\omega_l^2-\omega^2|\omega_u'^2}\bigg] + \bigg(\frac{-1}{\alpha_t^3} + \frac{3}{\alpha_t}\bigg)\log\bigg[\frac{|\omega_c^2-\omega^2|}{|\omega_u'^2 - \omega^2|}\bigg] - \frac{12}{|\alpha_t|^3}\bigg(\frac{\mu}{\hbar\omega}\bigg)^2\log\bigg[\frac{|\omega_c^2-\omega^2|\omega_u'^2}{|\omega_u'^2 - \omega^2|\omega_c^2}\bigg]\bigg)~. \nn
\eearr
\end{widetext}
Here, we have defined
\begin{equation}
\tau'(\alpha_t) = \frac{1}{2}\bigg(2 - \frac{1}{|\alpha_t|^3} + \frac{3}{|\alpha_t|}\bigg)~,
\end{equation}
\be
\Pi^{\prime}({\mu},\omega,\alpha_t) = \frac{4}{|\alpha_t|^3} + 3\bigg(\frac{\hbar\omega}{\mu}\bigg)^2\bigg(\frac{1}{|\alpha_t|^3} - \frac{1}{|\alpha_t|}\bigg)~.
\ee
The real and the imaginary parts of $\sigma_{zz}(\omega)$ are displayed in panels (e) and (f) of Fig.~\ref{fig1}. We recall that region I is completely Pauli blocked while region II is partially Pauli blocked for vertical optical transitions. 

\subsection{${\sigma}_{xy}(\omega)$}
A finite transverse optical conductivity $\sigma_{xy}(\omega)$ generally originates from the breaking of the time-reversal symmetry. In the case of TRS broken WSM, this manifests in the separation (in momentum space) of the two Weyl nodes of opposite chirality. Thus the energy scale of the transverse optical response is dictated by the inter-node separation, $\epsilon_Q = \hbar v_F Q$. 

\begin{figure}[t!]
\includegraphics[width = \linewidth]{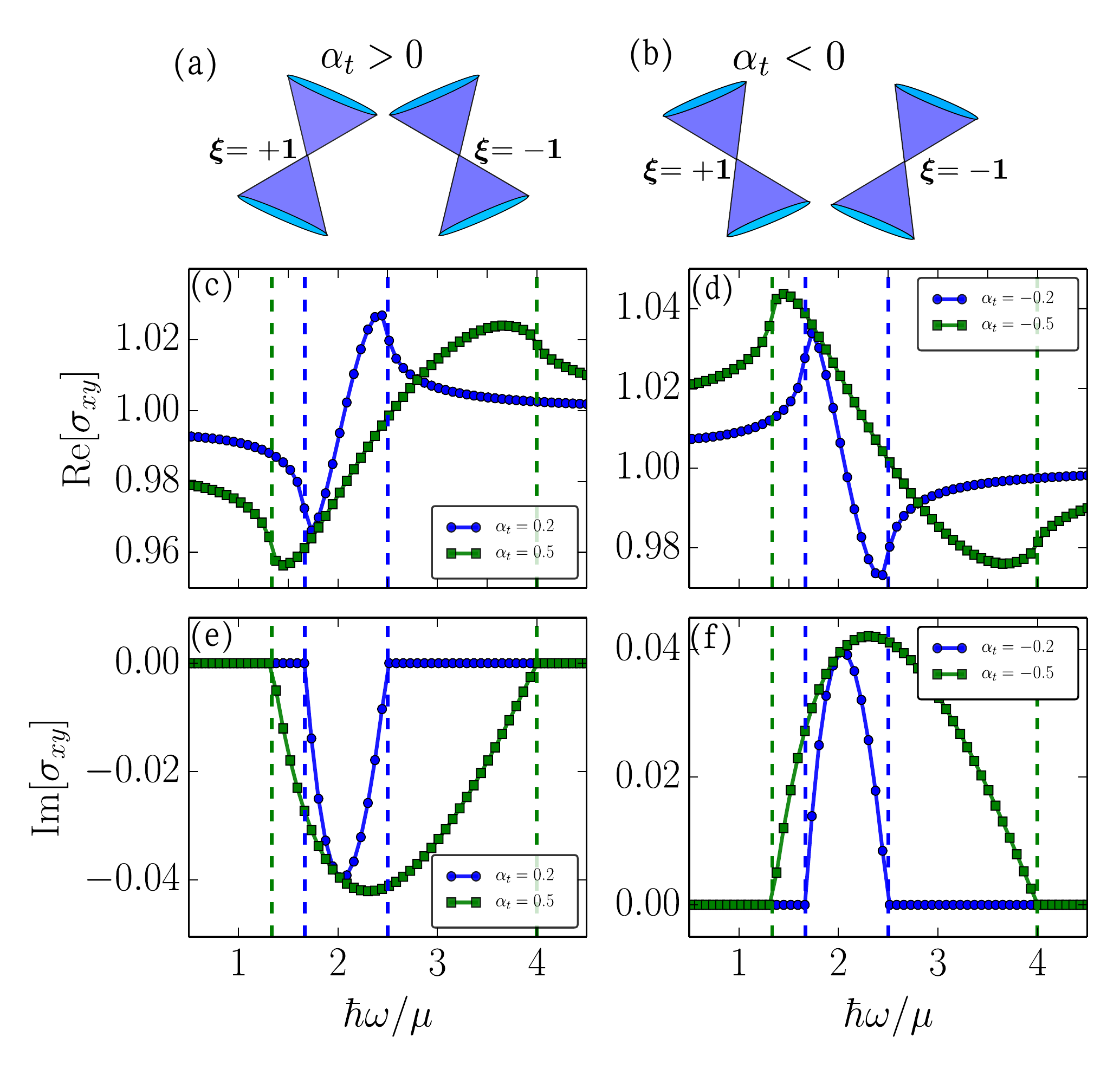}
\caption{(a), (b) The relative orientation of a pair of Weyl nodes for $\alpha_t > 0$ and for $\alpha_t < 0$, respectively. 
The sign of $\alpha_t$ becomes important in $\sigma_{xy}$. (c), (d) Real part of $\sigma_{xy}$ for $\alpha_t > 0$, and  $\alpha_t < 0$, respectively. 
Note that $\sigma_{xy}$ has a finite DC component, denoting the presence of quantum anomalous Hall conductivity. 
(e), (f), Imaginary parts of $\sigma_{xy}$ for $\alpha_t > 0$, and  $\alpha_t < 0$, respectively. Evidently, ${\rm Im}[ \sigma_{xy}] \propto {\rm sign}(\alpha_t)$.
Here, the conductivities are scaled in units of $\sigma_Q = e^2 Q/(\pi h)$, and other parameters are identical to those of Fig.~\ref{fig1}.
}
\label{fig2}
\end{figure}

The imaginary part of the dynamical transverse optical conductivity has already been calculated in Ref.~[\onlinecite{PhysRevB.96.085114}], and our calculations (see appendix \ref{TOC_cal} for details) reproduce their results. 
For a type-I WSM hosting a pair of oppositely tilted Weyl nodes with $|\alpha_t| < 1$, we obtain
\be \label{Imxy1}
    {\rm Im}[{\sigma}_{xy}(\omega)] = {\rm sign}(\alpha_t)\left\{\begin{array}{lr}
        0,& {\rm I}:~~ \omega < \omega_{l}\\
        3 \sigma_\omega\eta_3, & {\rm II}:~~\omega_{l} < \omega < \omega_{u}\\
        0, & {\rm III}:~~ \omega > \omega_{u}~.
        \end{array}\right.\,
\ee
Here we have defined 
\be
\eta_3 =  \frac{1}{\alpha_t^2}\bigg(\frac{1}{8}-\frac{{\mu}}{2 \hbar \omega} + \frac{{\mu}^2}{2 \hbar^2\omega^2} \bigg) - \frac{1}{8}~.
\ee 

For a type-II WSM with a pair of oppositely tilted Dirac nodes with $|\alpha_t| > 1$, we derive  
\be \label{Imxy2}
    {\rm Im}[{\sigma}_{xy}(\omega)] = {\rm sign}(\alpha_t)\left\{\begin{array}{lr}
        0,& {\rm I}:~~  \omega < \omega_{l}\\
         {3 \sigma_\omega\eta_3}, & {\rm II}:~~ \omega_{l} < \omega < \omega'_{u}\\
        \frac{-3{\mu}\sigma_\omega}{\hbar \omega \alpha_t^2}, &  {\rm III}:~~ \omega > \omega'_{u}~.
        \end{array}\right.\,
\ee
%

The real part of $\sigma_{xy}(\omega)$ can now be obtained by using Kramers Kronig relations on ${\rm Im}[{\sigma}_{xy}(\omega)]$. In general we find that 
${\rm Re}[\sigma_{xy}(\omega)] = {\rm Re}[\sigma_{xy}^{dc}] + {\rm Re}[\sigma_{xy}^{ac}]$, where ${\rm Re}[\sigma_{xy}^{ac} (\omega \to 0)] \to 0$ and 
${\rm Re}[\sigma_{xy}^{dc}] (\omega = 0)$ is finite. 
%
For the case of a type-I WSM, we obtain the anomalous Hall component, %
\be \label{Rxy_T1} %
{\rm Re}[\sigma_{xy}^{dc}] = \sigma_Q + \sigma_\mu \left[ \frac{2}{\alpha_t} + \frac{1}{\alpha_t^2} \log\left(\frac{1-\alpha_t}{1+\alpha_t}\right) \right]. 
\ee 
Here we have defined a chemical potential based  conductivity scale $\sigma_\mu = e^2 \mu/(h^2 v_F)$ and a node separation based conductivity scale $\sigma_Q = e^2Q/(\pi h)$. 
A similar calculation for the type-II WSM for the anomalous Hall component, leads to 
\be \label{RxyT2}
{\rm Re}[\sigma_{xy}^{dc}] = \dfrac{\sigma_Q}{|\alpha_t|} + \dfrac{\rm{sign(\alpha_t)}\sigma_{\mu}}{\alpha_t^2}\log\left[\dfrac{{\mu}^2}{\hbar^2 \omega_c^2 \alpha_t^2\left(\alpha_t^2-1\right)}\right]~. 
\ee

The AC component for a tilted type-I WSM is given by 
\begin{widetext}
\bearr \label{IxyT1}
& & {\rm Re}[\sigma_{xy}^{ac}] = {\rm sign}(\alpha_t)~{\sigma_\mu} \left\{ \frac{{-1}}{2\alpha_t^2}\log\bigg[\frac{|\omega_u^2 - \omega^2|}{|\omega_l^2 - \omega^2|} \frac{\omega_l^2}{\omega_u^2}\bigg] 
 + \bigg(\frac{{\mu}}{2\hbar\omega \alpha_t^2} + \frac{\hbar\omega}{8 \mu}\frac{1 -\alpha_t^2}{\alpha_t^2}\bigg) 
 \log\bigg[\frac{|\omega_u - \omega|(\omega_l + \omega)}{|\omega_l - \omega|(\omega_u + \omega)}\bigg] - \frac{1}{|\alpha_t|}\right\}~. 
\eearr
The corresponding AC component for a type-II WSM with a pair of oppositely tilted WSM nodes is given by 
\bearr \label{IxyT2}
{{\rm Re}[\sigma_{xy}^{ac}]} & = & {\rm sign}(\alpha_t)~ {\sigma_\mu}\left\{ \frac{{-1}}{2\alpha_t^2}\log\bigg[\frac{(\omega_c^2 - \omega^2)^2}{|\omega_l^2 - \omega^2| |\omega_{u'}^2 - \omega^2|}\frac{\omega_l^2 \omega_{u'}^2}{\omega_c^4}\bigg] 
 + \bigg(\frac{{\mu}}{2\hbar\omega \alpha_t^2} + \frac{\hbar\omega}{8 \mu}\frac{1 -\alpha_t^2}{\alpha_t^2}\bigg) \log\bigg[\frac{|\omega_{u'} - \omega|(\omega_l + \omega)}{(\omega_{u'} + \omega)|\omega_l - \omega|}\bigg] - \frac{2}{\alpha_t^2}
 \right\}. \nn \\
\eearr
\end{widetext}

\begin{figure}[t!]
\includegraphics[width = \linewidth]{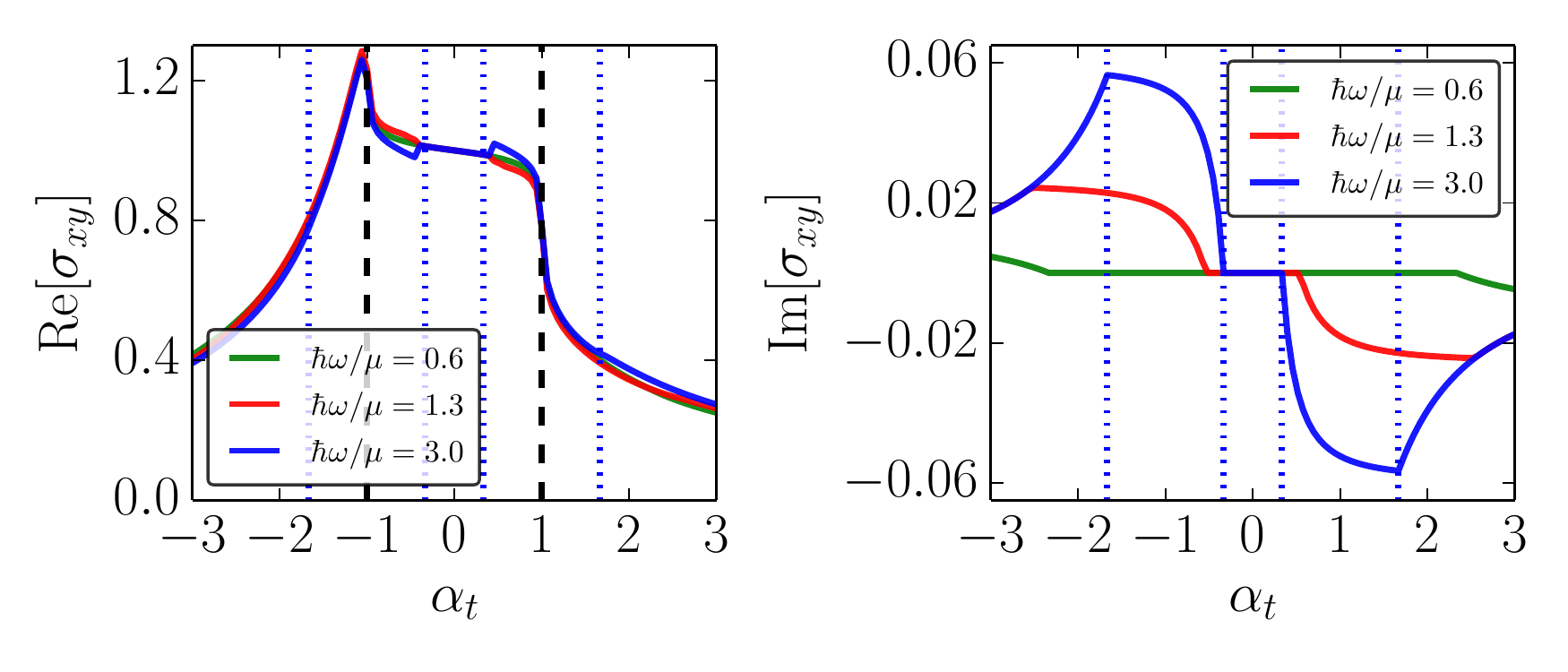}
\caption{The tilt ($\alpha_t$) dependence of the (a) real and (b) imaginary part of $\sigma_{xy}$ for different optical frequencies. 
The vertical black dashed lines simply mark $\alpha_t = \pm 1$. The vertical blue lines mark the boundary values for transition frequencies for region I and II, {\it i.e.}, $2 \hbar^{-1}\mu/(|\alpha_t| \pm 1)$ for $\hbar \omega/\mu = 3.0$. 
The curve for $\hbar \omega/\mu = 0.6$, lie completely within the Pauli blocked region, and thus they show only the QAH contribution. Clearly 
the behaviour of the QAH term changes significantly across the Lifshitz transition ($|\alpha_t| = \pm 1$) line, dividing the type-I and type-II regions.  All other parameters are identical to those of Fig.~\ref{fig2}, and the conductivities are in units of $\sigma_Q$. 
}
\label{fig3}
\end{figure}

The first term anomalous Hall conductivity in Eqs.~\eqref{Rxy_T1} and \eqref{RxyT2} is $\propto \sigma_Q$ and it denotes the `intrinsic contribution' which survives even if $\mu \to 0$. All other terms in the DC as well as the AC component of the off-diagonal conductivity ($\propto \sigma_\mu$)  are  `extrinsic contributions' which vanish in the limit 
$\mu \to 0$. The real and the imaginary part of $\sigma_{xy}(\omega)$ for a type-I WSM is shown in Fig.~\ref{fig2}, as a function of the optical frequency. The presence of the QAH contribution in the real part of $\sigma_{xy}$ is evident. The imaginary part of $\sigma_{xy}$ in Fig.~\ref{fig2}, reverses sign on changing the tilt orientation ($\alpha_t \to - \alpha_t$). Similar behaviour is also observed in the corresponding plots for type-II WSM (not shown here).
The tilt dependence of $\sigma_{xy}(\omega)$ is highlighted in Fig.~\ref{fig3}. In Fig.~\ref{fig3}(a), the dominant contribution comes from the AQH part and consequently the curves for different frequencies are very close to each other. The intrinsic part of the QAH conductivity ($\propto \sigma_Q$) in turn dominates in type-I WSM, while in type-II WSM the $\alpha_t$ dependent extrinsic part of the QAH ($\propto \sigma_\mu$) also contributes equally. 
   
Having obtained the full conductivity matrix, we now discuss the polarization rotation of a reflected beam in tilted type-I and type-II WSM, starting with the case of thin films in   the next section. The finite off-diagonal components of the conductivity matrix ($\sigma_{xy}$) will play an important role in the polarization rotation of the reflected (and transmitted) light and the corresponding ellipticity angle in WSM thin films. 

\section{Kerr rotation in thin films of WSM}\label{application1}

In this section, we consider an ultra-thin film of Weyl semimetal such that the thickness of the film ($d$) is much larger than the atomic separation ($a$) while being smaller than the wavelength ($\lambda$) of light, {\it i.e.},   $a \ll d \ll \lambda$. In these conditions the WSM film can be treated as a 2D surface, as far as its interaction with light is concerned. Thus the polarization  angle rotation can be obtained simply by matching the electromagnetic boundary conditions on the either sides of the thin film. 

The polarization angle $\Phi_{\rm Kerr}$, and the azimuth of the major axes of the polarization ellipse $\Psi_{\rm Kerr}$ (ellipticity angle) of the reflected beam, are given by Eq.~\eqref{Kerr_defn}. We need to distinguish between the two cases of the incident beam being $s$ (polarization $\parallel$ to the plane of incidence) or $p$ (polarization $\perp$ to the plane of incidence) polarized. For the case of a reflected beam, 
the corresponding dimensionless complex quantities ($\chi$ of Eq.~\eqref{Kerr_defn}) are defined as 
\be\label{Kerr}
\chi_{\rm Kerr}^s = \frac{r_{ps}}{r_{ss}}, ~~~~{\rm and}~~~~ \chi_{\rm Kerr}^p = -\frac{r_{sp}}{r_{pp}}~, 
\ee
where $r_{ij}$ with $(i,j)\in (s,p)$, are the corresponding reflection coefficients \cite{Yoshino:13}. For thin films ($d \ll \lambda$), these reflection coefficients in turn, depend on the {\it surface} conductivity matrix. In the case of a WSM, 
the surface conductivity matrix depends on the incident surface. For example, in a time-reversal symmetry broken WSM, the 
surface conductivity has finite off diagonal terms for the surface without Fermi arc states (surfaces $\perp {\bs Q}$) while it is diagonal 
for the surface hosting Fermi arcs states (surfaces $\parallel {\bs Q}$). 
We consider these two cases separately in the following subsections.

\subsection{Incidence on surface without Fermi arcs states ($\perp {\bf Q}$)} 

Let us start with the case where the linearly polarized incident beam propagates (say in direction $\mathbf{\hat{n}}$) parallel to $\bs{Q}$. 
In this case, both $\mathbf{\hat{n}}$ and ${\bs Q}$ are in $z$ direction and the electric polarization is in the $x-y$ plane, and thus the transverse conductivity $\sigma_{xy}$ will come into play.  

Lets assume that the interface of air and WSM thin film is located at the $z=0$ plane. The wave-vectors for initial, reflected and transmitted beam are,
\bearr\nn
{\bf k}_i &=& (0, k_i\sin\theta_i, k_i\cos\theta_i),\\\nn
{\bf k}_r &=& (0, k_r\sin\theta_r, -k_r\cos\theta_r),\\\nn
{\bf k}_t &=& (0, k_t\sin\theta_t, k_t\cos\theta_t).\nn
\eearr
Similarly the components of electric field ${\bf E}$ are, 
\bearr\nn
{\bf E}_i &=& (E_i^s, E_i^p\cos\theta_i, -E_i^p\sin\theta_i)e^{({\bf k_i}\cdot{\bf r}-\omega_it)},\\\nn
{\bf E}_r &=& (E_r^s, E_r^p\cos\theta_r, E_r^p\sin\theta_r)e^{({\bf k_r}\cdot{\bf r}-\omega_rt)},\\\nn
{\bf E}_t &=& (E_t^s, E_t^p\cos\theta_t, -E_t^p\sin\theta_t)e^{({\bf k_t}\cdot{\bf r}-\omega_tt)}.\nn
\eearr
The magnetic field ${\bf B}$ components can be simply obtained via  ${\bf B} = (n/c){\bf \hat k}\times{\bf E}$, with $n$ denoting the refractive index of the medium.
The fields on the two sides of the WSM thin film, are connected by the Maxwell's boundary condition: 
\be \label{BC}
 {\bf E}_1^{||} = {\bf E}_2^{||},~~~~{\rm and}~~~~~~~\bf{\hat{n}} \times \Big(\frac{{\bf B}_1^{||}}{\mu_1}-\frac{{\bf B}_2^{||}}{\mu_2}\Big) = {\bf J}~,
\ee
where ${\bf E}_1 = {\bf E}_i + {\bf E}_r$, ${\bf E}_2 = {\bf E}_t$, ${\bf B}_1 = {\bf B}_i + {\bf B}_r$, ${\bf B}_2 = {\bf B}_t$
and ${\bf J}$ is the surface current density generated by the incident electric field. The surface current density can be expressed in terms of the surface conductivity matrix, 
$J_i = \sigma^s_{ij} E_j$. The surface conductivity for a thin film of thickness $d$ can be approximated as $\sigma^s_{ij} = d~\sigma_{ij}$, with $\sigma_{ij}$ denoting the bulk conductivity \cite{Mehdi}.   

In general, the reflected electric field in the $p$- and $s-$ directions can be expressed as 
\be
\begin{pmatrix}
 E_r^p\\
 E_r^s\\
\end{pmatrix} = \begin{pmatrix}
                r_{pp} & r_{ps}\\
                r_{sp} & r_{ss}\\
                \end{pmatrix} 
\begin{pmatrix}
 E_i^p \\
 E_i^s\\
 \end{pmatrix}~.
\ee
Using this in Eq.~\eqref{BC}, we obtain the reflection coefficients to be,  
\be \label{eq:rsp}
r_{sp} = \left. \frac{E_r^s}{E_i^p}\right|_{E_i^s = 0} = \frac{2n_i}{c\mu_i}\frac{\sigma_{xy}^s\cos\theta_i\cos\theta_t}{\left(\sigma_{xy}^s\sigma_{yx}^s\cos\theta_i\cos\theta_t - \sigma_1^s\sigma_2^s\right)}~,
\ee 
\be \label{eq:tss}
r_{ss} = \left. \frac{E_t^s}{E_i^s}\right|_{E_i^p = 0} =  -1 + \frac{2n_i}{c\mu_i}\left[\frac{\sigma_2^s\cos\theta_i}{\sigma_1^s\sigma_2^s-\sigma_{xy}^s\sigma_{yx}^s\cos\theta_i\cos\theta_t}\right]~,
\ee
along with, 
\be \label{eq:tpp}
r_{pp} =  \left. \frac{E_t^p}{E_i^p}\right|_{E_i^s = 0} -1 + \frac{2n_i}{c\mu_i}\left[\frac{\sigma_1^s\cos\theta_t}{\sigma_1^s\sigma_2^s-\sigma_{xy}^s\sigma_{yx}^s\cos\theta_i\cos\theta_t}\right]~,
\ee 
and 
\be \label{eq:rps}
{r_{ps}}= \left. \frac{E_r^p}{E_i^s}\right|_{E_i^p = 0} = \frac{2n_i}{c\mu_i}\frac{\sigma_{yx}^s\cos\theta_i\cos\theta_t}{\left(\sigma_{xy}^s\sigma_{yx}^s\cos\theta_i\cos\theta_t - \sigma_1^s\sigma_2^s\right)}.
\ee
Here, we have defined  $\sigma_1^s \equiv n_i\cos\theta_i/(c\mu_i) + n_t\cos\theta_t/(c\mu_t) + \sigma_{xx}^s$ and 
$\sigma_2^s \equiv n_i\cos\theta_t/(c\mu_i) + n_t\cos\theta_i/(c\mu_t) + \sigma_{yy}^s\cos\theta_i\cos\theta_t.$ 

Clearly, the reflectivity coefficients that rotate the incoming polarization, $r_{sp}$ and $r_{ps}$, are proportional to the optical Hall conductivity and they 
vanish in the limit of $\sigma_{xy}^s \to 0$. In our case we find that $\sigma_{xy}^s = - \sigma_{yx}^s$ and thus $r_{sp} = -r_{ps}$. Furthermore for normal incidence we have $\theta_i = \theta_t = 0$ and thus $\sigma_1 = \sigma_2$ and  
$r_{ss} = r_{pp}$. As a consequence, for the case of normal incidence, we have $\Phi_s = \Phi_p$ \cite{Cserti}. 
As a consistency check, we note that for an isotropic WSM without any tilt, and for the case of normal incidence $(\theta_i = \theta_t = 0)$, we recover the results of 
Ref.~[\onlinecite{Mehdi}]. 


%

%
%
%

For the case of normal incidence, using Eq.~\eqref{eq:rsp}-\eqref{eq:rps}, in Eq.~\eqref{Kerr}, we obtain 
\begin{equation}\label{kerr_angle}
\chi_{\rm Kerr}^p = -\frac{\sigma_{xy}}{\sigma_{xx}}\left[1 +\frac{d \sigma_{xx}}{2 c\epsilon_0} \left( 1+ \frac{\sigma_{xy}^2}{\sigma_{xx}^2}\right) \right]^{-1}~. 
\end{equation}
This establishes that $\chi^p_{\rm Kerr} \propto Q$, to lowest order in $Q$, as $\sigma_{xy} \propto Q$, and $\sigma_{xx}$ is independent of ${Q}$.  
The polarization rotation is primarily determined by the ratio of $\sigma_{xy}/\sigma_{xx}$, and we get giant polarization rotation for 
$\sigma_{xy}/\sigma_{xx}\approx 1$, or alternately $\sigma_{Q}/\sigma_\omega  = 6 Q v_F/(\pi \omega) \approx 1$.  
In a typical WSM, we have $Q \approx 10^8$ m$^{-1}$ (for example, $Q = 3.2 \times 10^{8}$m$^{-1}$ in WTe$_2$ \cite{Peng}), and $v_F \approx 10^6$ m/s. Thus $\omega \approx 10^{14}$ rad/s or smaller is needed for observing the GKR in WSM thin films. 
In the other term in the denominator we have $d \sigma_{xx}/(c \epsilon_0) \approx \alpha_F \omega d / (3 v_F) = (2 \pi \alpha_F c/v_F) (d/\lambda)$, where $\alpha_F = e^2/4 \pi \hbar c \epsilon_0 \approx 1/137$ is the fine structure constant. Since $ 2 \pi \alpha_F c/v_F$ is of ${\cal O}(1)$ and we are working in the $d/\lambda \ll 1$, the $d$ dependence of $\chi^p_{\rm Kerr}$ in WSM thin films is insignificant. 

\begin{figure}[t!]
\includegraphics[width =\linewidth]{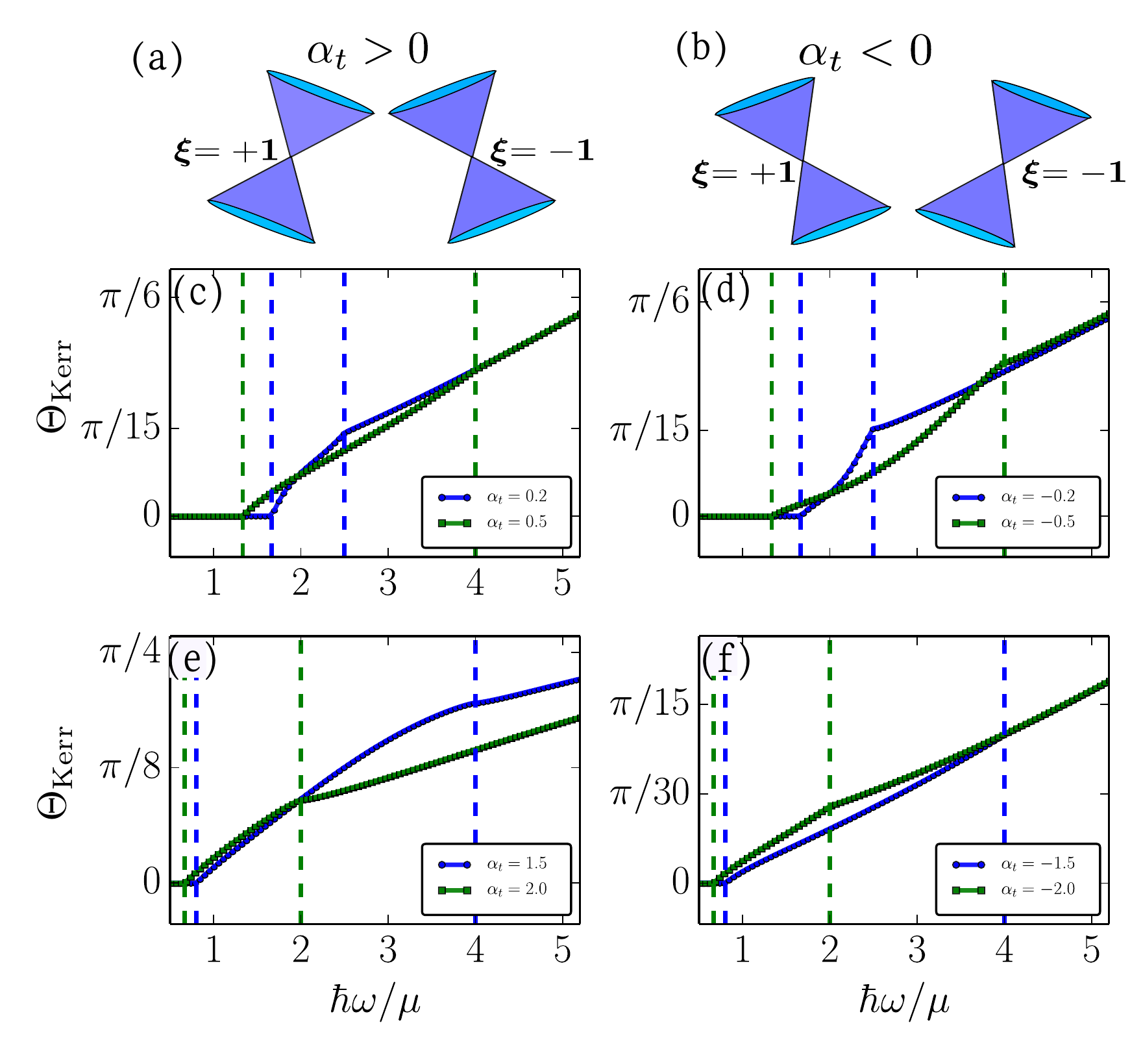}
\caption{ (a), (b) The relative orientation of a pair of Weyl nodes for $\alpha_t > 0$ and for $\alpha_t < 0$, respectively.  
The giant Kerr angle of the reflected optical beam as a function of the optical frequency for (c) type-I WSM with $\alpha_t > 0$, (d) type-II WSM with $\alpha_t <0$, 
(e) type-II WSM with $\alpha_t > 0$, (f) type-II WSM with $\alpha_t <0$. The scale of the GKR is fixed by the ratio $\sigma_{xy}/\sigma_{xx} \propto v_F Q/\omega$. 
Thus in crystalline systems, GKR of the order of a radian can be observed for $\omega \approx 10^{14}$ rad/s, and appropriate choice of $\mu$. 
The dashed vertical lines show the boundaries of region I and region II for the corresponding $\alpha_t$. 
Here we have chosen $\mu = 0.125$ eV and $Q = 10^{8}$ m$^{-1}$, $d = 10$ nm and $\Theta_{\rm Kerr}$ is in radians. }
\label{fig4}
\end{figure}

The dependence of the Kerr angle of rotation for normal incidence on a free standing tilted WSM thin film $(n_i = n_t \approx 1$ and $\theta_i = \theta_t)$, is shown in Fig.~\ref{fig4}.  As opposed to the typical values of micro-radians in Ferromagnetic systems and topological insulators \cite{}, in WSM the GKR can be of the order of a 
radian for reasonable choice of parameters. Furthermore, two distinct kinks in the GKR should be observable on scanning the optical frequencies (or the chemical potential/doping) across the Pauli blocking of region I and region II in tilted WSM. A similar behaviour will also be seen in the ellipticity angle measurement (not shown here). 

\subsection{Incidence on surface without with Fermi arcs states  $(\parallel \mathbf{Q})$}
In this scenario, light propagates ($\mathbf{\hat{n}}$) perpendicular to the $z$ axis, and the polarization of the incident electric field is in the $y-z$ (or $x-z$) plane. Consequently, the polarization rotating off diagonal reflection coefficients are 
\begin{equation}
r_{sp} \propto \sigma_{zy} = 0 \hspace{.8cm}{\rm or} \hspace{.8cm} r_{sp} \propto  \sigma_{zx} = 0~.
\end{equation}
Consequently, in this case there is no polarization rotation. This offers an optical probe to distinguish the surfaces of a WSM which host Fermi-arc surface states.


\section{Kerr rotation in semi-infinite WSM}\label{Application2}

Having explored the polarization rotation in thin films of WSM, we now focus on the polarization rotation due to a semi-infinite slab of WSM. 
In the bulk of a WSM, the Maxwell's equations are themselves modified by the presence of an axionic term\cite{Frank,Wu1124,Mehdi}. This axionic term in turn, has imprortant consequences in 
the  polarization rotation of the reflected light \cite{Mehdi}. It also results in the topological magnetoelectric effect (TME)\cite{Shou-cheng} which has been recently studied in topological insulators and WSM. As opposed to the case of topological insulators, where the axion field is a constant, the axion field in WSM has a nontrivial dependence on space and time owing to the breaking of time reversal and inversion symmetries.

The axionic term is added to the lagrangian of the electromagnetic field\cite{Frank} via the following term, $\delta {\cal L} = ~c \epsilon_0\alpha_F \vartheta {\bs E}\cdot{\bs B}/\pi$, with $\alpha_F$ being the fine structure constant, and $\vartheta$ the axionic field. In a WSM, we have $\vartheta({\bs r},t) = 2{\bs Q}\cdot{\bs r}-2{\mathcal Q}_0t$, with ${\mathcal Q}_0$ ($\bs Q$) denoting the separation of the two Weyl nodes in the energy (momentum) space. For a WSM with inversion symmetry we have, ${\mathcal Q}_0 = 0$.
In materials exhibiting the axionic response, the electric polarization and magnetisation have additional contributions arising from the topological terms \cite{Frank,Yang},
\bearr
{\bs D} &=& {\bs\epsilon} {\bs E} + c \epsilon_0\alpha_F\vartheta{\bs B}/\pi~,\\
{\bs H} &=& {\bs B}/{\mu_p}- c \epsilon_0 \alpha_F\vartheta{\bs E}/\pi \label{Max_H}~.
\eearr
Here ${\bs \epsilon}$ is the dielectric matrix and $\mu_p$ is the permeability.

Accordingly, the Maxwell's equation for the electric field propagation in a WSM is modified to be, 
\be
\label{wave_eq}
\nabla^2 {\bs E}-\nabla (\nabla\cdot{\bs E})
=\frac{1}{c^2\epsilon_0}\frac{\partial { ({\bs \sigma} \cdot \bs E})}{\partial t} + \frac{\epsilon_b}{c^2}\frac{\partial^2 {\bs E}}{\partial t^2} + \frac{2 \alpha_F}{\pi c }{{\bs Q}} \times \frac{\partial {\bs E}}{\partial t}~.
\ee
Here we have assumed the relative permeability ($\mu_p /\mu_{p0}$) to be unity, and $\epsilon_b$ is the static relative permittivity arising from the bound charge polarization. 
See Appendix \ref{Axion} for details of the derivation of Eq.~\eqref{wave_eq}. 
For the case of bulk WSM, with $d \gg \lambda$, the last term in Eq.~\eqref{wave_eq} plays an important role in obtaining the Fresnel coefficients. Again, we need to distinguish between the two cases when the light is incident on the surface hosting  Fermi arc surface states ($\hat{\bf n} \perp {\bs Q}$) 
and on a surface without the Fermi arc states ($\hat{\bf n} \parallel {\bs Q}$). 
We show that these two cases, qualitatively correspond to the Faraday and Voigt geometries discussed in the context of magneto-optic effects in magnetic systems such as (Ga,Mn)As.\cite{Tesarova} 
In a time reversal symmetry broken WSM, ${\bs Q}$ acts analogous to magnetization in a ferromagnet, whose relative orientation with respect to $\hat {\bf n}$, results in different effects described below. 

\subsection{Incidence on surface without Fermi arc states}\label{Kerr_bulk}
Let us consider a normal incidence of light on a surface without Fermi arcs, {\it i.e.}, $\hat {\bf n} \parallel {\bs Q} = Q\hat{z}$. 
In this case we have ${\bs Q}\cdot{\bs B} = 0$, and consequently 
the axion charge density,  $\rho_{\vartheta}= -2 \alpha_F c \epsilon_0 \bs Q\cdot\bs B /\pi = 0$ (see Eq.~\eqref{divE} in appendix \ref{Axion} for details).
Here, ${\bs Q}$ plays the role of effective magnetization, and light propagates parallel to it. This setting is similar to that of `Faraday geometry', which results in the magneto-optic polar Kerr effect in magnetic materials
\cite{visnovsky2006optics}. 

\begin{figure}[t!]
\includegraphics[width = \linewidth]{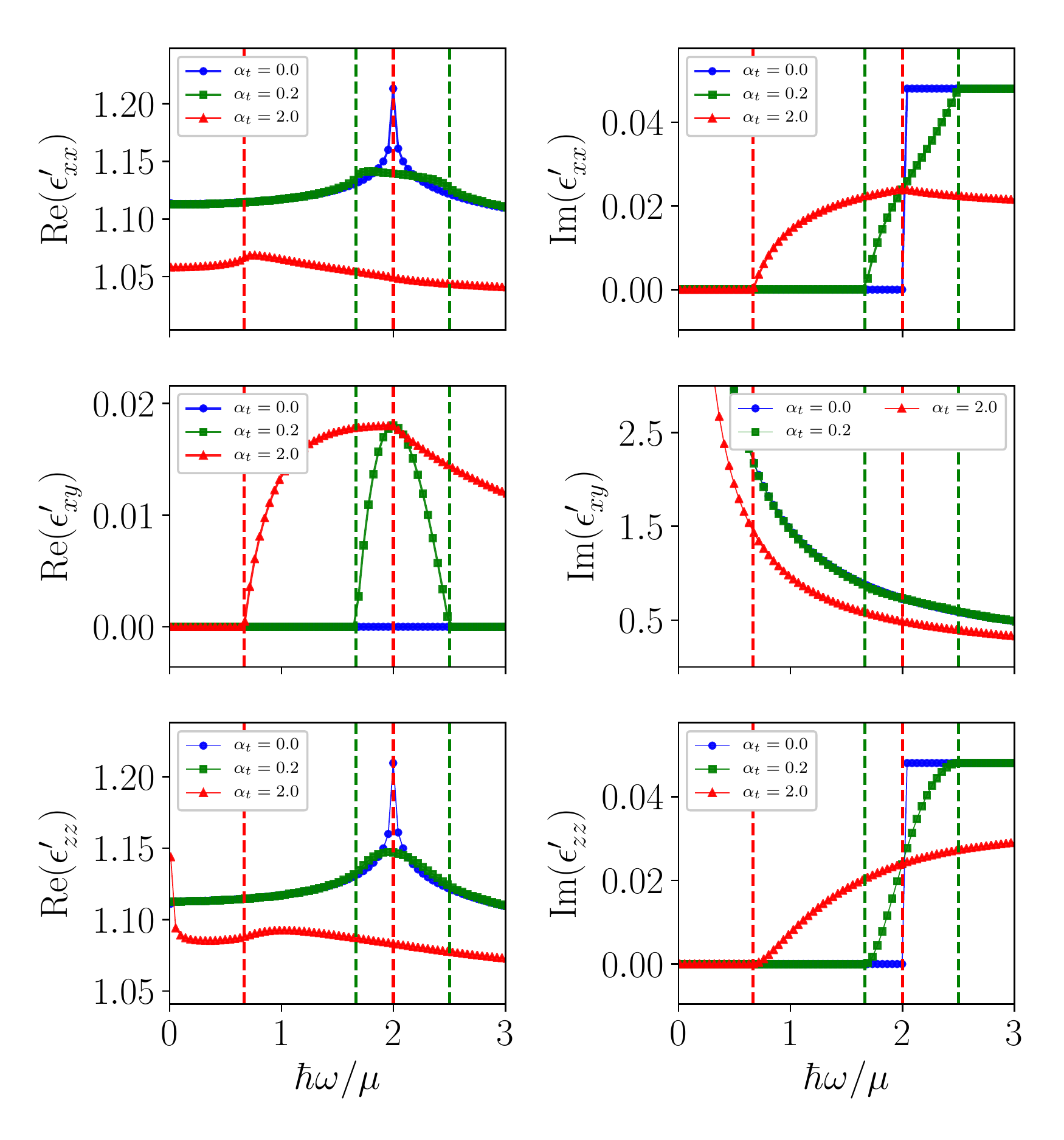}
\caption{The optical frequency dependence of the real and imaginary parts of the effective dielectric matrix $\epsilon'_{ij}$ (including the axion term) for different tilt parameters. 
(a) Re($\epsilon'_{xx}$), (b) Im($\epsilon'_{xx}$), (c) Re($\epsilon'_{xy}$), (d) Im($\epsilon'_{xy}$), (e) Re($\epsilon'_{zz}$), (f) Im($\epsilon'_{zz}$). 
The axion term appears in ${\rm Im} (\epsilon'_{xy}) \propto Q/\omega$, making it diverge in the 
low frequency regime. The vertical dashed lines mark the boundary of region I and II for $\alpha_t = 0.2$ in green, and $\alpha_t =2$ in red. Here we have chosen $\epsilon_b = 1$ and other parameters are identical to those of Fig.~\ref{fig4}.} 
\label{fig5}
\end{figure}

The propagation of light inside the WSM along the $\hat z$ direction can be obtained from Eq.~\eqref{wave_eq}, as detailed in Appendix~\ref{Axion}.
It is described by the following matrix equation, 
\be\label{tensor11a}
n^2\begin{pmatrix}
E_x\\
E_y\\
0\\
\end{pmatrix} = \begin{pmatrix}
                \epsilon'_{xx} & \epsilon'_{xy} & 0\\
                -\epsilon'_{xy} & \epsilon'_{yy} & 0\\
                0 & 0 & \epsilon'_{zz}\\
                \end{pmatrix} 
\begin{pmatrix}
 E_x \\
 E_y\\
 E_z\\
 \end{pmatrix}~.
\ee
Here, $n$ denotes the complex refractive index, and $\epsilon_{ij}'$ is the complex dielectric function including the axion terms. 
The elements of $\epsilon_{ij}'$ are readily expressed in terms of the optical conductivities and the internode separation as~\cite{Mehdi,Shibata}, $\epsilon_{yy}'=\epsilon_{xx}'$, where 
\be \label{eps_xx1}
\epsilon_{xx}' = \epsilon_b +  \frac{i}{\omega \epsilon_0}\sigma_{xx}~,\hspace{0.4cm}{\rm and}~\hspace{0.4cm}\epsilon_{zz}' = \epsilon_b +  \frac{i}{\omega \epsilon_0} \sigma_{zz}~.
\ee
The off-diagonal element of $\epsilon_{xy}'$ is what typically leads to optical activity, and it is given by,
\be\label{eps_xy1}
\epsilon_{xy}' =          \frac{i}{\omega \epsilon_0}\sigma_{xy} + \frac{2 i \alpha_F c}{\pi} \frac{Q}{\omega} = \frac{i}{\omega \epsilon_0} (\sigma_{xy} + \sigma_Q)~.
\ee
Here, $\alpha_F$ 
is the fine structure constant, and ${\sigma}_{ij}$ are the complex optical conductivities.

Interestingly, Eq.~\eqref{eps_xy1} comprises of two terms, one resulting from finite transverse conductivity, and the other owes its origin solely to the 
axionic term and modified dynamics of the electromagnetic waves. This `gyrotropic' term is not present in the conventional definition of the dielectric constant in metals: 
$\epsilon_{ij}(\omega) = \delta_{ij} \epsilon_b + i \sigma_{ij}(\omega)/\epsilon_0\omega$. 
The dielectric constant can also be expressed as $\epsilon_{ij}' = \epsilon_{ij} + \varepsilon_{ijk} Q_k \times 2i \alpha_F c/(\pi \omega)$, where $\varepsilon_{ijk}$ is the 
anti-symmetric Levi-Civita tensor, and $Q_k$ denotes the k'th component of the Weyl node separation vector ${\bs Q}~(= Q \hat{z})$. Thus, the gyrotropic constant in WSM $\propto Q/\omega$, making its optical response analogous in spirit to that of magnetic materials. 
The real and the imaginary part of the different components of the modified dielectric tensor are shown in Fig.~\ref{fig5}. The axionic term in ${\rm Im} (\epsilon'_{xy}) \propto Q/\omega$ makes it diverge in the low frequency regime. This will lead to anomalous optical activity even in the low frequency, Pauli blocked regime. Notice that in Fig.~\ref{fig5}, 
${\rm Im}(\epsilon'_{xx}) \ll {\rm Re}({\epsilon'_{xx}})$,  ${\rm Im}(\epsilon'_{zz}) \ll {\rm Re}({\epsilon'_{zz}})$, and ${\rm Re}(\epsilon'_{xy}) \ll {\rm Im}({\epsilon'_{xy}})$. 

Equation~\eqref{tensor11a} permits nontrivial solution for the elecrto-magnetic fields only for the following conditions\cite{visnovsky2006optics}: 
\begin{equation}\label{solution}
n_+^2 = \epsilon'_{xx} + i\epsilon'_{xy} ~~~~\text{and},~~~~ n_-^2 = \epsilon'_{xx}-i\epsilon'_{xy}~.
\end{equation}
Here, $n_{+}$ and $n_{-}$ are refractive indices of the left and right circularly polarised eigenmodes in the WSM. 
This becomes immediately clear on substituting  the two solutions obtained in Eq.~\eqref{solution} in \eqref{tensor1}, which yield ${\bs E}_x = \pm i {\bs E}_y$. 
The real (imaginary) part of the difference, $\delta n = n_+ - n_-$, leads to circular birefringence (circular dichroism).
A finite $\delta n$ solely arises from the $\epsilon'_{xy} \propto Q$ term (as $\sigma_{xy} \propto Q$), has both real and imaginary terms, and it vanishes in the  
$Q \to 0$ limit. 

Now the Fresnel reflection coefficient corresponding to these modes can be obtained via the well known relation, 
\be
r_{\pm} = (1-n_\pm)/(1+n_\pm)~.
\ee
Here, the dimensionless constant $\chi_{\rm PKE}$ is defined as 
\bearr\label{chi_2}
\chi_{\rm PKE} &=& i\frac{r_+ - r_-}{r_+ + r_-}~ = i\frac{n_+ - n_-}{n_+ n_- - 1}~. 
\eearr
As opposed to the thin film WSM geometry, where the polarization rotation solely arises due to a finite $\sigma_{xy}$, in bulk WSM, the axion term appearing in 
$\epsilon'_{xy}$ also plays an important role. Similar to the magneto-optic polar Kerr effect which is odd in the magnetization \cite{visnovsky2006optics}, 
the polarization rotation here also is an odd function of $Q$. This can be seen from the exact Eq.~\eqref{chi_2}, where the numerator is an odd function of $Q$ and the denominator is an even function of $Q$. Furthermore, this can also be seen from the limiting case of $|\epsilon'_{xy}| \ll |\epsilon'_{xx}|$, for which Eq.~\eqref{chi_2}, reduces to 
\be
 \chi_{\rm PKE} \approx \frac{\epsilon'_{xy}}{(1-\epsilon'_{xx}) \sqrt{\epsilon'_{xx}}}~.
\ee

\begin{figure}[t!]
\includegraphics[width =\linewidth]{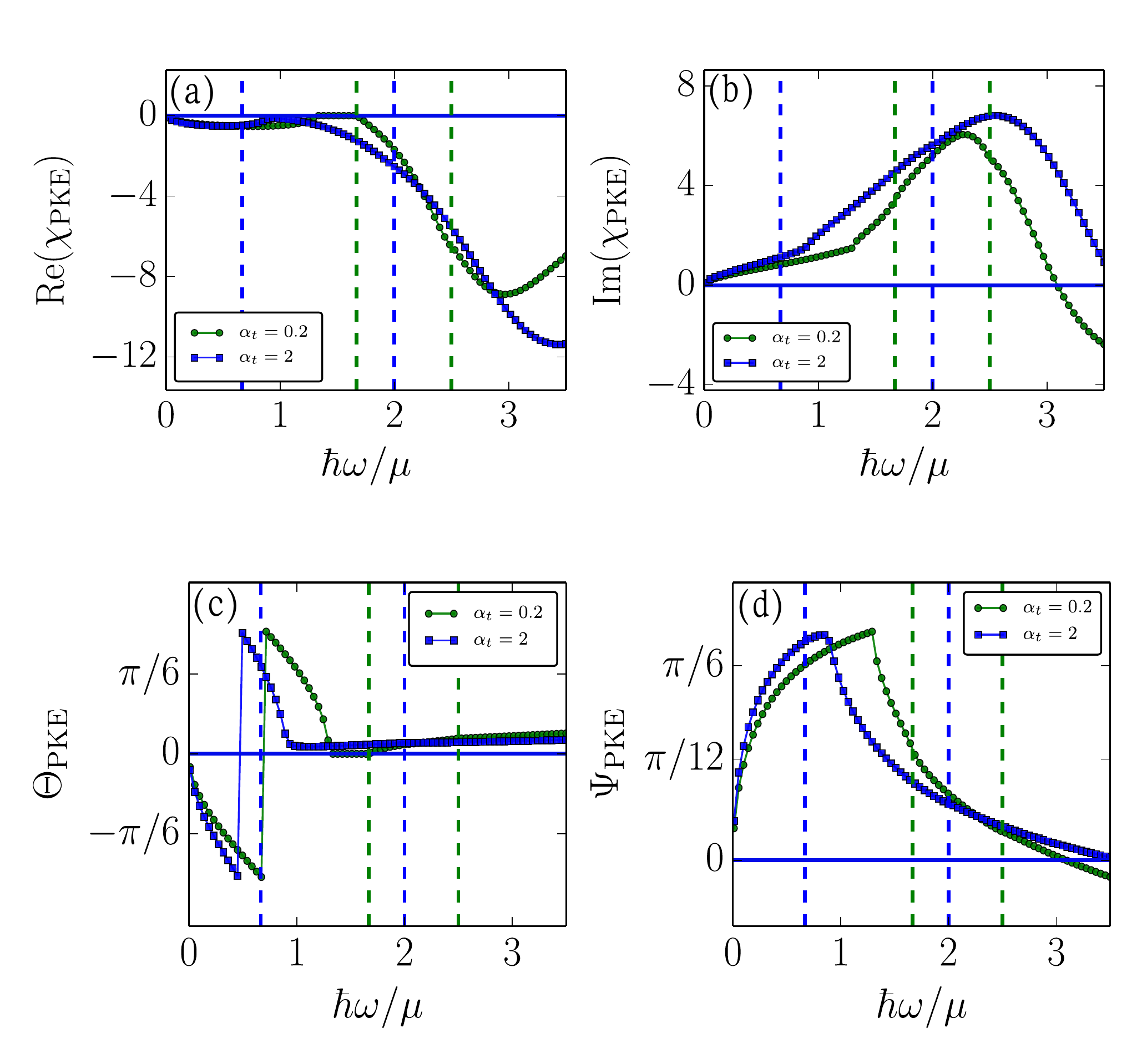}
\caption{(a) The real and (b) imaginary part of $\chi_{\rm PKE}$, as a function of the incident photon energy, for a semi infinite slab of type I and type II WSM. The resulting 
(c) giant polar Kerr effect ($\Theta_{\rm PKE}$), and ellipticity angle ($\Psi_{\rm PKE}$), in radian. In this `Faraday geometry', light propagates along the 
direction of the Weyl node separation, with large circular bifriengence and dichroism. The solid horizontal line in both (c) and (d) shows the vanishing small optical activity if the axion term is neglected. Thus the optical activity is caused predominantly by the axion term. Interestingly, it should be observable also in the Pauli blocked region where no optical transitions are allowed. Here, all the parameters are identical to those of Fig.~\ref{fig5}.
\label{fig6}}
\end{figure}

The exact polarization rotation ($\Theta_{\rm PKE}$) and ellipticity angle ($\Psi_{\rm PKE}$) can be obtained by using Eq.~\eqref{chi_2} in  Eq.~\eqref{Kerr_defn}. 
The real and imaginary part of $\chi_{\rm PKE}$, and the resulting angles are shown in Fig.~\ref{fig6}. We emphasize that 1) the optical activity is predominantly caused by the 
axion term, and 2) it persists even in the Pauli blocked frequency regime with no optical transitions. 

%
%
\begin{figure}[t!]
\includegraphics[width = \linewidth]{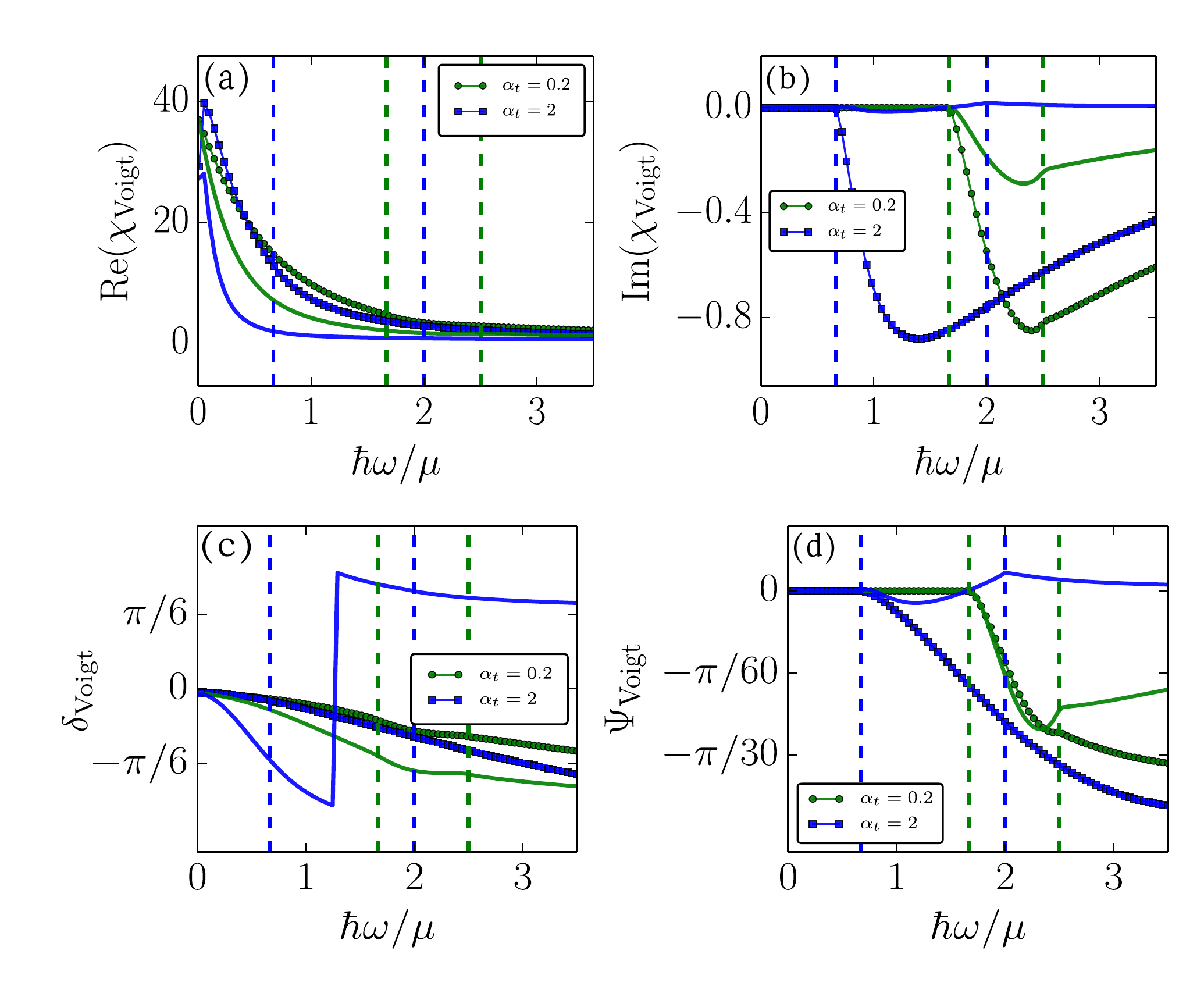}
\caption{(a) The real and (b) imaginary part of $\chi_{\rm Voigt}$, as a function of the incident photon energy, for a semi infinite slab of type I and type II WSM. The resulting (c) polarization rotation (or Voigt effect, $\delta_{\rm Voigt}=\Theta_{\rm Voigt} - \Theta_0$), and (d) ellipticity angle ($\Psi_{\rm Voigt}$), in radian. 
In this `Voigt' geometry, light propagates perpendicular to the Weyl node separation, with large linear bifriengence and dichroism. 
The solid lines in both (c) and (d) shows the corresponding optical activity if the axion term is neglected. Notice that unlike the case of `Faraday geometry', here $\Psi_{\rm Voigt}$ vanishes in the low frequency Pauli blocked regime where no optical transitions are allowed. Here we have chosen $\Theta_0 = \pi/4$ and all other parameters are identical to that of Fig.~\ref{fig5}. 
\label{fig7}}
\end{figure}

\subsection{Incidence on surface with Fermi arc states}
\label{VB}

As we have fixed the momentum space separation of the two Weyl nodes to be in the $z$-direction, the Fermi-arc surface states appear on the surfaces parallel to the $x-z$ or the $y-z$ plane. For definite-ness we consider the light propagating in the $x$-direction, and 
incident on the surface parallel $y-z$ plane. 
Thus we have $\hat{\bs n} = \hat{\bs x}$ and $\hat{\bs n} \perp {\bs Q}$ (recall that ${\bs Q} = Q{\hat{\bs z}}$).
This setting is similar to that of the `Voigt' geometry for polarization rotation in magnetic materials.  

In this case, for wave propagation in the $\hat{\bs x}$ direction, Eq.~\eqref{wave_eq} leads to (see Appendix~\ref{Voigt} for details of the calculation),  
\be\label{tensor2a}
n^2\begin{pmatrix}
0\\
E_y\\
E_z\\
\end{pmatrix} = \begin{pmatrix}
                \epsilon'_{xx} & \epsilon'_{xy} & 0\\
                -\epsilon'_{xy} & \epsilon'_{yy} & 0\\
                0 & 0 & \epsilon'_{zz}\\
                \end{pmatrix} 
\begin{pmatrix}
 E_x \\
 E_y\\
 E_z
 \end{pmatrix}~.
\ee
Here the effective dielectric constants have been defined in Eqs.~\eqref{eps_xx1}-\eqref{eps_xy1}. 
Similar to the case of Eq.~\eqref{tensor11a}, Eq.~\eqref{tensor2a} also permits two solutions for specific values of the refractive index, 
\begin{equation}\label{n_p}
n^2_{\parallel} = \epsilon_{zz}' ~~\text{and}, n^2_{\perp} = \epsilon_{yy}' - \frac{\epsilon_{xy}'^2}{\epsilon_{xx}'} ~,
\end{equation} 
which defines two linearly polarised modes: one travelling parallel to ${\bs Q}$, and another in the plane perpendicular to ${\bs Q}$.
Note that $n_{\perp}^2$ is an even function of $Q$ (of the form $a + bQ^2$, where both $a$ and $b$ are complex), while $n_{\parallel}^2$ is `almost' independent of $Q$. 
Consequently, $n_{\perp}$ is an even function of $Q$. 
The real (imaginary) part of the difference, $\delta n = n_\parallel - n_{\perp}$ leads to linear birefringence (linear dichroism).
The electric field corresponding to these modes is given by,
$
{\bs E}_{\parallel} = E_0\hat{\bs z}~e^{i \omega(t \mp n_{\parallel}x/c)}
$
and,
$
{\bs E}_{\perp} = E_0\left(\hat{\bs y} + \hat{\bs x}~{\epsilon'_{xy}}/{\epsilon'_{xx}}\right)e^{i \omega (t \mp n_{\perp}x/c)},
$
$E_0$ being the electric field amplitude. 
Analogous to the propagation of light in a `Voigt' geometry in a magnetic material, here also an electric field component parallel to the propagation direction ($E_x \propto \epsilon'_{xy}$) is also allowed.


Now let us consider the reflection of a linearly polarized optical beam from the infinte WSM slab. For an arbitrary incident polarization making an angle $\Theta_0$ with respect to ${\bs Q}$, the electric field can be decomposed as 
\be
E_{\parallel} = E_I \cos (\Theta_0)~, \hspace{.7cm} E_{\perp} = E_I \sin (\beta)~.
\ee
Thus, the initial polarization angle with respect to ${\bs Q}$ can be expressed as $\Theta_0 = \tan^{-1}({E_{\perp}}/{E_{\parallel}})$. The beam becomes elliptically polarized 
on reflection with from the WSM slab. The corresponding $\chi_{\rm Voigt}$ which determines the ellipticity and polarization angle, is given by 
%
%
\be \label{chi_Voigt}
\chi_{\rm Voigt} = \frac{r_{\perp}E_{\perp}}{r_{\parallel}E_{\parallel}}~ = \frac{r_{\perp}}{r_{\parallel}} \tan(\Theta_0).
\ee
Here $r_{\parallel}$ and $r_{\perp}$ are the Fresnel reflection coefficients for the parallel and perpendicular components, respectively. These are given by 
\be
r_{\parallel/\perp} = (1-n_{\parallel/\perp})/(1+n_{\parallel/\perp})~.
\ee
As $n_{\perp}$ is an even function of $Q$, $r_{\perp}$ is also an even function of $Q$, while $r_{\parallel}$ is `almost' independent of $Q$. Consequently $\chi_{\rm Voigt}$, and the resulting polarization rotation are also even function of $Q$. 
This quadratic in $Q$ polarization rotation in WSM is analogous to the `Voigt effect', which is also termed as the quadratic polar Kerr effect, or magnetic linear dichroism. 

The polarization ($\Theta_{\rm Voigt}$) and ellipticity ($\Psi_{\rm Voigt}$) angle of the reflected beam are easily calculated by substituting $\chi_{\rm Voigt}$ in Eq.~\eqref{Kerr_defn}. For the case of $\Theta_0 = 0$ or $\pi$, {\it i.e.} the incident polarization aligned along ${\bs Q}$, we have $\chi_{\rm Voigt} = 0$, and there is no polarization rotation. 
The reflected beam only suffers a phase change. For the general case, the relative polarization rotation angle $\delta_{\rm Voigt} = \Theta_{\rm Voigt} - \Theta_0$ can now calculated via the relation, 
\be \label{delta_beta}
\tan(2 \delta_{\rm Voigt})  = \frac{\tan(2 \Theta_{\rm Voigt}) - \tan (2 \Theta_0)}{1 + \tan(2 \Theta_{\rm Voigt})\tan (2 \Theta_0)}~.
\ee 
For the limiting case of $\Theta_0 = \pi/4$, Eq.~\eqref{delta_beta} reduces to $\delta_{\rm Voigt} = \pi/2 + 2 \Theta_{\rm Voigt}$. In another limiting case of $r_{\perp}/r_{\parallel}$ being real (as in the case of magnetic systems\cite{MLD}), we obtain, $\tan (\Theta_{\rm Voigt}) = r_{\perp} \tan(\Theta_0)/r_{\parallel}$, and Eq.~\eqref{delta_beta}
reduces to the known result \cite{MLD}
\be \label{eq_red}
\tan( \delta_{\rm Voigt}) = \frac{(r_{\parallel}-r_{\perp})\tan(\Theta_0)}{r_{\parallel} + r_{\perp}\tan^2 \Theta_0}~.
\ee
Furthermore if $r_{\parallel}/r_{\perp} \approx 1$, Eq.~\eqref{eq_red} reduces to 
%
$
\delta_{\rm Voigt} \approx 0.5 ({r_{\perp}}/{r_{\parallel}}-1)\sin(2\Theta).
$
%

Figure~\ref{fig7}, shows the dependence of the real and imaginary parts of $\chi_{\rm Voigt}$, along with $\Theta_{\rm Voigt}$ and $\Psi_{\rm Voigt}$, as a function of the 
optical frequency.  Notice the existence of finite $\Theta_{\rm Voigt}$ even in the Pauli blocked regime. However in contrast to the Faraday geometry, here the ellipticity angle 
vanishes in the Pauli blocked regime. 




\section{summary}\label{summary}
In this paper, we demonstrate the existence of giant Kerr effect in tilted WSM, which arises from the time reversal symmetry breaking in WSM thin films, and from axion electro-dynamics in bulk WSM. 
The time-reversal symmetry breaking in a WSM is captured by the Weyl node separation ($Q$) which explicitly determines the optical Hall conductivity ($\sigma_{xy} \propto Q$). The tilting of the Weyl nodes also plays a significant role in $\sigma_{xy}$, particularly in a type-II WSM. The existence of a finite $\sigma_{xy}$ in WSM is what leads to giant Kerr effect in the thin-film geometry of tilted WSM, as long as we are away from the Pauli blocked regime. 
In contrast to this, in bulk WSM, axion elctrodynamics plays an important role giving rise to optical activity and giant Kerr rotation even in the Pauli blocked low frequency regime ($\hbar \omega < 2 \mu$ in untilted WSM).  

In the case of WSM thin films (with thickness less than wavelength of light, $d \ll \lambda$), we find that the reflection and transmission coefficients are determined completely by the optical conductivity matrix. The optical Hall conductivity 
is finite only for surfaces perpendicular to ${\bs Q}$, and it plays an an important role in producing the giant Kerr effect when light is incident on WSM surfaces without Fermi arc states. 
We explicitly show that the polarization rotation in tilted WSM thin films $\Theta_{\rm Kerr}\propto \sigma_{xy}/\sigma_{xx} \propto v_F Q/\omega$. This yields giant Kerr rotation for light incident parallel to ${\bs Q}$ with $\omega \approx  10^{14}$ rad/s or less (frequencies in the optical range and below), as long as we are not in the Pauli blocked regime for optical transitions. 

In bulk WSM with broken time-reversal symmetry, we show that axion electrodynamics also plays an important role in the optical activity. This leads to giant Kerr effect {\it even in the Pauli blocked regime} for both type-I and type-II WSM. The effective dielectric constant including the axion term, can be expressed as $\epsilon_{ij}' = \epsilon_{ij} + \varepsilon_{ijk} Q_k \times 2i \alpha_F c/(\pi \omega)$ (see Sec.~\ref{Kerr_bulk} for details), where the second term establishes $Q/\omega$ to be the effective `gyrotropy' constant, analogous to magnetization in magnetic systems.  The case of normal incidence on surface without Fermi arc states is then identical to that of Faraday geometry with light propagating parallel to the node separation. Here, we show the existence of 
large circular bifriengence and circular dichroism, along with that of giant Kerr effect which is odd in $Q$. The case of normal incidence on surface with Fermi-arc states turns out to be identical to that of Voigt geometry in magnetic systems.  This leads to large 
linear bifriengence and dichrosim, along with a polarization angle dependent giant Kerr effect which is an even function of $Q$.  In both cases, the axion term leads to 
significant optical activity and giant polarization rotation, even in the Pauli blocked regime.

\section{ACKNOWLEDGEMENT}
We thank Prof. J. P. Carbotte for insightful discussions on the calculation of optical conductivity. 

\appendix
\label{PC}

\section{Longitudinal optical conductivity }\label{LOC_cal}
The Longitudinal optical conductivity can be obtained from Eq.~\eqref{Kubo}, after converting the momentum sum into an integral,
\be\label{LOC1}
\sigma_{ii}(\omega) = \frac{1}{8\pi^3}\int_{0}^{2\pi}d{\phi_{\bs k}}\int_0^{\infty}  k_{\perp}dk_{\perp}\int_{-k_c}^{k_c}dk_{z}~\sigma^{ii}_{\bs k}(\omega)~.
\ee
Here we have defined the kernel
\be\label{LOC2}
{\sigma^{ii}_{\bs k}(\omega)} = \frac{-i n^{\rm eq}_{\bs k} |M_i|^2}{\hbar\omega_k }\bigg(\frac{1}{\omega + \omega_k + i \gamma} + \frac{1}{\omega-\omega_k + i\gamma} \bigg)~.
\ee
The equilibrium population inversion (at $T=0$) is given by $n_{\bs k}^{\rm eq} = \Theta(\mu-\hbar v_f k - \hbar \xi v_t k_z^{\xi})-\Theta(\mu + \hbar v_f k -\hbar \xi v_t k_z^{\xi})$  and $\hbar\omega_k = 2 \hbar v_F k $ is the energy difference between the conduction and the valance band, which is independent of the tilt velocity. 
Here $k=\sqrt{k_{\perp}^2+(k_z^{\xi})^2}$. 
The real and the imaginary part can be separated in the limit $\gamma \to 0$, using the Dirac identity
\begin{equation}\label{dirac}
\lim_{\epsilon\to 0 } \frac{1}{x + i\epsilon} = {\cal{P}}\int_{-\infty}^{\infty} \Big(\frac{1}{x}\Big) - i \pi \delta (x)~.
\end{equation}
Here $\cal{P}$ denotes the principal value. We find that the longitudinal optical conductivity of time reversal broken Weyl semimetal is the sum of the optical conductivities of the two Weyl nodes. Thus we just calculate the conductivity of the $\xi =1$ node and multiply it by two. Thus allows us to substitute $k_z^\xi \to k_z$ and $\xi v_t \to v_t$ in the expressions below. 


\subsection{${\rm Re}[\sigma_{xx}(\omega)]$}

Thus, the real part of longitudinal optical conductivity is given by,

\bearr\label{LOC3}
{\rm Re}[\sigma_{ii}] & = & -\frac{1}{(2\pi)^3}\int_{0}^{2\pi}d{\phi_{\bs k}}\int_{0}^{\infty} ~
k_{\perp}dk_{\perp}\int_{-k_c}^{k_c}dk_{z} \nonumber \\
& & \frac{n_{\bs k}^{\rm eq}}{\hbar \omega_{\bs k}}|M_i|^2  \pi\delta(\omega-\omega_{\bs k})~.
\eearr
Here, only the optical matrix is dependent on $\phi_{\bs k}$, and the angular integration leads to  
\be\label{LOC4}
\int_{0}^{2\pi}|M_x|^2 d{\phi_{\bs k}}= \pi e^2 v_F^2
\bigg(1+\frac{k_z^2}{k^2}\bigg)~.
\ee
Thus we obtain 
\bearr\label{LOC5}\nonumber
{\rm Re}[\sigma_{xx}(\omega)] & & =  \frac{e^2 v_F^2}{8\pi}\int_{0}^{\infty}k_{\perp}dk_{\perp}\int_{-k_c}^{k_c} \left(1+\frac{k_z^2}{k^2}\right)dk_z\\\nonumber
 &\times&\frac{\delta(\omega - \omega_{\bs k})}{ \hbar v_F k}\bigg[\Theta(\mu-\hbar v_F k - \hbar |v_t| k_z)\\
 &-&\Theta(\mu + \hbar v_F k -\hbar |v_t| k_z)\bigg].
\eearr
Now, the integration for $k_{\perp}$ can easily be carried using the roots ($x_i$) of the argument in the Dirac delta function:
\be\label{LOC6}
\delta(f(x)) = \sum_{x_i} \frac{\delta(x - x_i)}{|f'(x_i)|}~.
\ee

Substituting the roots in the delta function leads to a much simpler one-dimensional integral,

\be\label{LOC7}
\begin{split}
{\rm Re}[\sigma_{xx}(\omega)] = {\frac{e^2}{16\pi \hbar}}\int_{-\omega/{2v_F}}^{\omega/{2v_F}}dk_z \left[\left(\frac{2v_Fk_z}{\omega}\right)^2 + 1\right]\times&\\
\Big[\Theta\left(\mu - \hbar v_t k_z + \hbar \omega/2\right)- \Theta\left(\mu - \hbar v_t k_z - \hbar \omega/2\right)\Big].
\end{split}
\ee
Equation~\eqref{LOC7} can be expressed in a dimensionless form:
\be\label{LOC8}
{\rm Re}[\sigma_{xx}(\omega)] = \sigma_\omega F_{xx},~{\rm where} \hspace{.3 cm} \sigma_\omega \equiv \frac{e^2}{6 h} \frac{\omega}{v_F},
\ee
and 
\be\label{LOC9}
F_{xx}  =  \frac{3}{8}\int_{-1}^1dx(1+x^2) (\Theta_+ - \Theta_-)~,
\ee
where we have defined $
\Theta_{\pm} =\Theta\left[\frac{2\mu}{\hbar \omega}-\alpha_t x \pm 1\right]$, and $x = 2 v_F k_z/\omega$.
Evaluating the integral in Eq.~\eqref{LOC9} for the two case of $|\alpha_t| <1$ and $|\alpha_t| >1$ leads to Eq.~\eqref{Resigmaxx_tl1}, and Eq.~\eqref{Resigmaxx_tl2}, of the main text, respectively. 

%

\subsection{${\rm Re}[\sigma_{zz}(\omega)]$}

Following the same procedure as above, the angular integration leads to 
\be\label{LOC14}
\int_{0}^{2\pi}|M_z|^2d\phi_{k} = 2\pi e^2 v_F^2 ~\frac{k_{\perp}^2}{k^2}~.
\ee
Thus the longitudinal optical conductivity in this case is given by,
\bearr\nn\label{LOC15}
{\rm Re}[\sigma_{zz}]&=&\frac{e^2 v_F^2}{4\pi}\int_{0}^{\infty}k_{\perp}dk_{\perp}\int_{-k_c}^{k_c} \frac{k_{\perp}^2 dk_z}{k_{\perp}^2 + k_z^2}
\frac{\delta(\omega - \omega_{\bs k})}{\hbar v_F k} \\\nn
&\times &\left[\Theta(\mu-\hbar v_F k - \hbar v_t k_z)- \Theta(\mu + \hbar v_F k- \hbar v_t k_z)\right]~.\\ 
\eearr
Performing the $k_{\perp}$ integral using the roots of the $\delta$-function, we have 
\bearr\label{LOC16}
& & {\rm Re}[\sigma_{zz}(\omega)] = {\frac{e^2}{8\pi \hbar}}\int_{-\omega/{2v_F}}^{\omega/{2v_F}}dk_z \left[\left(1-\frac{2v_Fk_z}{\omega}\right)^2 \right]\\\nn
 & &\times\Big[\Theta\left(\mu - \hbar |v_t| k_z + \hbar \omega/2\right)- \Theta\left(\mu - \hbar |v_t| k_z - \hbar \omega/2\right)\Big].
\eearr
Equation~\eqref{LOC16} can be expressed in a dimensionless form as 
\be\label{LOC17}
{{\rm Re}[{\sigma}_{zz}(\Omega)] =   2~\sigma_\omega F_{zz}}~,
\ee
where we have defined
\be\label{LOC17}
F_{zz}=\frac{3}{8}\int_{-1}^1dx(1-x^2) (\Theta_+ - \Theta_-)
\ee
and $\Theta_{\pm}$ are defined below Eq.~\eqref{LOC9}. Performing the integral in Eq.~\eqref{LOC17} for the two case of $|\alpha_t| <1$ and $|\alpha_t| >1$ leads to Eq.~\eqref{Resigmazz_tl1}, and Eq.~\eqref{Resigmazz_tl2}, respectively. 

\subsection{${\rm Im}[\sigma_{xx}(\omega)]~~{\rm and} ~~{\rm Im}[\sigma_{zz}(\omega)]$}
\label{Im_LOC_Cal}
Once the real parts are obtained and they converge to a constant as $\omega \to \infty$, the 
imaginary parts are computed using Kramers-Kroing relation for the imaginary part of the optical conductivity, 
\be\label{KK}
{\rm Im}[{\sigma}_{ii}(\omega)] = \frac{-2 \omega}{\pi}\int\limits_{0}^{\omega_c}\frac{{\rm Re}[{\sigma}_{xx}(\omega') - \sigma_{xx}(0)] }{\omega'^2 - \omega^2}d\omega'~.
\ee
Note that since the low energy Weyl Hamiltonian has an infinite bandwith, with an infinite filled sea of quasiparticles even in the ground state, 
we are forced to use an upper frequency cutoff ($\hbar \omega_c = \hbar v_F k_c$) for the  integral. This cutoff was not needed in the calculation of the 
real part of the longitudinal conductivity due to the presence of the delta function. 

\section{Transverse optical conductivity}\label{TOC_cal}

The transverse optical Hall conductivity is given by,
\be
\sigma_{ij}(\omega) = \frac{1}{8\pi^3}\int_{0}^{2\pi}d{\phi_{\bs k}}\int_0^{\infty}  k_{\perp}dk_{\perp}\int_{-k_c}^{k_c} dk_{z}~\sigma^{ij}_{\bs k}(\omega)~,
\ee
where the conductivity kernel is given by 
\be
\sigma^{ij}_{\bs k} = \frac{-i n^{\rm eq}_{\bs k}}{\hbar \omega_{\bs k}}\bigg(\frac{M_j M_i^*}{\omega + \omega_{\bs k} + i \gamma} + \frac{M_j^*M_i}{\omega - \omega_{\bs k}  + i\gamma}\bigg)~.
\ee

Writing in terms of real and imaginary parts of $M^{vc}$, we have
\be \label{T1}
\sigma_{xy} = \int {\color{blue}\frac{d^3k}{8\pi^3}} \frac{n^{\rm eq}_{\bs k}}{\hbar \omega_{\bs k}} (M_R^x M_I^y - M_I^x M_R^y)\bigg(\frac{1}{\omega^+ + \omega_{\bs k} } - \frac{1}{\omega^+ - \omega_{\bs k} }\bigg).
\ee
Here $\omega^+ = \omega + i \gamma$. As in the previous cases, performing the angular integral over the $\phi_{\bs k}$ dependent optical matrix elements yields, 
\be
\int_{0}^{2 \pi} \left(M_R^x M_I^y - M_I^x M_R^y \right)d \phi_{\bs k} = \frac{2 \pi \xi k_z^{\xi}}{k}~.
\ee
Again, the real and imaginary parts can now be evaluated separately in the limit $\gamma \to 0$ using the Dirac identity. 
Before proceeding, we note that Eq.~\eqref{T1} differs in form from Eq.~\eqref{LOC4} by a factor of $i$. As a consequence now $\delta$-functions appear in 
${\rm Im} [\sigma_{xy}(\omega)]$. In contrast to the longitudinal conductivity, for the transverse case, 
the two nodes have to be treated together and the inter-node separation becomes important.  

\subsection{${\rm Im}[{\sigma_{xy}}]$}
Using the Dirac identity in Eq.~\eqref{dirac}, the imaginary part of $\sigma_{xy}(\omega)$ can be expressed as 
{\begin{equation}
{\rm Im}[\sigma_{xy}]  = \frac{e^2v_F^2}{4\pi}\sum_{\xi=\pm1} \xi\int k_{\perp}dk_{\perp}dk_{z}^{\xi}\frac{k_{z}^{\xi}}{k}\frac{ n_{\bs k}^{\rm eq}}{\hbar\omega_{\bs k}} \delta(\omega-\omega_{\bs k}).
\end{equation}}
%
Performing the $k_{\perp}$ integral, using the $\delta$-function lead to the following one dimensional integral, 
\bearr\label{e8}
& & {\rm Im[\sigma_{xy}(\omega)]} = \frac{e^2 v_F}{8\pi\hbar\omega} \sum_{\xi=\pm1} \xi\int_{- \omega/2v_F}^{\omega/2v_F}k_z^{\xi} dk_z^{\xi} \\ 
& & \left[\Theta\left(1-\frac{\hbar\omega}{2\mu} - \frac{\hbar \xi v_t k_z^\xi}{\mu}\right)-\Theta\left(1+\frac{\hbar\omega}{2\mu} - \frac{\hbar {\xi} v_tk_z^\xi}{\mu}\right)\right]. \nn
\eearr

Equation~\eqref{e8} can be expressed as 

\be 
{\rm Im}[\sigma_{xy}(\omega)] = \frac{3 \sigma_\omega}{2} \frac{\mu^2}{\hbar^2 \omega^2} F_{xy}, 
\ee
where we have defined 
\be \label{Fxy}
F_{xy} = \sum_{\xi=\pm1} \xi\int_{- \frac{\hbar \omega}{2\mu}}^{\frac{\hbar \omega}{2\mu}} x dx  \sum_{p=\pm} 
\Theta\left(1 - \xi \alpha_t x -p\frac{\hbar \omega}{2 \mu}\right). 
\ee
Solving Eq.~\eqref{Fxy} leads to Eq.\eqref{Imxy1} [Eq.~\eqref{Imxy2}] for the cases of $|\alpha_t| <1$ ($|\alpha_t| >1$). 

\subsection{${\rm Re}[{\sigma_{xy}}]$}

Having obtained the imaginary part of the transverse conductivity, its real part can be obtained from the Kramers-Kronig relation for conductivity, 
\be \label{eqKK}
{\rm Re}[\sigma_{ij}(\omega) - \sigma_{ij}(0)] = \frac{2 \omega^2}{\pi}  \int_{0}^{\omega_c} d \omega' \frac{{\rm Im}[\sigma_{ij}(\omega')]}{\omega' [\omega'^2 - \omega^2]}~.
\ee
Here the use of an upper frequency cutoff is forced to cutoff the contribution from the infinite filled Fermi sea in the low energy model of Weyl (or Dirac) semi-metals.  
The AC (finite $\omega$) component of the real part of the transverse conductivity calculated from Eq.~\eqref{eqKK} for type-I and type-II WSM is presented in Eq.~\eqref{IxyT1} and Eq.~\eqref{IxyT2}, respectively. 

It turns out that the dc component of the transverse conductivity in a WSM is finite, and it has to be calculated separately as Eq.~\eqref{eqKK} only gives the finite frequency contributions. 
The DC component (real part) of the transverse conductivity is given by:
\bearr\label{e3}
{\rm Re}[\sigma_{xy}^{dc}] &=& {\frac{e^2 v_F}{4\pi^2} }\sum_{\xi=\pm1} \int_0^{\infty} k_{\perp}dk_{\perp} \int_{-k_c}^{k_c}\Bigg(\frac{n_{\bs k}^{\rm eq}}{\hbar\omega_k}\Bigg)\frac{\xi k_z^{\xi}}{k^2}dk_z^{\xi}~\nn \\
\eearr
where $\xi k_z^{\xi}/k$ results from the optical matrix element, and $1/k$ from $1/\omega_{k}$. Further change of variable from $k_z^\xi \to k_z$ lead to 
\be\label{e4}
{\rm Re}[\sigma_{xy}^{dc}] = {\frac{e^2}{4\pi h} }\sum_{\xi=\pm1} \int_0^{\infty} k_{\perp}dk_{\perp} \int_{-k_c-\xi Q}^{k_c-\xi Q} \frac{\xi n_{\bs k}^{\rm eq} k_z dk_z}{(k_\perp^2 + k_z^2)^{3/2}}.
\ee
It turns out that Eq.~\eqref{e4} has an \emph{intrinsic} contribution (finite for  $\mu \to 0$) and an \emph{extrinsic} contribution (which vanishes in the $\mu \to 0$ limit). To evaluate them 
it is useful to express the population inversion as a sum of extrinsic (in square brackets) and intrinsic contribution (independent of $\mu$), 
$n_{\bs k}^{\rm eq} = [\Theta(\mu-\hbar v_F k_{\xi} - \hbar {\xi} v_t k_z^{\xi})-\Theta(\mu + \hbar v_F k_{\xi} -\hbar \xi v_t k_z^{\xi}) + 1] -1$. 
Evaluating these two terms separately, we obtain the intrinsic contribution $\propto \sigma_{Q}$, and the extrinsic contribution $\propto \sigma_\mu$. Their explicit expressions for the two cases of $|\alpha_t| <1$ and $|\alpha_t| >1$ 
are presented in Eq.~\eqref{Rxy_T1}, and Eq.~\eqref{RxyT2}, respectively. 

\section{Axion electrodynamics}\label{Axion}
The Maxwell's equation inside the bulk of a material are modified by the presence of an axionic term. This modification is derived by adding the additional topological magneto-electric term in the lagrangian\cite{Frank}, $\delta {\cal L} = ~\alpha_F c \epsilon_0\vartheta {\bs E}\cdot{\bs B}/\pi$, with $\alpha_F$ being the fine structure constant, and $\vartheta$ is the  axion field. One experimentally demonstrated manifestation of this axionic term in materials is the topological magnetoelectric effect. Its impact on the material properties is primarily based on additional topological contributions to the electric polarization and magnetization \cite{Frank},  
\bearr\label{polarisations}
{\bs D} &=& {\bs\epsilon} {\bs E} + c \epsilon_0\alpha_F\vartheta{\bs B}/\pi,\\
{\bs H} &=& {\bs B/\mu_p} - c \epsilon_0 \alpha_F\vartheta{\bs E}/\pi \label{Max_H}~,
\eearr
where ${\bs \epsilon}$ is the dielectric tensor, and $\mu_p$ is the permeability.
The Maxwell's field equations with sources, are thus modified as:
\be\label{divE}
\nabla\cdot\Big(\bs \epsilon {\bs E} + c \epsilon_0\alpha_F\vartheta{\bs B}/\pi\Big) = \rho 
\Rightarrow \nabla\cdot(\bs \epsilon{\bs E}) = \left(\rho + \rho_{\vartheta}\right)~.
\ee
where $\rho_{\vartheta} = - c \epsilon_0\alpha_F\nabla\vartheta.{\bs B}/(\pi)$ is the axion charge density. 
From Eq.~\ref{Max_H}, we have,
\bearr\nn \label{crosB}
\nabla \times ({\bs B/ \mu_p}-c \epsilon_0\alpha_F\vartheta{\bs E}/\pi) &=& {\bs J} + \frac{\partial}{\partial t}(\bs \epsilon {\bs E} + c \bs \epsilon_0\alpha_F\vartheta{\bs B}/\pi), \\
\Rightarrow \nabla \times {\bs B} &=& \mu_p {\bs J} + {\bs \epsilon \mu_p}\frac{\partial {\bs E}}{\partial t} + {\bs J}_{\vartheta}~.
\eearr
Here,  ${\bs J}_{\vartheta} =  \alpha_F c \epsilon_0 \mu_p(\dot{\vartheta}{\bs B} + \nabla\vartheta \times {\bs E})/\pi$ is the axion field dependent current density. The other two source-less Maxwell's equations, remain unchanged as:
\bearr
\nabla\cdot{\bs B} &=& 0~, \\
~\nabla \times {\bs E} + {\partial_t {\bs B}} &=& 0.
\eearr

In a WSM, the axion field can be expressed in terms of internodal seperation in the momentum (${\bs Q}$) and energy (${\mathcal Q}_0$) space,  $\vartheta({\bs r},t) = 2{\bs Q}\cdot{\bs r}-2{\mathcal Q}_0t$. For a WSM which preserves the inversion symmetry, we have ${\mathcal Q}_0 = 0$, and thus $\nabla \vartheta = {\bs Q}$ and $\dot{\vartheta} = 0$. 
Now eliminating ${\bf B}$ from Eq.~\eqref{crosB} using the other modified Maxwell equations and using ${\bs J} = {\bs \sigma}\cdot {\bs E}$, we obtain the wave propagation equation in a time-reversal symmetry broken WSM, 
\begin{equation}
\label{wp}
\nabla (\nabla\cdot{\bs E}) -\nabla^2 {\bs E}
=-{\mu_p \bs \sigma}\frac{\partial {\bs E}}{\partial t} - \bs \epsilon \mu_p \frac{\partial^2 {\bs E}}{\partial t^2} - \frac{2\alpha_F c \epsilon_0 \mu_p }{ \pi}{{\bs Q}} \times \frac{\partial {\bs E}}{\partial t}.
\end{equation}

Using Eq.~\eqref{wp}, we now explore the wave propagation in the two geometries with ${\bs n} \parallel {\bs Q}$ (``Faraday" geometry) and ${\bs n} \perp {\bs Q}$ (``Voigt'' geometry) in the next two subsections.

\subsection{The ``Faraday" geometry} 

For an electric field $\mathbf{E} = E_o e^{i(\bs k.\bs r-\omega t)}$, with amplitude $E_o$, propagating in the $\hat{z}$ direction$(\bs k = k \hat{z})$, the LHS of equation \eqref{wave_eq} yields,
\be
\begin{pmatrix} 
{\partial^2}/{\partial z^2} & 0 & 0\\
0 &  {\partial^2}/{\partial z^2} & 0\\ 
0 & 0 & 0
\end{pmatrix}
\mathbf{E} = -k^2\mathbf{E}~.
\ee
The RHS of equation \eqref{wave_eq} results in,
\be\label{RHS}
- \epsilon_{ij}\frac{\omega^2}{c^2} \bs E  -\frac{2i\omega \alpha_F\bs Q \cdot\bs E}{c \pi},
\ee
where $\epsilon_{ij} = \epsilon_b\delta_{ij} + \frac{i\bs \sigma}{\epsilon_0 \omega}$.
Substituting $k = n \omega /c$, the wave equation becomes:
\begin{widetext}
	\be\label{wave_eq1}
	n^2 \begin{pmatrix}
		E_x\\
		 E_y\\
		0
	\end{pmatrix}
	= \frac{ i }{\omega \epsilon_0}\begin{pmatrix} 
		\sigma_{xx} & \sigma_{xy} & 0\\
		-\sigma_{xy} &  \sigma_{yy} & 0\\ 
		0 & 0 & \sigma_{zz}
	\end{pmatrix}
	\begin{pmatrix}
		E_x\\
		E_y\\
		E_z
	\end{pmatrix}
	+ \begin{pmatrix} 
		0 & \frac{2i\alpha_F Q c}{\pi\omega} & 0\\
		-\frac{2i\alpha_F Q c}{\pi\omega} &  0 & 0\\ 
		0 & 0 & 0
	\end{pmatrix}
      \begin{pmatrix}                                         
		E_x\\
		E_y\\
		E_z
	\end{pmatrix}
	+ \epsilon_b \mathbb{I}
		\begin{pmatrix}
		E_x\\
		E_y\\
		E_z
	\end{pmatrix}~.
	\ee
Here $\mathbb{I}$ denotes the identity matrix.
Thus, the effective dielectric tensor inside the WSM becomes:
\be\label{tensor1}
n^2\begin{pmatrix}
	E_x\\
	E_y\\
	0\\
\end{pmatrix} = \begin{pmatrix}
	\epsilon'_{xx} & \epsilon'_{xy} & 0\\
	-\epsilon'_{xy} & \epsilon'_{yy} & 0\\
	0 & 0 & \epsilon'_{zz}\\
\end{pmatrix} 
\begin{pmatrix}
	E_x \\
	E_y\\
	E_z\\
\end{pmatrix}~,
\ee
\end{widetext}
where the diagonal components are given by:
\be \label{eps_xx}
\epsilon_{xx}'(\epsilon_{zz}') = \epsilon_b +  \frac{i}{\omega \epsilon_0}\sigma_{xx}(\sigma_{zz})
\ee
and the off-diagonal elements responsible for optical activity are given by:  
\be\label{eps_xy}
\epsilon_{xy}' = \frac{2i}{\omega}\bigg(\frac{ \alpha_F Q c}{\pi}+ \frac{\sigma_{xy}}{\epsilon_0}\bigg )
\ee

\subsection{The ``Voigt" geometry}
\label{Voigt}
In this case, the electric field propagates in the $\hat{x}$ direction $(\bs k = k \hat{x})$.

Since the RHS of the wave equation contains only time derivatives, the terms in the RHS remain unchanged, and is given by equation \eqref{RHS} in this geometry too.

The LHS of the wave equation, containing spatial derivatives modifies for this geometry as:
\be
\begin{pmatrix} 
0 & 0 & 0\\
0 &  {\partial^2}/{\partial x^2} & 0\\ 
0 & 0 & {\partial^2}/{\partial x^2}
\end{pmatrix}
\mathbf{E} = -k^2\mathbf{E}
\ee
with the second term $\nabla (\nabla\cdot{\bs E}) = 0$ in this case too.
For this geometry, equations \eqref{wave_eq1} and \eqref{tensor1} are modified to be 
\begin{widetext}
	\be
	n^2 \begin{pmatrix}
		0\\
		 E_y\\
		E_z\\
	\end{pmatrix}
	= \frac{i }{\omega \epsilon_0}\begin{pmatrix} 
		\sigma_{xx} & \sigma_{xy} & 0\\
		-\sigma_{xy} &  \sigma_{yy} & 0\\ 
		0 & 0 & \sigma_{zz}
	\end{pmatrix}
	\begin{pmatrix}
		E_x\\
		E_y\\
		E_z
	\end{pmatrix}
	+ \begin{pmatrix} 
		0 & \frac{2i\alpha_F Q c}{\pi\omega} & 0\\
		-\frac{2i\alpha_F Q c}{\pi\omega} &  0 & 0\\ 
		0 & 0 & 0
	\end{pmatrix}
      \begin{pmatrix}                                         
		E_x\\
		E_y\\
		E_z
	\end{pmatrix}
	+ \epsilon_b \mathbb{I}
		\begin{pmatrix}
		E_x\\
		E_y\\
		E_z
	\end{pmatrix}~,
	\ee
\be
n^2\begin{pmatrix}
	0\\
	E_y\\
	E_z\\
\end{pmatrix} = \begin{pmatrix}
	\epsilon'_{xx} & \epsilon'_{xy} & 0\\
	-\epsilon'_{xy} & \epsilon'_{yy} & 0\\
	0 & 0 & \epsilon'_{zz}\\
\end{pmatrix} 
\begin{pmatrix}
	E_x \\
	E_y\\
	E_z\\
\end{pmatrix}~.
\ee
\end{widetext}
Here, the elements of the modified dielectric tensor are given by equations \eqref{eps_xx} and \eqref{eps_xy}.

\bibliography{Ref2.bib}

\begin{thebibliography}{73}%
\makeatletter
\providecommand \@ifxundefined [1]{%
 \@ifx{#1\undefined}
}%
\providecommand \@ifnum [1]{%
 \ifnum #1\expandafter \@firstoftwo
 \else \expandafter \@secondoftwo
 \fi
}%
\providecommand \@ifx [1]{%
 \ifx #1\expandafter \@firstoftwo
 \else \expandafter \@secondoftwo
 \fi
}%
\providecommand \natexlab [1]{#1}%
\providecommand \enquote  [1]{``#1''}%
\providecommand \bibnamefont  [1]{#1}%
\providecommand \bibfnamefont [1]{#1}%
\providecommand \citenamefont [1]{#1}%
\providecommand \href@noop [0]{\@secondoftwo}%
\providecommand \href [0]{\begingroup \@sanitize@url \@href}%
\providecommand \@href[1]{\@@startlink{#1}\@@href}%
\providecommand \@@href[1]{\endgroup#1\@@endlink}%
\providecommand \@sanitize@url [0]{\catcode `\\12\catcode `\$12\catcode
  `\&12\catcode `\#12\catcode `\^12\catcode `\_12\catcode `\%12\relax}%
\providecommand \@@startlink[1]{}%
\providecommand \@@endlink[0]{}%
\providecommand \url  [0]{\begingroup\@sanitize@url \@url }%
\providecommand \@url [1]{\endgroup\@href {#1}{\urlprefix }}%
\providecommand \urlprefix  [0]{URL }%
\providecommand \Eprint [0]{\href }%
\providecommand \doibase [0]{http://dx.doi.org/}%
\providecommand \selectlanguage [0]{\@gobble}%
\providecommand \bibinfo  [0]{\@secondoftwo}%
\providecommand \bibfield  [0]{\@secondoftwo}%
\providecommand \translation [1]{[#1]}%
\providecommand \BibitemOpen [0]{}%
\providecommand \bibitemStop [0]{}%
\providecommand \bibitemNoStop [0]{.\EOS\space}%
\providecommand \EOS [0]{\spacefactor3000\relax}%
\providecommand \BibitemShut  [1]{\csname bibitem#1\endcsname}%
\let\auto@bib@innerbib\@empty
\bibitem [{\citenamefont {Nielsen}\ and\ \citenamefont
  {Ninomiya}(1983)}]{Nielsen83}%
  \BibitemOpen
  \bibfield  {author} {\bibinfo {author} {\bibfnamefont {H.B.}\ \bibnamefont
  {Nielsen}}\ and\ \bibinfo {author} {\bibfnamefont {Masao}\ \bibnamefont
  {Ninomiya}},\ }\bibfield  {title} {\enquote {\bibinfo {title} {The
  adler-bell-jackiw anomaly and weyl fermions in a crystal},}\ }\href {\doibase
  https://doi.org/10.1016/0370-2693(83)91529-0} {\bibfield  {journal} {\bibinfo
   {journal} {Physics Letters B}\ }\textbf {\bibinfo {volume} {130}},\ \bibinfo
  {pages} {389 -- 396} (\bibinfo {year} {1983})}\BibitemShut {NoStop}%
\bibitem [{\citenamefont {Wan}\ \emph {et~al.}(2011)\citenamefont {Wan},
  \citenamefont {Turner}, \citenamefont {Vishwanath},\ and\ \citenamefont
  {Savrasov}}]{PhysRevB.83.205101}%
  \BibitemOpen
  \bibfield  {author} {\bibinfo {author} {\bibfnamefont {Xiangang}\
  \bibnamefont {Wan}}, \bibinfo {author} {\bibfnamefont {Ari~M.}\ \bibnamefont
  {Turner}}, \bibinfo {author} {\bibfnamefont {Ashvin}\ \bibnamefont
  {Vishwanath}}, \ and\ \bibinfo {author} {\bibfnamefont {Sergey~Y.}\
  \bibnamefont {Savrasov}},\ }\bibfield  {title} {\enquote {\bibinfo {title}
  {Topological semimetal and fermi-arc surface states in the electronic
  structure of pyrochlore iridates},}\ }\href {\doibase
  10.1103/PhysRevB.83.205101} {\bibfield  {journal} {\bibinfo  {journal} {Phys.
  Rev. B}\ }\textbf {\bibinfo {volume} {83}},\ \bibinfo {pages} {205101}
  (\bibinfo {year} {2011})}\BibitemShut {NoStop}%
\bibitem [{\citenamefont {{Soluyanov Alexey A.}}\ \emph
  {et~al.}(2015)\citenamefont {{Soluyanov Alexey A.}}, \citenamefont {{Gresch
  Dominik}}, \citenamefont {{Wang Zhijun}}, \citenamefont {{Wu QuanSheng}},
  \citenamefont {{Troyer Matthias}}, \citenamefont {{Dai Xi}},\ and\
  \citenamefont {{Bernevig B. Andrei}}}]{alexey}%
  \BibitemOpen
  \bibfield  {author} {\bibinfo {author} {\bibnamefont {{Soluyanov Alexey
  A.}}}, \bibinfo {author} {\bibnamefont {{Gresch Dominik}}}, \bibinfo {author}
  {\bibnamefont {{Wang Zhijun}}}, \bibinfo {author} {\bibnamefont {{Wu
  QuanSheng}}}, \bibinfo {author} {\bibnamefont {{Troyer Matthias}}}, \bibinfo
  {author} {\bibnamefont {{Dai Xi}}}, \ and\ \bibinfo {author} {\bibnamefont
  {{Bernevig B. Andrei}}},\ }\bibfield  {title} {\enquote {\bibinfo {title}
  {{Type-II Weyl semimetals}},}\ }\href {\doibase
  http://dx.doi.org/10.1038/nature15768 10.1038/nature15768} {\bibfield
  {journal} {\bibinfo  {journal} {Nature}\ }\textbf {\bibinfo {volume} {527}},\
  \bibinfo {pages} {495} (\bibinfo {year} {2015})}\BibitemShut {NoStop}%
\bibitem [{\citenamefont {{Jia Shuang}}\ \emph {et~al.}(2016)\citenamefont
  {{Jia Shuang}}, \citenamefont {{Xu Su-Yang}},\ and\ \citenamefont {{Hasan M.
  Zahid}}}]{Jia}%
  \BibitemOpen
  \bibfield  {author} {\bibinfo {author} {\bibnamefont {{Jia Shuang}}},
  \bibinfo {author} {\bibnamefont {{Xu Su-Yang}}}, \ and\ \bibinfo {author}
  {\bibnamefont {{Hasan M. Zahid}}},\ }\bibfield  {title} {\enquote {\bibinfo
  {title} {{Weyl semimetals, Fermi arcs and chiral anomalies}},}\ }\href
  {\doibase 10.1038/nmat4787} {\bibfield  {journal} {\bibinfo  {journal}
  {Nature Materials}\ }\textbf {\bibinfo {volume} {15}},\ \bibinfo {pages}
  {1140} (\bibinfo {year} {2016})}\BibitemShut {NoStop}%
\bibitem [{\citenamefont {Bansil}\ \emph {et~al.}(2016)\citenamefont {Bansil},
  \citenamefont {Lin},\ and\ \citenamefont {Das}}]{Bansil16}%
  \BibitemOpen
  \bibfield  {author} {\bibinfo {author} {\bibfnamefont {A.}~\bibnamefont
  {Bansil}}, \bibinfo {author} {\bibfnamefont {Hsin}\ \bibnamefont {Lin}}, \
  and\ \bibinfo {author} {\bibfnamefont {Tanmoy}\ \bibnamefont {Das}},\
  }\bibfield  {title} {\enquote {\bibinfo {title} {\textit{Colloquium} :
  Topological band theory},}\ }\href {\doibase 10.1103/RevModPhys.88.021004}
  {\bibfield  {journal} {\bibinfo  {journal} {Rev. Mod. Phys.}\ }\textbf
  {\bibinfo {volume} {88}},\ \bibinfo {pages} {021004} (\bibinfo {year}
  {2016})}\BibitemShut {NoStop}%
\bibitem [{\citenamefont {Yan}\ and\ \citenamefont {Felser}(2017)}]{Yan}%
  \BibitemOpen
  \bibfield  {author} {\bibinfo {author} {\bibfnamefont {Binghai}\ \bibnamefont
  {Yan}}\ and\ \bibinfo {author} {\bibfnamefont {Claudia}\ \bibnamefont
  {Felser}},\ }\bibfield  {title} {\enquote {\bibinfo {title} {{Topological
  Materials: Weyl Semimetals}},}\ }\href {\doibase
  10.1146/annurev-conmatphys-031016-025458} {\bibfield  {journal} {\bibinfo
  {journal} {Annual Review of Condensed Matter Physics}\ }\textbf {\bibinfo
  {volume} {8}},\ \bibinfo {pages} {337--354} (\bibinfo {year}
  {2017})}\BibitemShut {NoStop}%
\bibitem [{\citenamefont {Armitage}\ \emph {et~al.}(2018)\citenamefont
  {Armitage}, \citenamefont {Mele},\ and\ \citenamefont
  {Vishwanath}}]{RevModPhys.90.015001}%
  \BibitemOpen
  \bibfield  {author} {\bibinfo {author} {\bibfnamefont {N.~P.}\ \bibnamefont
  {Armitage}}, \bibinfo {author} {\bibfnamefont {E.~J.}\ \bibnamefont {Mele}},
  \ and\ \bibinfo {author} {\bibfnamefont {Ashvin}\ \bibnamefont
  {Vishwanath}},\ }\bibfield  {title} {\enquote {\bibinfo {title} {Weyl and
  dirac semimetals in three-dimensional solids},}\ }\href {\doibase
  10.1103/RevModPhys.90.015001} {\bibfield  {journal} {\bibinfo  {journal}
  {Rev. Mod. Phys.}\ }\textbf {\bibinfo {volume} {90}},\ \bibinfo {pages}
  {015001} (\bibinfo {year} {2018})}\BibitemShut {NoStop}%
\bibitem [{\citenamefont {Burkov}\ and\ \citenamefont
  {Balents}(2011)}]{PhysRevLett.107.127205}%
  \BibitemOpen
  \bibfield  {author} {\bibinfo {author} {\bibfnamefont {A.~A.}\ \bibnamefont
  {Burkov}}\ and\ \bibinfo {author} {\bibfnamefont {Leon}\ \bibnamefont
  {Balents}},\ }\bibfield  {title} {\enquote {\bibinfo {title} {{Weyl Semimetal
  in a Topological Insulator Multilayer}},}\ }\href {\doibase
  10.1103/PhysRevLett.107.127205} {\bibfield  {journal} {\bibinfo  {journal}
  {Phys. Rev. Lett.}\ }\textbf {\bibinfo {volume} {107}},\ \bibinfo {pages}
  {127205} (\bibinfo {year} {2011})}\BibitemShut {NoStop}%
\bibitem [{\citenamefont {Juyal}\ \emph {et~al.}(2018)\citenamefont {Juyal},
  \citenamefont {Agarwal},\ and\ \citenamefont
  {Mukhopadhyay}}]{PhysRevLett.120.096801}%
  \BibitemOpen
  \bibfield  {author} {\bibinfo {author} {\bibfnamefont {Abhishek}\
  \bibnamefont {Juyal}}, \bibinfo {author} {\bibfnamefont {Amit}\ \bibnamefont
  {Agarwal}}, \ and\ \bibinfo {author} {\bibfnamefont {Soumik}\ \bibnamefont
  {Mukhopadhyay}},\ }\bibfield  {title} {\enquote {\bibinfo {title} {Negative
  longitudinal magnetoresistance in the density wave phase of
  ${\mathrm{y}}_{2}{\mathrm{ir}}_{2}{\mathrm{o}}_{7}$},}\ }\href {\doibase
  10.1103/PhysRevLett.120.096801} {\bibfield  {journal} {\bibinfo  {journal}
  {Phys. Rev. Lett.}\ }\textbf {\bibinfo {volume} {120}},\ \bibinfo {pages}
  {096801} (\bibinfo {year} {2018})}\BibitemShut {NoStop}%
\bibitem [{\citenamefont {Wang}\ \emph
  {et~al.}(2016{\natexlab{a}})\citenamefont {Wang}, \citenamefont {Zhang},
  \citenamefont {Huang}, \citenamefont {Nie}, \citenamefont {Liu},
  \citenamefont {Liang}, \citenamefont {Zhang}, \citenamefont {Shen},
  \citenamefont {Liu}, \citenamefont {Hu}, \citenamefont {Ding}, \citenamefont
  {Liu}, \citenamefont {Hu}, \citenamefont {He}, \citenamefont {Zhao},
  \citenamefont {Yu}, \citenamefont {Hu}, \citenamefont {Wei}, \citenamefont
  {Mao}, \citenamefont {Shi}, \citenamefont {Jia}, \citenamefont {Zhang},
  \citenamefont {Zhang}, \citenamefont {Yang}, \citenamefont {Wang},
  \citenamefont {Peng}, \citenamefont {Weng}, \citenamefont {Dai},
  \citenamefont {Fang}, \citenamefont {Xu}, \citenamefont {Chen},\ and\
  \citenamefont {Zhou}}]{PhysRevB.94.241119}%
  \BibitemOpen
  \bibfield  {author} {\bibinfo {author} {\bibfnamefont {Chenlu}\ \bibnamefont
  {Wang}}, \bibinfo {author} {\bibfnamefont {Yan}\ \bibnamefont {Zhang}},
  \bibinfo {author} {\bibfnamefont {Jianwei}\ \bibnamefont {Huang}}, \bibinfo
  {author} {\bibfnamefont {Simin}\ \bibnamefont {Nie}}, \bibinfo {author}
  {\bibfnamefont {Guodong}\ \bibnamefont {Liu}}, \bibinfo {author}
  {\bibfnamefont {Aiji}\ \bibnamefont {Liang}}, \bibinfo {author}
  {\bibfnamefont {Yuxiao}\ \bibnamefont {Zhang}}, \bibinfo {author}
  {\bibfnamefont {Bing}\ \bibnamefont {Shen}}, \bibinfo {author} {\bibfnamefont
  {Jing}\ \bibnamefont {Liu}}, \bibinfo {author} {\bibfnamefont {Cheng}\
  \bibnamefont {Hu}}, \bibinfo {author} {\bibfnamefont {Ying}\ \bibnamefont
  {Ding}}, \bibinfo {author} {\bibfnamefont {Defa}\ \bibnamefont {Liu}},
  \bibinfo {author} {\bibfnamefont {Yong}\ \bibnamefont {Hu}}, \bibinfo
  {author} {\bibfnamefont {Shaolong}\ \bibnamefont {He}}, \bibinfo {author}
  {\bibfnamefont {Lin}\ \bibnamefont {Zhao}}, \bibinfo {author} {\bibfnamefont
  {Li}~\bibnamefont {Yu}}, \bibinfo {author} {\bibfnamefont {Jin}\ \bibnamefont
  {Hu}}, \bibinfo {author} {\bibfnamefont {Jiang}\ \bibnamefont {Wei}},
  \bibinfo {author} {\bibfnamefont {Zhiqiang}\ \bibnamefont {Mao}}, \bibinfo
  {author} {\bibfnamefont {Youguo}\ \bibnamefont {Shi}}, \bibinfo {author}
  {\bibfnamefont {Xiaowen}\ \bibnamefont {Jia}}, \bibinfo {author}
  {\bibfnamefont {Fengfeng}\ \bibnamefont {Zhang}}, \bibinfo {author}
  {\bibfnamefont {Shenjin}\ \bibnamefont {Zhang}}, \bibinfo {author}
  {\bibfnamefont {Feng}\ \bibnamefont {Yang}}, \bibinfo {author} {\bibfnamefont
  {Zhimin}\ \bibnamefont {Wang}}, \bibinfo {author} {\bibfnamefont {Qinjun}\
  \bibnamefont {Peng}}, \bibinfo {author} {\bibfnamefont {Hongming}\
  \bibnamefont {Weng}}, \bibinfo {author} {\bibfnamefont {Xi}~\bibnamefont
  {Dai}}, \bibinfo {author} {\bibfnamefont {Zhong}\ \bibnamefont {Fang}},
  \bibinfo {author} {\bibfnamefont {Zuyan}\ \bibnamefont {Xu}}, \bibinfo
  {author} {\bibfnamefont {Chuangtian}\ \bibnamefont {Chen}}, \ and\ \bibinfo
  {author} {\bibfnamefont {X.~J.}\ \bibnamefont {Zhou}},\ }\bibfield  {title}
  {\enquote {\bibinfo {title} {Observation of fermi arc and its connection with
  bulk states in the candidate type-ii weyl semimetal ${\mathrm{wte}}_{2}$},}\
  }\href {\doibase 10.1103/PhysRevB.94.241119} {\bibfield  {journal} {\bibinfo
  {journal} {Phys. Rev. B}\ }\textbf {\bibinfo {volume} {94}},\ \bibinfo
  {pages} {241119} (\bibinfo {year} {2016}{\natexlab{a}})}\BibitemShut
  {NoStop}%
\bibitem [{\citenamefont {{Li Peng}}\ \emph {et~al.}(2017)\citenamefont {{Li
  Peng}}, \citenamefont {{Wen Yan}}, \citenamefont {{He Xin}}, \citenamefont
  {{Zhang Qiang}}, \citenamefont {{Xia Chuan}}, \citenamefont {{Yu Zhi-Ming}},
  \citenamefont {{Yang Shengyuan A.}}, \citenamefont {{Zhu Zhiyong}},
  \citenamefont {{Alshareef Husam N.}},\ and\ \citenamefont {{Zhang
  Xi-Xiang}}}]{Li2017}%
  \BibitemOpen
  \bibfield  {author} {\bibinfo {author} {\bibnamefont {{Li Peng}}}, \bibinfo
  {author} {\bibnamefont {{Wen Yan}}}, \bibinfo {author} {\bibnamefont {{He
  Xin}}}, \bibinfo {author} {\bibnamefont {{Zhang Qiang}}}, \bibinfo {author}
  {\bibnamefont {{Xia Chuan}}}, \bibinfo {author} {\bibnamefont {{Yu
  Zhi-Ming}}}, \bibinfo {author} {\bibnamefont {{Yang Shengyuan A.}}}, \bibinfo
  {author} {\bibnamefont {{Zhu Zhiyong}}}, \bibinfo {author} {\bibnamefont
  {{Alshareef Husam N.}}}, \ and\ \bibinfo {author} {\bibnamefont {{Zhang
  Xi-Xiang}}},\ }\bibfield  {title} {\enquote {\bibinfo {title} {{Evidence for
  topological type-II Weyl semimetal WTe2}},}\ }\href
  {https://doi.org/10.1038/s41467-017-02237-1} {\bibfield  {journal} {\bibinfo
  {journal} {Nature Communications}\ }\textbf {\bibinfo {volume} {8}},\
  \bibinfo {pages} {2150} (\bibinfo {year} {2017})}\BibitemShut {NoStop}%
\bibitem [{\citenamefont {Sun}\ \emph {et~al.}(2015{\natexlab{a}})\citenamefont
  {Sun}, \citenamefont {Wu}, \citenamefont {Ali}, \citenamefont {Felser},\ and\
  \citenamefont {Yan}}]{PhysRevB.92.161107}%
  \BibitemOpen
  \bibfield  {author} {\bibinfo {author} {\bibfnamefont {Yan}\ \bibnamefont
  {Sun}}, \bibinfo {author} {\bibfnamefont {Shu-Chun}\ \bibnamefont {Wu}},
  \bibinfo {author} {\bibfnamefont {Mazhar~N.}\ \bibnamefont {Ali}}, \bibinfo
  {author} {\bibfnamefont {Claudia}\ \bibnamefont {Felser}}, \ and\ \bibinfo
  {author} {\bibfnamefont {Binghai}\ \bibnamefont {Yan}},\ }\bibfield  {title}
  {\enquote {\bibinfo {title} {Prediction of weyl semimetal in orthorhombic
  ${\mathrm{mote}}_{2}$},}\ }\href {\doibase 10.1103/PhysRevB.92.161107}
  {\bibfield  {journal} {\bibinfo  {journal} {Phys. Rev. B}\ }\textbf {\bibinfo
  {volume} {92}},\ \bibinfo {pages} {161107} (\bibinfo {year}
  {2015}{\natexlab{a}})}\BibitemShut {NoStop}%
\bibitem [{\citenamefont {{Deng Ke}}\ \emph {et~al.}(2016)\citenamefont {{Deng
  Ke}}, \citenamefont {{Wan Guoliang}}, \citenamefont {{Deng Peng}},
  \citenamefont {{Zhang Kenan}}, \citenamefont {{Ding Shijie}}, \citenamefont
  {{Wang Eryin}}, \citenamefont {{Yan Mingzhe}}, \citenamefont {{Huang
  Huaqing}}, \citenamefont {{Zhang Hongyun}}, \citenamefont {{Xu Zhilin}},
  \citenamefont {{Denlinger Jonathan}}, \citenamefont {{Fedorov Alexei}},
  \citenamefont {{Yang Haitao}}, \citenamefont {{Duan Wenhui}}, \citenamefont
  {{Yao Hong}}, \citenamefont {{Wu Yang}}, \citenamefont {{Fan Shoushan}},
  \citenamefont {{Zhang Haijun}}, \citenamefont {{Chen Xi}},\ and\
  \citenamefont {{Zhou Shuyun}}}]{Deng}%
  \BibitemOpen
  \bibfield  {author} {\bibinfo {author} {\bibnamefont {{Deng Ke}}}, \bibinfo
  {author} {\bibnamefont {{Wan Guoliang}}}, \bibinfo {author} {\bibnamefont
  {{Deng Peng}}}, \bibinfo {author} {\bibnamefont {{Zhang Kenan}}}, \bibinfo
  {author} {\bibnamefont {{Ding Shijie}}}, \bibinfo {author} {\bibnamefont
  {{Wang Eryin}}}, \bibinfo {author} {\bibnamefont {{Yan Mingzhe}}}, \bibinfo
  {author} {\bibnamefont {{Huang Huaqing}}}, \bibinfo {author} {\bibnamefont
  {{Zhang Hongyun}}}, \bibinfo {author} {\bibnamefont {{Xu Zhilin}}}, \bibinfo
  {author} {\bibnamefont {{Denlinger Jonathan}}}, \bibinfo {author}
  {\bibnamefont {{Fedorov Alexei}}}, \bibinfo {author} {\bibnamefont {{Yang
  Haitao}}}, \bibinfo {author} {\bibnamefont {{Duan Wenhui}}}, \bibinfo
  {author} {\bibnamefont {{Yao Hong}}}, \bibinfo {author} {\bibnamefont {{Wu
  Yang}}}, \bibinfo {author} {\bibnamefont {{Fan Shoushan}}}, \bibinfo {author}
  {\bibnamefont {{Zhang Haijun}}}, \bibinfo {author} {\bibnamefont {{Chen
  Xi}}}, \ and\ \bibinfo {author} {\bibnamefont {{Zhou Shuyun}}},\ }\bibfield
  {title} {\enquote {\bibinfo {title} {{Experimental observation of topological
  Fermi arcs in type-II Weyl semimetal MoTe2}},}\ }\href {\doibase
  http://dx.doi.org/10.1038/nphys3871 10.1038/nphys3871} {\bibfield  {journal}
  {\bibinfo  {journal} {Nature Physics}\ }\textbf {\bibinfo {volume} {12}},\
  \bibinfo {pages} {1105} (\bibinfo {year} {2016})}\BibitemShut {NoStop}%
\bibitem [{\citenamefont {{Huang Lunan}}\ \emph {et~al.}(2016)\citenamefont
  {{Huang Lunan}}, \citenamefont {{McCormick Timothy M.}}, \citenamefont {{Ochi
  Masayuki}}, \citenamefont {{Zhao Zhiying}}, \citenamefont {{Suzuki
  Michi-To}}, \citenamefont {{Arita Ryotaro}}, \citenamefont {{Wu Yun}},
  \citenamefont {{Mou Daixiang}}, \citenamefont {{Cao Huibo}}, \citenamefont
  {{Yan Jiaqiang}}, \citenamefont {{Trivedi Nandini}},\ and\ \citenamefont
  {{Kaminski Adam}}}]{Lunan}%
  \BibitemOpen
  \bibfield  {author} {\bibinfo {author} {\bibnamefont {{Huang Lunan}}},
  \bibinfo {author} {\bibnamefont {{McCormick Timothy M.}}}, \bibinfo {author}
  {\bibnamefont {{Ochi Masayuki}}}, \bibinfo {author} {\bibnamefont {{Zhao
  Zhiying}}}, \bibinfo {author} {\bibnamefont {{Suzuki Michi-To}}}, \bibinfo
  {author} {\bibnamefont {{Arita Ryotaro}}}, \bibinfo {author} {\bibnamefont
  {{Wu Yun}}}, \bibinfo {author} {\bibnamefont {{Mou Daixiang}}}, \bibinfo
  {author} {\bibnamefont {{Cao Huibo}}}, \bibinfo {author} {\bibnamefont {{Yan
  Jiaqiang}}}, \bibinfo {author} {\bibnamefont {{Trivedi Nandini}}}, \ and\
  \bibinfo {author} {\bibnamefont {{Kaminski Adam}}},\ }\bibfield  {title}
  {\enquote {\bibinfo {title} {{Spectroscopic evidence for a type II Weyl
  semimetallic state in MoTe2}},}\ }\href {\doibase
  http://dx.doi.org/10.1038/nmat4685 10.1038/nmat4685} {\bibfield  {journal}
  {\bibinfo  {journal} {Nature Materials}\ }\textbf {\bibinfo {volume} {15}},\
  \bibinfo {pages} {1155} (\bibinfo {year} {2016})}\BibitemShut {NoStop}%
\bibitem [{\citenamefont {Wang}\ \emph
  {et~al.}(2016{\natexlab{b}})\citenamefont {Wang}, \citenamefont {Gresch},
  \citenamefont {Soluyanov}, \citenamefont {Xie}, \citenamefont {Kushwaha},
  \citenamefont {Dai}, \citenamefont {Troyer}, \citenamefont {Cava},\ and\
  \citenamefont {Bernevig}}]{Wang}%
  \BibitemOpen
  \bibfield  {author} {\bibinfo {author} {\bibfnamefont {Zhijun}\ \bibnamefont
  {Wang}}, \bibinfo {author} {\bibfnamefont {Dominik}\ \bibnamefont {Gresch}},
  \bibinfo {author} {\bibfnamefont {Alexey~A.}\ \bibnamefont {Soluyanov}},
  \bibinfo {author} {\bibfnamefont {Weiwei}\ \bibnamefont {Xie}}, \bibinfo
  {author} {\bibfnamefont {S.}~\bibnamefont {Kushwaha}}, \bibinfo {author}
  {\bibfnamefont {Xi}~\bibnamefont {Dai}}, \bibinfo {author} {\bibfnamefont
  {Matthias}\ \bibnamefont {Troyer}}, \bibinfo {author} {\bibfnamefont
  {Robert~J.}\ \bibnamefont {Cava}}, \ and\ \bibinfo {author} {\bibfnamefont
  {B.~Andrei}\ \bibnamefont {Bernevig}},\ }\bibfield  {title} {\enquote
  {\bibinfo {title} {${\mathrm{mote}}_{2}$: A type-ii weyl topological
  metal},}\ }\href {\doibase 10.1103/PhysRevLett.117.056805} {\bibfield
  {journal} {\bibinfo  {journal} {Phys. Rev. Lett.}\ }\textbf {\bibinfo
  {volume} {117}},\ \bibinfo {pages} {056805} (\bibinfo {year}
  {2016}{\natexlab{b}})}\BibitemShut {NoStop}%
\bibitem [{\citenamefont {Tamai}\ \emph {et~al.}(2016)\citenamefont {Tamai},
  \citenamefont {Wu}, \citenamefont {Cucchi}, \citenamefont {Bruno},
  \citenamefont {Ricc\`o}, \citenamefont {Kim}, \citenamefont {Hoesch},
  \citenamefont {Barreteau}, \citenamefont {Giannini}, \citenamefont {Besnard},
  \citenamefont {Soluyanov},\ and\ \citenamefont {Baumberger}}]{Tamai}%
  \BibitemOpen
  \bibfield  {author} {\bibinfo {author} {\bibfnamefont {A.}~\bibnamefont
  {Tamai}}, \bibinfo {author} {\bibfnamefont {Q.~S.}\ \bibnamefont {Wu}},
  \bibinfo {author} {\bibfnamefont {I.}~\bibnamefont {Cucchi}}, \bibinfo
  {author} {\bibfnamefont {F.~Y.}\ \bibnamefont {Bruno}}, \bibinfo {author}
  {\bibfnamefont {S.}~\bibnamefont {Ricc\`o}}, \bibinfo {author} {\bibfnamefont
  {T.~K.}\ \bibnamefont {Kim}}, \bibinfo {author} {\bibfnamefont
  {M.}~\bibnamefont {Hoesch}}, \bibinfo {author} {\bibfnamefont
  {C.}~\bibnamefont {Barreteau}}, \bibinfo {author} {\bibfnamefont
  {E.}~\bibnamefont {Giannini}}, \bibinfo {author} {\bibfnamefont
  {C.}~\bibnamefont {Besnard}}, \bibinfo {author} {\bibfnamefont {A.~A.}\
  \bibnamefont {Soluyanov}}, \ and\ \bibinfo {author} {\bibfnamefont
  {F.}~\bibnamefont {Baumberger}},\ }\bibfield  {title} {\enquote {\bibinfo
  {title} {Fermi arcs and their topological character in the candidate type-ii
  weyl semimetal ${\mathrm{mote}}_{2}$},}\ }\href {\doibase
  10.1103/PhysRevX.6.031021} {\bibfield  {journal} {\bibinfo  {journal} {Phys.
  Rev. X}\ }\textbf {\bibinfo {volume} {6}},\ \bibinfo {pages} {031021}
  (\bibinfo {year} {2016})}\BibitemShut {NoStop}%
\bibitem [{\citenamefont {{Jiang J.}}\ \emph {et~al.}(2017)\citenamefont
  {{Jiang J.}}, \citenamefont {{Liu Z.K.}}, \citenamefont {{Sun Y.}},
  \citenamefont {{Yang H.F.}}, \citenamefont {{Rajamathi C.R.}}, \citenamefont
  {{Qi Y.P.}}, \citenamefont {{Yang L.X.}}, \citenamefont {{Chen C.}},
  \citenamefont {{Peng H.}}, \citenamefont {{Hwang C-C.}}, \citenamefont {{Sun
  S.Z.}}, \citenamefont {{Mo S-K.}}, \citenamefont {{Vobornik I.}},
  \citenamefont {{Fujii J.}}, \citenamefont {{Parkin S.S.P.}}, \citenamefont
  {{Felser C.}}, \citenamefont {{Yan B.H.}},\ and\ \citenamefont {{Chen
  Y.L.}}}]{Jiang}%
  \BibitemOpen
  \bibfield  {author} {\bibinfo {author} {\bibnamefont {{Jiang J.}}}, \bibinfo
  {author} {\bibnamefont {{Liu Z.K.}}}, \bibinfo {author} {\bibnamefont {{Sun
  Y.}}}, \bibinfo {author} {\bibnamefont {{Yang H.F.}}}, \bibinfo {author}
  {\bibnamefont {{Rajamathi C.R.}}}, \bibinfo {author} {\bibnamefont {{Qi
  Y.P.}}}, \bibinfo {author} {\bibnamefont {{Yang L.X.}}}, \bibinfo {author}
  {\bibnamefont {{Chen C.}}}, \bibinfo {author} {\bibnamefont {{Peng H.}}},
  \bibinfo {author} {\bibnamefont {{Hwang C-C.}}}, \bibinfo {author}
  {\bibnamefont {{Sun S.Z.}}}, \bibinfo {author} {\bibnamefont {{Mo S-K.}}},
  \bibinfo {author} {\bibnamefont {{Vobornik I.}}}, \bibinfo {author}
  {\bibnamefont {{Fujii J.}}}, \bibinfo {author} {\bibnamefont {{Parkin
  S.S.P.}}}, \bibinfo {author} {\bibnamefont {{Felser C.}}}, \bibinfo {author}
  {\bibnamefont {{Yan B.H.}}}, \ and\ \bibinfo {author} {\bibnamefont {{Chen
  Y.L.}}},\ }\bibfield  {title} {\enquote {\bibinfo {title} {{Signature of
  type-II Weyl semimetal phase in MoTe2}},}\ }\href {\doibase
  http://dx.doi.org/10.1038/ncomms13973 10.1038/ncomms13973} {\bibfield
  {journal} {\bibinfo  {journal} {Nature Communications}\ }\textbf {\bibinfo
  {volume} {8}},\ \bibinfo {pages} {13973} (\bibinfo {year}
  {2017})}\BibitemShut {NoStop}%
\bibitem [{\citenamefont {Belopolski}\ \emph {et~al.}(2016)\citenamefont
  {Belopolski}, \citenamefont {Xu}, \citenamefont {Ishida}, \citenamefont
  {Pan}, \citenamefont {Yu}, \citenamefont {Sanchez}, \citenamefont {Zheng},
  \citenamefont {Neupane}, \citenamefont {Alidoust}, \citenamefont {Chang},
  \citenamefont {Chang}, \citenamefont {Wu}, \citenamefont {Bian},
  \citenamefont {Huang}, \citenamefont {Lee}, \citenamefont {Mou},
  \citenamefont {Huang}, \citenamefont {Song}, \citenamefont {Wang},
  \citenamefont {Wang}, \citenamefont {Yeh}, \citenamefont {Yao}, \citenamefont
  {Rault}, \citenamefont {{Le F{\`e}vre}}, \citenamefont {Bertran},
  \citenamefont {Jeng}, \citenamefont {Kondo}, \citenamefont {Kaminski},
  \citenamefont {Lin}, \citenamefont {Liu}, \citenamefont {Song}, \citenamefont
  {Shin},\ and\ \citenamefont {Hasan}}]{PhysRevB.94.085127}%
  \BibitemOpen
  \bibfield  {author} {\bibinfo {author} {\bibfnamefont {Ilya}\ \bibnamefont
  {Belopolski}}, \bibinfo {author} {\bibfnamefont {Su-Yang}\ \bibnamefont
  {Xu}}, \bibinfo {author} {\bibfnamefont {Yukiaki}\ \bibnamefont {Ishida}},
  \bibinfo {author} {\bibfnamefont {Xingchen}\ \bibnamefont {Pan}}, \bibinfo
  {author} {\bibfnamefont {Peng}\ \bibnamefont {Yu}}, \bibinfo {author}
  {\bibfnamefont {Daniel~S.}\ \bibnamefont {Sanchez}}, \bibinfo {author}
  {\bibfnamefont {Hao}\ \bibnamefont {Zheng}}, \bibinfo {author} {\bibfnamefont
  {Madhab}\ \bibnamefont {Neupane}}, \bibinfo {author} {\bibfnamefont {Nasser}\
  \bibnamefont {Alidoust}}, \bibinfo {author} {\bibfnamefont {Guoqing}\
  \bibnamefont {Chang}}, \bibinfo {author} {\bibfnamefont {Tay-Rong}\
  \bibnamefont {Chang}}, \bibinfo {author} {\bibfnamefont {Yun}\ \bibnamefont
  {Wu}}, \bibinfo {author} {\bibfnamefont {Guang}\ \bibnamefont {Bian}},
  \bibinfo {author} {\bibfnamefont {Shin-Ming}\ \bibnamefont {Huang}}, \bibinfo
  {author} {\bibfnamefont {Chi-Cheng}\ \bibnamefont {Lee}}, \bibinfo {author}
  {\bibfnamefont {Daixiang}\ \bibnamefont {Mou}}, \bibinfo {author}
  {\bibfnamefont {Lunan}\ \bibnamefont {Huang}}, \bibinfo {author}
  {\bibfnamefont {You}\ \bibnamefont {Song}}, \bibinfo {author} {\bibfnamefont
  {Baigeng}\ \bibnamefont {Wang}}, \bibinfo {author} {\bibfnamefont {Guanghou}\
  \bibnamefont {Wang}}, \bibinfo {author} {\bibfnamefont {Yao-Wen}\
  \bibnamefont {Yeh}}, \bibinfo {author} {\bibfnamefont {Nan}\ \bibnamefont
  {Yao}}, \bibinfo {author} {\bibfnamefont {Julien~E.}\ \bibnamefont {Rault}},
  \bibinfo {author} {\bibfnamefont {Patrick}\ \bibnamefont {{Le F{\`e}vre}}},
  \bibinfo {author} {\bibfnamefont {Fran\ifmmode
  \mbox\c{c}\else~\c{c}\fi{}ois}\ \bibnamefont {Bertran}}, \bibinfo {author}
  {\bibfnamefont {Horng-Tay}\ \bibnamefont {Jeng}}, \bibinfo {author}
  {\bibfnamefont {Takeshi}\ \bibnamefont {Kondo}}, \bibinfo {author}
  {\bibfnamefont {Adam}\ \bibnamefont {Kaminski}}, \bibinfo {author}
  {\bibfnamefont {Hsin}\ \bibnamefont {Lin}}, \bibinfo {author} {\bibfnamefont
  {Zheng}\ \bibnamefont {Liu}}, \bibinfo {author} {\bibfnamefont {Fengqi}\
  \bibnamefont {Song}}, \bibinfo {author} {\bibfnamefont {Shik}\ \bibnamefont
  {Shin}}, \ and\ \bibinfo {author} {\bibfnamefont {M.~Zahid}\ \bibnamefont
  {Hasan}},\ }\bibfield  {title} {\enquote {\bibinfo {title} {{Fermi arc
  electronic structure and Chern numbers in the type-II Weyl semimetal
  candidate
  ${\mathrm{Mo}}_{x}{\mathrm{W}}_{1\ensuremath{-}x}{\mathrm{Te}}_{2}$}},}\
  }\href {\doibase 10.1103/PhysRevB.94.085127} {\bibfield  {journal} {\bibinfo
  {journal} {Phys. Rev. B}\ }\textbf {\bibinfo {volume} {94}},\ \bibinfo
  {pages} {085127} (\bibinfo {year} {2016})}\BibitemShut {NoStop}%
\bibitem [{\citenamefont {{Chang Tay-Rong}}\ \emph {et~al.}(2016)\citenamefont
  {{Chang Tay-Rong}}, \citenamefont {{Xu Su-Yang}}, \citenamefont {{Chang
  Guoqing}}, \citenamefont {{Lee Chi-Cheng}}, \citenamefont {{Huang
  Shin-Ming}}, \citenamefont {{Wang BaoKai}}, \citenamefont {{Bian Guang}},
  \citenamefont {{Zheng Hao}}, \citenamefont {{Sanchez Daniel S.}},
  \citenamefont {{Belopolski Ilya}}, \citenamefont {{Alidoust Nasser}},
  \citenamefont {{Neupane Madhab}}, \citenamefont {{Bansil Arun}},
  \citenamefont {{Jeng Horng-Tay}}, \citenamefont {{Lin Hsin}},\ and\
  \citenamefont {{Zahid Hasan M.}}}]{Chang}%
  \BibitemOpen
  \bibfield  {author} {\bibinfo {author} {\bibnamefont {{Chang Tay-Rong}}},
  \bibinfo {author} {\bibnamefont {{Xu Su-Yang}}}, \bibinfo {author}
  {\bibnamefont {{Chang Guoqing}}}, \bibinfo {author} {\bibnamefont {{Lee
  Chi-Cheng}}}, \bibinfo {author} {\bibnamefont {{Huang Shin-Ming}}}, \bibinfo
  {author} {\bibnamefont {{Wang BaoKai}}}, \bibinfo {author} {\bibnamefont
  {{Bian Guang}}}, \bibinfo {author} {\bibnamefont {{Zheng Hao}}}, \bibinfo
  {author} {\bibnamefont {{Sanchez Daniel S.}}}, \bibinfo {author}
  {\bibnamefont {{Belopolski Ilya}}}, \bibinfo {author} {\bibnamefont
  {{Alidoust Nasser}}}, \bibinfo {author} {\bibnamefont {{Neupane Madhab}}},
  \bibinfo {author} {\bibnamefont {{Bansil Arun}}}, \bibinfo {author}
  {\bibnamefont {{Jeng Horng-Tay}}}, \bibinfo {author} {\bibnamefont {{Lin
  Hsin}}}, \ and\ \bibinfo {author} {\bibnamefont {{Zahid Hasan M.}}},\
  }\bibfield  {title} {\enquote {\bibinfo {title} {{Prediction of an
  arc-tunable Weyl Fermion metallic state in MoxW1−xTe2}},}\ }\href {\doibase
  http://dx.doi.org/10.1038/ncomms10639 10.1038/ncomms10639} {\bibfield
  {journal} {\bibinfo  {journal} {Nature Communications}\ }\textbf {\bibinfo
  {volume} {7}},\ \bibinfo {pages} {10639} (\bibinfo {year}
  {2016})}\BibitemShut {NoStop}%
\bibitem [{\citenamefont {Lv}\ \emph {et~al.}(2015)\citenamefont {Lv},
  \citenamefont {Weng}, \citenamefont {Fu}, \citenamefont {Wang}, \citenamefont
  {Miao}, \citenamefont {Ma}, \citenamefont {Richard}, \citenamefont {Huang},
  \citenamefont {Zhao}, \citenamefont {Chen}, \citenamefont {Fang},
  \citenamefont {Dai}, \citenamefont {Qian},\ and\ \citenamefont {Ding}}]{Lv}%
  \BibitemOpen
  \bibfield  {author} {\bibinfo {author} {\bibfnamefont {B.~Q.}\ \bibnamefont
  {Lv}}, \bibinfo {author} {\bibfnamefont {H.~M.}\ \bibnamefont {Weng}},
  \bibinfo {author} {\bibfnamefont {B.~B.}\ \bibnamefont {Fu}}, \bibinfo
  {author} {\bibfnamefont {X.~P.}\ \bibnamefont {Wang}}, \bibinfo {author}
  {\bibfnamefont {H.}~\bibnamefont {Miao}}, \bibinfo {author} {\bibfnamefont
  {J.}~\bibnamefont {Ma}}, \bibinfo {author} {\bibfnamefont {P.}~\bibnamefont
  {Richard}}, \bibinfo {author} {\bibfnamefont {X.~C.}\ \bibnamefont {Huang}},
  \bibinfo {author} {\bibfnamefont {L.~X.}\ \bibnamefont {Zhao}}, \bibinfo
  {author} {\bibfnamefont {G.~F.}\ \bibnamefont {Chen}}, \bibinfo {author}
  {\bibfnamefont {Z.}~\bibnamefont {Fang}}, \bibinfo {author} {\bibfnamefont
  {X.}~\bibnamefont {Dai}}, \bibinfo {author} {\bibfnamefont {T.}~\bibnamefont
  {Qian}}, \ and\ \bibinfo {author} {\bibfnamefont {H.}~\bibnamefont {Ding}},\
  }\bibfield  {title} {\enquote {\bibinfo {title} {Experimental discovery of
  weyl semimetal taas},}\ }\href {\doibase 10.1103/PhysRevX.5.031013}
  {\bibfield  {journal} {\bibinfo  {journal} {Phys. Rev. X}\ }\textbf {\bibinfo
  {volume} {5}},\ \bibinfo {pages} {031013} (\bibinfo {year}
  {2015})}\BibitemShut {NoStop}%
\bibitem [{\citenamefont {{Lv B. Q.}}\ \emph {et~al.}(2015)\citenamefont {{Lv
  B. Q.}}, \citenamefont {{Xu N.}}, \citenamefont {{Weng H. M.}}, \citenamefont
  {{Ma J. Z.}}, \citenamefont {{Richard P.}}, \citenamefont {{Huang X. C.}},
  \citenamefont {{Zhao L. X.}}, \citenamefont {{Chen G. F.}}, \citenamefont
  {{Matt C. E.}}, \citenamefont {{Bisti F.}}, \citenamefont {{Strocov V. N.}},
  \citenamefont {{Mesot J.}}, \citenamefont {{Fang Z.}}, \citenamefont {{Dai
  X.}}, \citenamefont {{Qian T.}}, \citenamefont {{Shi M.}},\ and\
  \citenamefont {{Ding H.}}}]{Lv1}%
  \BibitemOpen
  \bibfield  {author} {\bibinfo {author} {\bibnamefont {{Lv B. Q.}}}, \bibinfo
  {author} {\bibnamefont {{Xu N.}}}, \bibinfo {author} {\bibnamefont {{Weng H.
  M.}}}, \bibinfo {author} {\bibnamefont {{Ma J. Z.}}}, \bibinfo {author}
  {\bibnamefont {{Richard P.}}}, \bibinfo {author} {\bibnamefont {{Huang X.
  C.}}}, \bibinfo {author} {\bibnamefont {{Zhao L. X.}}}, \bibinfo {author}
  {\bibnamefont {{Chen G. F.}}}, \bibinfo {author} {\bibnamefont {{Matt C.
  E.}}}, \bibinfo {author} {\bibnamefont {{Bisti F.}}}, \bibinfo {author}
  {\bibnamefont {{Strocov V. N.}}}, \bibinfo {author} {\bibnamefont {{Mesot
  J.}}}, \bibinfo {author} {\bibnamefont {{Fang Z.}}}, \bibinfo {author}
  {\bibnamefont {{Dai X.}}}, \bibinfo {author} {\bibnamefont {{Qian T.}}},
  \bibinfo {author} {\bibnamefont {{Shi M.}}}, \ and\ \bibinfo {author}
  {\bibnamefont {{Ding H.}}},\ }\bibfield  {title} {\enquote {\bibinfo {title}
  {{Observation of Weyl nodes in TaAs}},}\ }\href {\doibase
  http://dx.doi.org/10.1038/nphys3426 10.1038/nphys3426} {\bibfield  {journal}
  {\bibinfo  {journal} {Nature Physics}\ }\textbf {\bibinfo {volume} {11}},\
  \bibinfo {pages} {724} (\bibinfo {year} {2015})}\BibitemShut {NoStop}%
\bibitem [{\citenamefont {{Yang L. X.}}\ \emph {et~al.}(2015)\citenamefont
  {{Yang L. X.}}, \citenamefont {{Liu Z. K.}}, \citenamefont {{Sun Y.}},
  \citenamefont {{Peng H.}}, \citenamefont {{Yang H. F.}}, \citenamefont
  {{Zhang T.}}, \citenamefont {{Zhou B.}}, \citenamefont {{Zhang Y.}},
  \citenamefont {{Guo Y. F.}}, \citenamefont {{Rahn M.}}, \citenamefont
  {{Prabhakaran D.}}, \citenamefont {{Hussain Z.}}, \citenamefont {{Mo S.-K.}},
  \citenamefont {{Felser C.}}, \citenamefont {{Yan B.}},\ and\ \citenamefont
  {{Chen Y. L.}}}]{Yang}%
  \BibitemOpen
  \bibfield  {author} {\bibinfo {author} {\bibnamefont {{Yang L. X.}}},
  \bibinfo {author} {\bibnamefont {{Liu Z. K.}}}, \bibinfo {author}
  {\bibnamefont {{Sun Y.}}}, \bibinfo {author} {\bibnamefont {{Peng H.}}},
  \bibinfo {author} {\bibnamefont {{Yang H. F.}}}, \bibinfo {author}
  {\bibnamefont {{Zhang T.}}}, \bibinfo {author} {\bibnamefont {{Zhou B.}}},
  \bibinfo {author} {\bibnamefont {{Zhang Y.}}}, \bibinfo {author}
  {\bibnamefont {{Guo Y. F.}}}, \bibinfo {author} {\bibnamefont {{Rahn M.}}},
  \bibinfo {author} {\bibnamefont {{Prabhakaran D.}}}, \bibinfo {author}
  {\bibnamefont {{Hussain Z.}}}, \bibinfo {author} {\bibnamefont {{Mo S.-K.}}},
  \bibinfo {author} {\bibnamefont {{Felser C.}}}, \bibinfo {author}
  {\bibnamefont {{Yan B.}}}, \ and\ \bibinfo {author} {\bibnamefont {{Chen Y.
  L.}}},\ }\bibfield  {title} {\enquote {\bibinfo {title} {{Weyl semimetal
  phase in the non-centrosymmetric compound TaAs}},}\ }\href {\doibase
  http://dx.doi.org/10.1038/nphys3425 10.1038/nphys3425} {\bibfield  {journal}
  {\bibinfo  {journal} {Nature Physics}\ }\textbf {\bibinfo {volume} {11}},\
  \bibinfo {pages} {728} (\bibinfo {year} {2015})}\BibitemShut {NoStop}%
\bibitem [{\citenamefont {{Xu N.}}\ \emph {et~al.}(2016)\citenamefont {{Xu
  N.}}, \citenamefont {{Weng H. M.}}, \citenamefont {{Lv B. Q.}}, \citenamefont
  {{Matt C. E.}}, \citenamefont {{Park J.}}, \citenamefont {{Bisti F.}},
  \citenamefont {{Strocov V. N.}}, \citenamefont {{Gawryluk D.}}, \citenamefont
  {{Pomjakushina E.}}, \citenamefont {{Conder K.}}, \citenamefont {{Plumb N.
  C.}}, \citenamefont {{Radovic M.}}, \citenamefont {{Autès G.}},
  \citenamefont {{Yazyev O. V.}}, \citenamefont {{Fang Z.}}, \citenamefont
  {{Dai X.}}, \citenamefont {{Qian T.}}, \citenamefont {{Mesot J.}},
  \citenamefont {{Ding H.}},\ and\ \citenamefont {{Shi M.}}}]{Xu}%
  \BibitemOpen
  \bibfield  {author} {\bibinfo {author} {\bibnamefont {{Xu N.}}}, \bibinfo
  {author} {\bibnamefont {{Weng H. M.}}}, \bibinfo {author} {\bibnamefont {{Lv
  B. Q.}}}, \bibinfo {author} {\bibnamefont {{Matt C. E.}}}, \bibinfo {author}
  {\bibnamefont {{Park J.}}}, \bibinfo {author} {\bibnamefont {{Bisti F.}}},
  \bibinfo {author} {\bibnamefont {{Strocov V. N.}}}, \bibinfo {author}
  {\bibnamefont {{Gawryluk D.}}}, \bibinfo {author} {\bibnamefont
  {{Pomjakushina E.}}}, \bibinfo {author} {\bibnamefont {{Conder K.}}},
  \bibinfo {author} {\bibnamefont {{Plumb N. C.}}}, \bibinfo {author}
  {\bibnamefont {{Radovic M.}}}, \bibinfo {author} {\bibnamefont {{Autès
  G.}}}, \bibinfo {author} {\bibnamefont {{Yazyev O. V.}}}, \bibinfo {author}
  {\bibnamefont {{Fang Z.}}}, \bibinfo {author} {\bibnamefont {{Dai X.}}},
  \bibinfo {author} {\bibnamefont {{Qian T.}}}, \bibinfo {author} {\bibnamefont
  {{Mesot J.}}}, \bibinfo {author} {\bibnamefont {{Ding H.}}}, \ and\ \bibinfo
  {author} {\bibnamefont {{Shi M.}}},\ }\bibfield  {title} {\enquote {\bibinfo
  {title} {{Observation of Weyl nodes and Fermi arcs in tantalum phosphide}},}\
  }\href {\doibase http://dx.doi.org/10.1038/ncomms11006 10.1038/ncomms11006}
  {\bibfield  {journal} {\bibinfo  {journal} {Nature Communications}\ }\textbf
  {\bibinfo {volume} {7}},\ \bibinfo {pages} {11006} (\bibinfo {year}
  {2016})}\BibitemShut {NoStop}%
\bibitem [{\citenamefont {Xu}\ \emph {et~al.}(2015)\citenamefont {Xu},
  \citenamefont {Belopolski}, \citenamefont {Sanchez}, \citenamefont {Zhang},
  \citenamefont {Chang}, \citenamefont {Guo}, \citenamefont {Bian},
  \citenamefont {Yuan}, \citenamefont {Lu}, \citenamefont {Chang},
  \citenamefont {Shibayev}, \citenamefont {Prokopovych}, \citenamefont
  {Alidoust}, \citenamefont {Zheng}, \citenamefont {Lee}, \citenamefont
  {Huang}, \citenamefont {Sankar}, \citenamefont {Chou}, \citenamefont {Hsu},
  \citenamefont {Jeng}, \citenamefont {Bansil}, \citenamefont {Neupert},
  \citenamefont {Strocov}, \citenamefont {Lin}, \citenamefont {Jia},\ and\
  \citenamefont {Hasan}}]{Xue1501092}%
  \BibitemOpen
  \bibfield  {author} {\bibinfo {author} {\bibfnamefont {Su-Yang}\ \bibnamefont
  {Xu}}, \bibinfo {author} {\bibfnamefont {Ilya}\ \bibnamefont {Belopolski}},
  \bibinfo {author} {\bibfnamefont {Daniel~S.}\ \bibnamefont {Sanchez}},
  \bibinfo {author} {\bibfnamefont {Chenglong}\ \bibnamefont {Zhang}}, \bibinfo
  {author} {\bibfnamefont {Guoqing}\ \bibnamefont {Chang}}, \bibinfo {author}
  {\bibfnamefont {Cheng}\ \bibnamefont {Guo}}, \bibinfo {author} {\bibfnamefont
  {Guang}\ \bibnamefont {Bian}}, \bibinfo {author} {\bibfnamefont {Zhujun}\
  \bibnamefont {Yuan}}, \bibinfo {author} {\bibfnamefont {Hong}\ \bibnamefont
  {Lu}}, \bibinfo {author} {\bibfnamefont {Tay-Rong}\ \bibnamefont {Chang}},
  \bibinfo {author} {\bibfnamefont {Pavel~P.}\ \bibnamefont {Shibayev}},
  \bibinfo {author} {\bibfnamefont {Mykhailo~L.}\ \bibnamefont {Prokopovych}},
  \bibinfo {author} {\bibfnamefont {Nasser}\ \bibnamefont {Alidoust}}, \bibinfo
  {author} {\bibfnamefont {Hao}\ \bibnamefont {Zheng}}, \bibinfo {author}
  {\bibfnamefont {Chi-Cheng}\ \bibnamefont {Lee}}, \bibinfo {author}
  {\bibfnamefont {Shin-Ming}\ \bibnamefont {Huang}}, \bibinfo {author}
  {\bibfnamefont {Raman}\ \bibnamefont {Sankar}}, \bibinfo {author}
  {\bibfnamefont {Fangcheng}\ \bibnamefont {Chou}}, \bibinfo {author}
  {\bibfnamefont {Chuang-Han}\ \bibnamefont {Hsu}}, \bibinfo {author}
  {\bibfnamefont {Horng-Tay}\ \bibnamefont {Jeng}}, \bibinfo {author}
  {\bibfnamefont {Arun}\ \bibnamefont {Bansil}}, \bibinfo {author}
  {\bibfnamefont {Titus}\ \bibnamefont {Neupert}}, \bibinfo {author}
  {\bibfnamefont {Vladimir~N.}\ \bibnamefont {Strocov}}, \bibinfo {author}
  {\bibfnamefont {Hsin}\ \bibnamefont {Lin}}, \bibinfo {author} {\bibfnamefont
  {Shuang}\ \bibnamefont {Jia}}, \ and\ \bibinfo {author} {\bibfnamefont
  {M.~Zahid}\ \bibnamefont {Hasan}},\ }\bibfield  {title} {\enquote {\bibinfo
  {title} {Experimental discovery of a topological weyl semimetal state in
  tap},}\ }\href {\doibase 10.1126/sciadv.1501092} {\bibfield  {journal}
  {\bibinfo  {journal} {Science Advances}\ }\textbf {\bibinfo {volume} {1}},\
  \bibinfo {pages} {e1501092} (\bibinfo {year} {2015})}\BibitemShut {NoStop}%
\bibitem [{\citenamefont {Sun}\ \emph {et~al.}(2015{\natexlab{b}})\citenamefont
  {Sun}, \citenamefont {Wu},\ and\ \citenamefont {Yan}}]{Sun}%
  \BibitemOpen
  \bibfield  {author} {\bibinfo {author} {\bibfnamefont {Yan}\ \bibnamefont
  {Sun}}, \bibinfo {author} {\bibfnamefont {Shu-Chun}\ \bibnamefont {Wu}}, \
  and\ \bibinfo {author} {\bibfnamefont {Binghai}\ \bibnamefont {Yan}},\
  }\bibfield  {title} {\enquote {\bibinfo {title} {Topological surface states
  and fermi arcs of the noncentrosymmetric weyl semimetals taas, tap, nbas, and
  nbp},}\ }\href {\doibase 10.1103/PhysRevB.92.115428} {\bibfield  {journal}
  {\bibinfo  {journal} {Phys. Rev. B}\ }\textbf {\bibinfo {volume} {92}},\
  \bibinfo {pages} {115428} (\bibinfo {year} {2015}{\natexlab{b}})}\BibitemShut
  {NoStop}%
\bibitem [{\citenamefont {Haubold}\ \emph {et~al.}(2017)\citenamefont
  {Haubold}, \citenamefont {Koepernik}, \citenamefont {Efremov}, \citenamefont
  {Khim}, \citenamefont {Fedorov}, \citenamefont {Kushnirenko}, \citenamefont
  {van~den Brink}, \citenamefont {Wurmehl}, \citenamefont {B\"uchner},
  \citenamefont {Kim}, \citenamefont {Hoesch}, \citenamefont {Sumida},
  \citenamefont {Taguchi}, \citenamefont {Yoshikawa}, \citenamefont {Kimura},
  \citenamefont {Okuda},\ and\ \citenamefont {Borisenko}}]{Haubold}%
  \BibitemOpen
  \bibfield  {author} {\bibinfo {author} {\bibfnamefont {E.}~\bibnamefont
  {Haubold}}, \bibinfo {author} {\bibfnamefont {K.}~\bibnamefont {Koepernik}},
  \bibinfo {author} {\bibfnamefont {D.}~\bibnamefont {Efremov}}, \bibinfo
  {author} {\bibfnamefont {S.}~\bibnamefont {Khim}}, \bibinfo {author}
  {\bibfnamefont {A.}~\bibnamefont {Fedorov}}, \bibinfo {author} {\bibfnamefont
  {Y.}~\bibnamefont {Kushnirenko}}, \bibinfo {author} {\bibfnamefont
  {J.}~\bibnamefont {van~den Brink}}, \bibinfo {author} {\bibfnamefont
  {S.}~\bibnamefont {Wurmehl}}, \bibinfo {author} {\bibfnamefont
  {B.}~\bibnamefont {B\"uchner}}, \bibinfo {author} {\bibfnamefont {T.~K.}\
  \bibnamefont {Kim}}, \bibinfo {author} {\bibfnamefont {M.}~\bibnamefont
  {Hoesch}}, \bibinfo {author} {\bibfnamefont {K.}~\bibnamefont {Sumida}},
  \bibinfo {author} {\bibfnamefont {K.}~\bibnamefont {Taguchi}}, \bibinfo
  {author} {\bibfnamefont {T.}~\bibnamefont {Yoshikawa}}, \bibinfo {author}
  {\bibfnamefont {A.}~\bibnamefont {Kimura}}, \bibinfo {author} {\bibfnamefont
  {T.}~\bibnamefont {Okuda}}, \ and\ \bibinfo {author} {\bibfnamefont {S.~V.}\
  \bibnamefont {Borisenko}},\ }\bibfield  {title} {\enquote {\bibinfo {title}
  {Experimental realization of type-ii weyl state in noncentrosymmetric
  ${\mathrm{tairte}}_{4}$},}\ }\href {\doibase 10.1103/PhysRevB.95.241108}
  {\bibfield  {journal} {\bibinfo  {journal} {Phys. Rev. B}\ }\textbf {\bibinfo
  {volume} {95}},\ \bibinfo {pages} {241108} (\bibinfo {year}
  {2017})}\BibitemShut {NoStop}%
\bibitem [{\citenamefont {Xu}\ \emph {et~al.}(2017)\citenamefont {Xu},
  \citenamefont {Alidoust}, \citenamefont {Chang}, \citenamefont {Lu},
  \citenamefont {Singh}, \citenamefont {Belopolski}, \citenamefont {Sanchez},
  \citenamefont {Zhang}, \citenamefont {Bian}, \citenamefont {Zheng},
  \citenamefont {Husanu}, \citenamefont {Bian}, \citenamefont {Huang},
  \citenamefont {Hsu}, \citenamefont {Chang}, \citenamefont {Jeng},
  \citenamefont {Bansil}, \citenamefont {Neupert}, \citenamefont {Strocov},
  \citenamefont {Lin}, \citenamefont {Jia},\ and\ \citenamefont
  {Hasan}}]{Xue1603266}%
  \BibitemOpen
  \bibfield  {author} {\bibinfo {author} {\bibfnamefont {Su-Yang}\ \bibnamefont
  {Xu}}, \bibinfo {author} {\bibfnamefont {Nasser}\ \bibnamefont {Alidoust}},
  \bibinfo {author} {\bibfnamefont {Guoqing}\ \bibnamefont {Chang}}, \bibinfo
  {author} {\bibfnamefont {Hong}\ \bibnamefont {Lu}}, \bibinfo {author}
  {\bibfnamefont {Bahadur}\ \bibnamefont {Singh}}, \bibinfo {author}
  {\bibfnamefont {Ilya}\ \bibnamefont {Belopolski}}, \bibinfo {author}
  {\bibfnamefont {Daniel~S.}\ \bibnamefont {Sanchez}}, \bibinfo {author}
  {\bibfnamefont {Xiao}\ \bibnamefont {Zhang}}, \bibinfo {author}
  {\bibfnamefont {Guang}\ \bibnamefont {Bian}}, \bibinfo {author}
  {\bibfnamefont {Hao}\ \bibnamefont {Zheng}}, \bibinfo {author} {\bibfnamefont
  {Marious-Adrian}\ \bibnamefont {Husanu}}, \bibinfo {author} {\bibfnamefont
  {Yi}~\bibnamefont {Bian}}, \bibinfo {author} {\bibfnamefont {Shin-Ming}\
  \bibnamefont {Huang}}, \bibinfo {author} {\bibfnamefont {Chuang-Han}\
  \bibnamefont {Hsu}}, \bibinfo {author} {\bibfnamefont {Tay-Rong}\
  \bibnamefont {Chang}}, \bibinfo {author} {\bibfnamefont {Horng-Tay}\
  \bibnamefont {Jeng}}, \bibinfo {author} {\bibfnamefont {Arun}\ \bibnamefont
  {Bansil}}, \bibinfo {author} {\bibfnamefont {Titus}\ \bibnamefont {Neupert}},
  \bibinfo {author} {\bibfnamefont {Vladimir~N.}\ \bibnamefont {Strocov}},
  \bibinfo {author} {\bibfnamefont {Hsin}\ \bibnamefont {Lin}}, \bibinfo
  {author} {\bibfnamefont {Shuang}\ \bibnamefont {Jia}}, \ and\ \bibinfo
  {author} {\bibfnamefont {M.~Zahid}\ \bibnamefont {Hasan}},\ }\bibfield
  {title} {\enquote {\bibinfo {title} {Discovery of lorentz-violating type ii
  weyl fermions in laalge},}\ }\href {\doibase 10.1126/sciadv.1603266}
  {\bibfield  {journal} {\bibinfo  {journal} {Science Advances}\ }\textbf
  {\bibinfo {volume} {3}},\ \bibinfo {pages} {e1603266} (\bibinfo {year}
  {2017})}\BibitemShut {NoStop}%
\bibitem [{\citenamefont {Ferreiros}\ \emph {et~al.}(2017)\citenamefont
  {Ferreiros}, \citenamefont {Zyuzin},\ and\ \citenamefont
  {Bardarson}}]{PhysRevB.96.115202}%
  \BibitemOpen
  \bibfield  {author} {\bibinfo {author} {\bibfnamefont {Yago}\ \bibnamefont
  {Ferreiros}}, \bibinfo {author} {\bibfnamefont {A.~A.}\ \bibnamefont
  {Zyuzin}}, \ and\ \bibinfo {author} {\bibfnamefont {Jens~H.}\ \bibnamefont
  {Bardarson}},\ }\bibfield  {title} {\enquote {\bibinfo {title} {{Anomalous
  Nernst and thermal Hall effects in tilted Weyl semimetals}},}\ }\href
  {\doibase 10.1103/PhysRevB.96.115202} {\bibfield  {journal} {\bibinfo
  {journal} {Phys. Rev. B}\ }\textbf {\bibinfo {volume} {96}},\ \bibinfo
  {pages} {115202} (\bibinfo {year} {2017})}\BibitemShut {NoStop}%
\bibitem [{\citenamefont {Hosur}\ \emph {et~al.}(2012)\citenamefont {Hosur},
  \citenamefont {Parameswaran},\ and\ \citenamefont
  {Vishwanath}}]{PhysRevLett.108.046602}%
  \BibitemOpen
  \bibfield  {author} {\bibinfo {author} {\bibfnamefont {Pavan}\ \bibnamefont
  {Hosur}}, \bibinfo {author} {\bibfnamefont {S.~A.}\ \bibnamefont
  {Parameswaran}}, \ and\ \bibinfo {author} {\bibfnamefont {Ashvin}\
  \bibnamefont {Vishwanath}},\ }\bibfield  {title} {\enquote {\bibinfo {title}
  {Charge transport in weyl semimetals},}\ }\href {\doibase
  10.1103/PhysRevLett.108.046602} {\bibfield  {journal} {\bibinfo  {journal}
  {Phys. Rev. Lett.}\ }\textbf {\bibinfo {volume} {108}},\ \bibinfo {pages}
  {046602} (\bibinfo {year} {2012})}\BibitemShut {NoStop}%
\bibitem [{\citenamefont {Wang}\ \emph {et~al.}(2017)\citenamefont {Wang},
  \citenamefont {Lin}, \citenamefont {Wang}, \citenamefont {Yu},\ and\
  \citenamefont {Liao}}]{Shou}%
  \BibitemOpen
  \bibfield  {author} {\bibinfo {author} {\bibfnamefont {Shuo}\ \bibnamefont
  {Wang}}, \bibinfo {author} {\bibfnamefont {Ben-Chuan}\ \bibnamefont {Lin}},
  \bibinfo {author} {\bibfnamefont {An-Qi}\ \bibnamefont {Wang}}, \bibinfo
  {author} {\bibfnamefont {Da-Peng}\ \bibnamefont {Yu}}, \ and\ \bibinfo
  {author} {\bibfnamefont {Zhi-Min}\ \bibnamefont {Liao}},\ }\bibfield  {title}
  {\enquote {\bibinfo {title} {Quantum transport in dirac and weyl semimetals:
  a review},}\ }\href {\doibase 10.1080/23746149.2017.1327329} {\bibfield
  {journal} {\bibinfo  {journal} {Advances in Physics: X}\ }\textbf {\bibinfo
  {volume} {2}},\ \bibinfo {pages} {518--544} (\bibinfo {year}
  {2017})}\BibitemShut {NoStop}%
\bibitem [{\citenamefont {Das}\ and\ \citenamefont
  {Agarwal}(2019)}]{Kamal2018}%
  \BibitemOpen
  \bibfield  {author} {\bibinfo {author} {\bibfnamefont {Kamal}\ \bibnamefont
  {Das}}\ and\ \bibinfo {author} {\bibfnamefont {Amit}\ \bibnamefont
  {Agarwal}},\ }\bibfield  {title} {\enquote {\bibinfo {title} {Linear
  magnetochiral transport in tilted type-i and type-ii weyl semimetals},}\
  }\href {\doibase 10.1103/PhysRevB.99.085405} {\bibfield  {journal} {\bibinfo
  {journal} {Phys. Rev. B}\ }\textbf {\bibinfo {volume} {99}},\ \bibinfo
  {pages} {085405} (\bibinfo {year} {2019})}\BibitemShut {NoStop}%
\bibitem [{\citenamefont {Zyuzin}\ and\ \citenamefont
  {Burkov}(2012)}]{Zyuzin_Burkov12}%
  \BibitemOpen
  \bibfield  {author} {\bibinfo {author} {\bibfnamefont {A.~A.}\ \bibnamefont
  {Zyuzin}}\ and\ \bibinfo {author} {\bibfnamefont {A.~A.}\ \bibnamefont
  {Burkov}},\ }\bibfield  {title} {\enquote {\bibinfo {title} {Topological
  response in weyl semimetals and the chiral anomaly},}\ }\href {\doibase
  10.1103/PhysRevB.86.115133} {\bibfield  {journal} {\bibinfo  {journal} {Phys.
  Rev. B}\ }\textbf {\bibinfo {volume} {86}},\ \bibinfo {pages} {115133}
  (\bibinfo {year} {2012})}\BibitemShut {NoStop}%
\bibitem [{\citenamefont {Son}\ and\ \citenamefont {Yamamoto}(2012)}]{Son12}%
  \BibitemOpen
  \bibfield  {author} {\bibinfo {author} {\bibfnamefont {Dam~Thanh}\
  \bibnamefont {Son}}\ and\ \bibinfo {author} {\bibfnamefont {Naoki}\
  \bibnamefont {Yamamoto}},\ }\bibfield  {title} {\enquote {\bibinfo {title}
  {Berry curvature, triangle anomalies, and the chiral magnetic effect in fermi
  liquids},}\ }\href {\doibase 10.1103/PhysRevLett.109.181602} {\bibfield
  {journal} {\bibinfo  {journal} {Phys. Rev. Lett.}\ }\textbf {\bibinfo
  {volume} {109}},\ \bibinfo {pages} {181602} (\bibinfo {year}
  {2012})}\BibitemShut {NoStop}%
\bibitem [{\citenamefont {Burkov}(2014)}]{Burkov14}%
  \BibitemOpen
  \bibfield  {author} {\bibinfo {author} {\bibfnamefont {A.~A.}\ \bibnamefont
  {Burkov}},\ }\bibfield  {title} {\enquote {\bibinfo {title} {Anomalous hall
  effect in weyl metals},}\ }\href {\doibase 10.1103/PhysRevLett.113.187202}
  {\bibfield  {journal} {\bibinfo  {journal} {Phys. Rev. Lett.}\ }\textbf
  {\bibinfo {volume} {113}},\ \bibinfo {pages} {187202} (\bibinfo {year}
  {2014})}\BibitemShut {NoStop}%
\bibitem [{\citenamefont {Kim}\ \emph {et~al.}(2014)\citenamefont {Kim},
  \citenamefont {Kim},\ and\ \citenamefont {Sasaki}}]{Kim14}%
  \BibitemOpen
  \bibfield  {author} {\bibinfo {author} {\bibfnamefont {Ki-Seok}\ \bibnamefont
  {Kim}}, \bibinfo {author} {\bibfnamefont {Heon-Jung}\ \bibnamefont {Kim}}, \
  and\ \bibinfo {author} {\bibfnamefont {M.}~\bibnamefont {Sasaki}},\
  }\bibfield  {title} {\enquote {\bibinfo {title} {Boltzmann equation approach
  to anomalous transport in a weyl metal},}\ }\href {\doibase
  10.1103/PhysRevB.89.195137} {\bibfield  {journal} {\bibinfo  {journal} {Phys.
  Rev. B}\ }\textbf {\bibinfo {volume} {89}},\ \bibinfo {pages} {195137}
  (\bibinfo {year} {2014})}\BibitemShut {NoStop}%
\bibitem [{\citenamefont {Steiner}\ \emph {et~al.}(2017)\citenamefont
  {Steiner}, \citenamefont {Andreev},\ and\ \citenamefont
  {Pesin}}]{PhysRevLett.119.036601}%
  \BibitemOpen
  \bibfield  {author} {\bibinfo {author} {\bibfnamefont {J.~F.}\ \bibnamefont
  {Steiner}}, \bibinfo {author} {\bibfnamefont {A.~V.}\ \bibnamefont
  {Andreev}}, \ and\ \bibinfo {author} {\bibfnamefont {D.~A.}\ \bibnamefont
  {Pesin}},\ }\bibfield  {title} {\enquote {\bibinfo {title} {{Anomalous Hall
  Effect in Type-I Weyl Metals}},}\ }\href {\doibase
  10.1103/PhysRevLett.119.036601} {\bibfield  {journal} {\bibinfo  {journal}
  {Phys. Rev. Lett.}\ }\textbf {\bibinfo {volume} {119}},\ \bibinfo {pages}
  {036601} (\bibinfo {year} {2017})}\BibitemShut {NoStop}%
\bibitem [{\citenamefont {Zhong}\ \emph {et~al.}(2015)\citenamefont {Zhong},
  \citenamefont {Orenstein},\ and\ \citenamefont
  {Moore}}]{PhysRevLett.115.117403}%
  \BibitemOpen
  \bibfield  {author} {\bibinfo {author} {\bibfnamefont {Shudan}\ \bibnamefont
  {Zhong}}, \bibinfo {author} {\bibfnamefont {Joseph}\ \bibnamefont
  {Orenstein}}, \ and\ \bibinfo {author} {\bibfnamefont {Joel~E.}\ \bibnamefont
  {Moore}},\ }\bibfield  {title} {\enquote {\bibinfo {title} {Optical gyrotropy
  from axion electrodynamics in momentum space},}\ }\href {\doibase
  10.1103/PhysRevLett.115.117403} {\bibfield  {journal} {\bibinfo  {journal}
  {Phys. Rev. Lett.}\ }\textbf {\bibinfo {volume} {115}},\ \bibinfo {pages}
  {117403} (\bibinfo {year} {2015})}\BibitemShut {NoStop}%
\bibitem [{\citenamefont {Goswami}\ \emph {et~al.}(2015)\citenamefont
  {Goswami}, \citenamefont {Sharma},\ and\ \citenamefont
  {Tewari}}]{PhysRevB.92.161110}%
  \BibitemOpen
  \bibfield  {author} {\bibinfo {author} {\bibfnamefont {Pallab}\ \bibnamefont
  {Goswami}}, \bibinfo {author} {\bibfnamefont {Girish}\ \bibnamefont
  {Sharma}}, \ and\ \bibinfo {author} {\bibfnamefont {Sumanta}\ \bibnamefont
  {Tewari}},\ }\bibfield  {title} {\enquote {\bibinfo {title} {Optical activity
  as a test for dynamic chiral magnetic effect of weyl semimetals},}\ }\href
  {\doibase 10.1103/PhysRevB.92.161110} {\bibfield  {journal} {\bibinfo
  {journal} {Phys. Rev. B}\ }\textbf {\bibinfo {volume} {92}},\ \bibinfo
  {pages} {161110} (\bibinfo {year} {2015})}\BibitemShut {NoStop}%
\bibitem [{\citenamefont {Trescher}\ \emph {et~al.}(2015)\citenamefont
  {Trescher}, \citenamefont {Sbierski}, \citenamefont {Brouwer},\ and\
  \citenamefont {Bergholtz}}]{PhysRevB.91.115135}%
  \BibitemOpen
  \bibfield  {author} {\bibinfo {author} {\bibfnamefont {Maximilian}\
  \bibnamefont {Trescher}}, \bibinfo {author} {\bibfnamefont {Bj\"orn}\
  \bibnamefont {Sbierski}}, \bibinfo {author} {\bibfnamefont {Piet~W.}\
  \bibnamefont {Brouwer}}, \ and\ \bibinfo {author} {\bibfnamefont {Emil~J.}\
  \bibnamefont {Bergholtz}},\ }\bibfield  {title} {\enquote {\bibinfo {title}
  {Quantum transport in dirac materials: Signatures of tilted and anisotropic
  dirac and weyl cones},}\ }\href {\doibase 10.1103/PhysRevB.91.115135}
  {\bibfield  {journal} {\bibinfo  {journal} {Phys. Rev. B}\ }\textbf {\bibinfo
  {volume} {91}},\ \bibinfo {pages} {115135} (\bibinfo {year}
  {2015})}\BibitemShut {NoStop}%
\bibitem [{\citenamefont {Hosur}\ and\ \citenamefont
  {Qi}(2015)}]{PhysRevB.91.081106}%
  \BibitemOpen
  \bibfield  {author} {\bibinfo {author} {\bibfnamefont {Pavan}\ \bibnamefont
  {Hosur}}\ and\ \bibinfo {author} {\bibfnamefont {Xiao-Liang}\ \bibnamefont
  {Qi}},\ }\bibfield  {title} {\enquote {\bibinfo {title} {{Tunable circular
  dichroism due to the chiral anomaly in Weyl semimetals}},}\ }\href {\doibase
  10.1103/PhysRevB.91.081106} {\bibfield  {journal} {\bibinfo  {journal} {Phys.
  Rev. B}\ }\textbf {\bibinfo {volume} {91}},\ \bibinfo {pages} {081106}
  (\bibinfo {year} {2015})}\BibitemShut {NoStop}%
\bibitem [{\citenamefont {Burkov}(2017)}]{Burkov17}%
  \BibitemOpen
  \bibfield  {author} {\bibinfo {author} {\bibfnamefont {A.~A.}\ \bibnamefont
  {Burkov}},\ }\bibfield  {title} {\enquote {\bibinfo {title} {Giant planar
  hall effect in topological metals},}\ }\href {\doibase
  10.1103/PhysRevB.96.041110} {\bibfield  {journal} {\bibinfo  {journal} {Phys.
  Rev. B}\ }\textbf {\bibinfo {volume} {96}},\ \bibinfo {pages} {041110}
  (\bibinfo {year} {2017})}\BibitemShut {NoStop}%
\bibitem [{\citenamefont {Thakur}\ \emph {et~al.}(2018)\citenamefont {Thakur},
  \citenamefont {Sadhukhan},\ and\ \citenamefont
  {Agarwal}}]{PhysRevB.97.035403}%
  \BibitemOpen
  \bibfield  {author} {\bibinfo {author} {\bibfnamefont {Anmol}\ \bibnamefont
  {Thakur}}, \bibinfo {author} {\bibfnamefont {Krishanu}\ \bibnamefont
  {Sadhukhan}}, \ and\ \bibinfo {author} {\bibfnamefont {Amit}\ \bibnamefont
  {Agarwal}},\ }\bibfield  {title} {\enquote {\bibinfo {title} {Dynamic
  current-current susceptibility in three-dimensional dirac and weyl
  semimetals},}\ }\href {\doibase 10.1103/PhysRevB.97.035403} {\bibfield
  {journal} {\bibinfo  {journal} {Phys. Rev. B}\ }\textbf {\bibinfo {volume}
  {97}},\ \bibinfo {pages} {035403} (\bibinfo {year} {2018})}\BibitemShut
  {NoStop}%
\bibitem [{\citenamefont {Wu}\ \emph {et~al.}(2016{\natexlab{a}})\citenamefont
  {Wu}, \citenamefont {Patankar}, \citenamefont {Morimoto}, \citenamefont
  {Nair}, \citenamefont {Thewalt}, \citenamefont {Little}, \citenamefont
  {Analytis}, \citenamefont {Moore},\ and\ \citenamefont {Orenstein}}]{Wu2016}%
  \BibitemOpen
  \bibfield  {author} {\bibinfo {author} {\bibfnamefont {Liang}\ \bibnamefont
  {Wu}}, \bibinfo {author} {\bibfnamefont {S.}~\bibnamefont {Patankar}},
  \bibinfo {author} {\bibfnamefont {T.}~\bibnamefont {Morimoto}}, \bibinfo
  {author} {\bibfnamefont {N.~L.}\ \bibnamefont {Nair}}, \bibinfo {author}
  {\bibfnamefont {E.}~\bibnamefont {Thewalt}}, \bibinfo {author} {\bibfnamefont
  {A.}~\bibnamefont {Little}}, \bibinfo {author} {\bibfnamefont {J.~G.}\
  \bibnamefont {Analytis}}, \bibinfo {author} {\bibfnamefont {J.~E.}\
  \bibnamefont {Moore}}, \ and\ \bibinfo {author} {\bibfnamefont
  {J.}~\bibnamefont {Orenstein}},\ }\bibfield  {title} {\enquote {\bibinfo
  {title} {Giant anisotropic nonlinear optical response in transition metal
  monopnictide weyl semimetals},}\ }\href {https://doi.org/10.1038/nphys3969}
  {\bibfield  {journal} {\bibinfo  {journal} {Nature Physics}\ }\textbf
  {\bibinfo {volume} {13}},\ \bibinfo {pages} {350} (\bibinfo {year}
  {2016}{\natexlab{a}})}\BibitemShut {NoStop}%
\bibitem [{\citenamefont {Ma}\ \emph {et~al.}(2017)\citenamefont {Ma},
  \citenamefont {Xu}, \citenamefont {Chan}, \citenamefont {Zhang},
  \citenamefont {Chang}, \citenamefont {Lin}, \citenamefont {Xie},
  \citenamefont {Palacios}, \citenamefont {Lin}, \citenamefont {Jia},
  \citenamefont {Lee}, \citenamefont {Jarillo-Herrero},\ and\ \citenamefont
  {Gedik}}]{Ma2017}%
  \BibitemOpen
  \bibfield  {author} {\bibinfo {author} {\bibfnamefont {Qiong}\ \bibnamefont
  {Ma}}, \bibinfo {author} {\bibfnamefont {Su-Yang}\ \bibnamefont {Xu}},
  \bibinfo {author} {\bibfnamefont {Ching-Kit}\ \bibnamefont {Chan}}, \bibinfo
  {author} {\bibfnamefont {Cheng-Long}\ \bibnamefont {Zhang}}, \bibinfo
  {author} {\bibfnamefont {Guoqing}\ \bibnamefont {Chang}}, \bibinfo {author}
  {\bibfnamefont {Yuxuan}\ \bibnamefont {Lin}}, \bibinfo {author}
  {\bibfnamefont {Weiwei}\ \bibnamefont {Xie}}, \bibinfo {author}
  {\bibfnamefont {Tom{\'a}s}\ \bibnamefont {Palacios}}, \bibinfo {author}
  {\bibfnamefont {Hsin}\ \bibnamefont {Lin}}, \bibinfo {author} {\bibfnamefont
  {Shuang}\ \bibnamefont {Jia}}, \bibinfo {author} {\bibfnamefont {Patrick~A.}\
  \bibnamefont {Lee}}, \bibinfo {author} {\bibfnamefont {Pablo}\ \bibnamefont
  {Jarillo-Herrero}}, \ and\ \bibinfo {author} {\bibfnamefont {Nuh}\
  \bibnamefont {Gedik}},\ }\bibfield  {title} {\enquote {\bibinfo {title}
  {Direct optical detection of weyl fermion chirality in a topological
  semimetal},}\ }\href {https://doi.org/10.1038/nphys4146} {\bibfield
  {journal} {\bibinfo  {journal} {Nature Physics}\ }\textbf {\bibinfo {volume}
  {13}},\ \bibinfo {pages} {842} (\bibinfo {year} {2017})}\BibitemShut
  {NoStop}%
\bibitem [{\citenamefont {Barnes}\ \emph {et~al.}(2016)\citenamefont {Barnes},
  \citenamefont {Heremans},\ and\ \citenamefont
  {Minic}}]{PhysRevLett.117.217204}%
  \BibitemOpen
  \bibfield  {author} {\bibinfo {author} {\bibfnamefont {Edwin}\ \bibnamefont
  {Barnes}}, \bibinfo {author} {\bibfnamefont {J.~J.}\ \bibnamefont
  {Heremans}}, \ and\ \bibinfo {author} {\bibfnamefont {Djordje}\ \bibnamefont
  {Minic}},\ }\bibfield  {title} {\enquote {\bibinfo {title} {Electromagnetic
  signatures of the chiral anomaly in weyl semimetals},}\ }\href {\doibase
  10.1103/PhysRevLett.117.217204} {\bibfield  {journal} {\bibinfo  {journal}
  {Phys. Rev. Lett.}\ }\textbf {\bibinfo {volume} {117}},\ \bibinfo {pages}
  {217204} (\bibinfo {year} {2016})}\BibitemShut {NoStop}%
\bibitem [{\citenamefont {{Kargarian Mehdi}}\ \emph {et~al.}(2015)\citenamefont
  {{Kargarian Mehdi}}, \citenamefont {{Randeria Mohit}},\ and\ \citenamefont
  {{Trivedi Nandini}}}]{Mehdi}%
  \BibitemOpen
  \bibfield  {author} {\bibinfo {author} {\bibnamefont {{Kargarian Mehdi}}},
  \bibinfo {author} {\bibnamefont {{Randeria Mohit}}}, \ and\ \bibinfo {author}
  {\bibnamefont {{Trivedi Nandini}}},\ }\bibfield  {title} {\enquote {\bibinfo
  {title} {{Theory of Kerr and Faraday rotations and linear dichroism in
  Topological Weyl Semimetals}},}\ }\href {\doibase 10.1038/srep12683}
  {\bibfield  {journal} {\bibinfo  {journal} {Scientific Reports}\ }\textbf
  {\bibinfo {volume} {5}},\ \bibinfo {pages} {12683} (\bibinfo {year}
  {2015})}\BibitemShut {NoStop}%
\bibitem [{\citenamefont {Yang}\ \emph {et~al.}(2018)\citenamefont {Yang},
  \citenamefont {Kim},\ and\ \citenamefont {Kim}}]{Yang1}%
  \BibitemOpen
  \bibfield  {author} {\bibinfo {author} {\bibfnamefont {Jinho}\ \bibnamefont
  {Yang}}, \bibinfo {author} {\bibfnamefont {Jeehoon}\ \bibnamefont {Kim}}, \
  and\ \bibinfo {author} {\bibfnamefont {Ki-Seok}\ \bibnamefont {Kim}},\
  }\bibfield  {title} {\enquote {\bibinfo {title} {Transmission and reflection
  coefficients and faraday/kerr rotations as a function of applied magnetic
  fields in spin-orbit coupled dirac metals},}\ }\href {\doibase
  10.1103/PhysRevB.98.075203} {\bibfield  {journal} {\bibinfo  {journal} {Phys.
  Rev. B}\ }\textbf {\bibinfo {volume} {98}},\ \bibinfo {pages} {075203}
  (\bibinfo {year} {2018})}\BibitemShut {NoStop}%
\bibitem [{\citenamefont {Chen}\ \emph {et~al.}(2018)\citenamefont {Chen},
  \citenamefont {Mi}, \citenamefont {Wu}, \citenamefont {Zhang}, \citenamefont
  {Shu}, \citenamefont {Luo},\ and\ \citenamefont {Wen}}]{Chen2018}%
  \BibitemOpen
  \bibfield  {author} {\bibinfo {author} {\bibfnamefont {Shizhen}\ \bibnamefont
  {Chen}}, \bibinfo {author} {\bibfnamefont {Chengquan}\ \bibnamefont {Mi}},
  \bibinfo {author} {\bibfnamefont {Weijie}\ \bibnamefont {Wu}}, \bibinfo
  {author} {\bibfnamefont {Wenshuai}\ \bibnamefont {Zhang}}, \bibinfo {author}
  {\bibfnamefont {Weixing}\ \bibnamefont {Shu}}, \bibinfo {author}
  {\bibfnamefont {Hailu}\ \bibnamefont {Luo}}, \ and\ \bibinfo {author}
  {\bibfnamefont {Shuangchun}\ \bibnamefont {Wen}},\ }\bibfield  {title}
  {\enquote {\bibinfo {title} {Weak-value amplification for weyl-point
  separation in momentum space},}\ }\href {\doibase 10.1088/1367-2630/aae2d5}
  {\bibfield  {journal} {\bibinfo  {journal} {New Journal of Physics}\ }\textbf
  {\bibinfo {volume} {20}},\ \bibinfo {pages} {103050} (\bibinfo {year}
  {2018})}\BibitemShut {NoStop}%
\bibitem [{\citenamefont {Halterman}\ \emph {et~al.}(2018)\citenamefont
  {Halterman}, \citenamefont {Alidoust},\ and\ \citenamefont
  {Zyuzin}}]{Klaus2018}%
  \BibitemOpen
  \bibfield  {author} {\bibinfo {author} {\bibfnamefont {Klaus}\ \bibnamefont
  {Halterman}}, \bibinfo {author} {\bibfnamefont {Mohammad}\ \bibnamefont
  {Alidoust}}, \ and\ \bibinfo {author} {\bibfnamefont {Alexander}\
  \bibnamefont {Zyuzin}},\ }\bibfield  {title} {\enquote {\bibinfo {title}
  {Epsilon-near-zero response and tunable perfect absorption in weyl
  semimetals},}\ }\href {\doibase 10.1103/PhysRevB.98.085109} {\bibfield
  {journal} {\bibinfo  {journal} {Phys. Rev. B}\ }\textbf {\bibinfo {volume}
  {98}},\ \bibinfo {pages} {085109} (\bibinfo {year} {2018})}\BibitemShut
  {NoStop}%
\bibitem [{\citenamefont {Kotov}\ and\ \citenamefont {Lozovik}(2016)}]{Kotov2}%
  \BibitemOpen
  \bibfield  {author} {\bibinfo {author} {\bibfnamefont {O.~V.}\ \bibnamefont
  {Kotov}}\ and\ \bibinfo {author} {\bibfnamefont {Yu.~E.}\ \bibnamefont
  {Lozovik}},\ }\bibfield  {title} {\enquote {\bibinfo {title} {Dielectric
  response and novel electromagnetic modes in three-dimensional dirac semimetal
  films},}\ }\href {\doibase 10.1103/PhysRevB.93.235417} {\bibfield  {journal}
  {\bibinfo  {journal} {Phys. Rev. B}\ }\textbf {\bibinfo {volume} {93}},\
  \bibinfo {pages} {235417} (\bibinfo {year} {2016})}\BibitemShut {NoStop}%
\bibitem [{\citenamefont {{Singh}}\ and\ \citenamefont
  {{Carbotte}}(2019)}]{2019arXiv190303072S}%
  \BibitemOpen
  \bibfield  {author} {\bibinfo {author} {\bibfnamefont {Ashutosh}\
  \bibnamefont {{Singh}}}\ and\ \bibinfo {author} {\bibfnamefont {J.~P.}\
  \bibnamefont {{Carbotte}}},\ }\bibfield  {title} {\enquote {\bibinfo {title}
  {{Effect of chiral anomaly on the circular dichroism and Hall angle in doped
  and tilted Weyl semimetals}},}\ }\href@noop {} {\bibfield  {journal}
  {\bibinfo  {journal} {arXiv e-prints}\ ,\ \bibinfo {eid} {arXiv:1903.03072}}
  (\bibinfo {year} {2019})},\ \Eprint {http://arxiv.org/abs/1903.03072}
  {arXiv:1903.03072 [cond-mat.mes-hall]} \BibitemShut {NoStop}%
\bibitem [{\citenamefont {{Das}}\ and\ \citenamefont
  {{Agarwal}}(2019)}]{2019arXiv190301205D}%
  \BibitemOpen
  \bibfield  {author} {\bibinfo {author} {\bibfnamefont {Kamal}\ \bibnamefont
  {{Das}}}\ and\ \bibinfo {author} {\bibfnamefont {Amit}\ \bibnamefont
  {{Agarwal}}},\ }\bibfield  {title} {\enquote {\bibinfo {title} {{Berry
  curvature induced thermopower in type-I and type-II Weyl Semimetals}},}\
  }\href@noop {} {\bibfield  {journal} {\bibinfo  {journal} {arXiv e-prints}\
  ,\ \bibinfo {eid} {arXiv:1903.01205}} (\bibinfo {year} {2019})},\ \Eprint
  {http://arxiv.org/abs/1903.01205} {arXiv:1903.01205 [cond-mat.mes-hall]}
  \BibitemShut {NoStop}%
\bibitem [{\citenamefont {Wilczek}(1987)}]{Frank}%
  \BibitemOpen
  \bibfield  {author} {\bibinfo {author} {\bibfnamefont {Frank}\ \bibnamefont
  {Wilczek}},\ }\bibfield  {title} {\enquote {\bibinfo {title} {Two
  applications of axion electrodynamics},}\ }\href {\doibase
  10.1103/PhysRevLett.58.1799} {\bibfield  {journal} {\bibinfo  {journal}
  {Phys. Rev. Lett.}\ }\textbf {\bibinfo {volume} {58}},\ \bibinfo {pages}
  {1799--1802} (\bibinfo {year} {1987})}\BibitemShut {NoStop}%
\bibitem [{\citenamefont {Wu}\ \emph {et~al.}(2016{\natexlab{b}})\citenamefont
  {Wu}, \citenamefont {Salehi}, \citenamefont {Koirala}, \citenamefont {Moon},
  \citenamefont {Oh},\ and\ \citenamefont {Armitage}}]{Wu1124}%
  \BibitemOpen
  \bibfield  {author} {\bibinfo {author} {\bibfnamefont {Liang}\ \bibnamefont
  {Wu}}, \bibinfo {author} {\bibfnamefont {M.}~\bibnamefont {Salehi}}, \bibinfo
  {author} {\bibfnamefont {N.}~\bibnamefont {Koirala}}, \bibinfo {author}
  {\bibfnamefont {J.}~\bibnamefont {Moon}}, \bibinfo {author} {\bibfnamefont
  {S.}~\bibnamefont {Oh}}, \ and\ \bibinfo {author} {\bibfnamefont {N.~P.}\
  \bibnamefont {Armitage}},\ }\bibfield  {title} {\enquote {\bibinfo {title}
  {{Quantized Faraday and Kerr rotation and axion electrodynamics of a 3D
  topological insulator}},}\ }\href {\doibase 10.1126/science.aaf5541}
  {\bibfield  {journal} {\bibinfo  {journal} {Science}\ }\textbf {\bibinfo
  {volume} {354}},\ \bibinfo {pages} {1124--1127} (\bibinfo {year}
  {2016}{\natexlab{b}})}\BibitemShut {NoStop}%
\bibitem [{\citenamefont {Tesa\ifmmode~\check{r}\else \v{r}\fi{}ov\'a}\ \emph
  {et~al.}(2014)\citenamefont {Tesa\ifmmode~\check{r}\else \v{r}\fi{}ov\'a},
  \citenamefont {Ostatnick\'y}, \citenamefont {Nov\'ak}, \citenamefont
  {Olejnik}, \citenamefont {\ifmmode~\check{S}\else \v{S}\fi{}ubrt},
  \citenamefont {Reichlov\'a}, \citenamefont {Ellis}, \citenamefont
  {Mukherjee}, \citenamefont {Lee}, \citenamefont {Sipahi}, \citenamefont
  {Sinova}, \citenamefont {Hamrle}, \citenamefont {Jungwirth}, \citenamefont
  {N\ifmmode~\check{e}\else \v{e}\fi{}mec}, \citenamefont
  {\ifmmode~\check{C}\else \v{C}\fi{}erne},\ and\ \citenamefont
  {V\'yborn\'y}}]{Tesarova}%
  \BibitemOpen
  \bibfield  {author} {\bibinfo {author} {\bibfnamefont {N.}~\bibnamefont
  {Tesa\ifmmode~\check{r}\else \v{r}\fi{}ov\'a}}, \bibinfo {author}
  {\bibfnamefont {T.}~\bibnamefont {Ostatnick\'y}}, \bibinfo {author}
  {\bibfnamefont {V.}~\bibnamefont {Nov\'ak}}, \bibinfo {author} {\bibfnamefont
  {K.}~\bibnamefont {Olejnik}}, \bibinfo {author} {\bibfnamefont
  {J.}~\bibnamefont {\ifmmode~\check{S}\else \v{S}\fi{}ubrt}}, \bibinfo
  {author} {\bibfnamefont {H.}~\bibnamefont {Reichlov\'a}}, \bibinfo {author}
  {\bibfnamefont {C.~T.}\ \bibnamefont {Ellis}}, \bibinfo {author}
  {\bibfnamefont {A.}~\bibnamefont {Mukherjee}}, \bibinfo {author}
  {\bibfnamefont {J.}~\bibnamefont {Lee}}, \bibinfo {author} {\bibfnamefont
  {G.~M.}\ \bibnamefont {Sipahi}}, \bibinfo {author} {\bibfnamefont
  {J.}~\bibnamefont {Sinova}}, \bibinfo {author} {\bibfnamefont
  {J.}~\bibnamefont {Hamrle}}, \bibinfo {author} {\bibfnamefont
  {T.}~\bibnamefont {Jungwirth}}, \bibinfo {author} {\bibfnamefont
  {P.}~\bibnamefont {N\ifmmode~\check{e}\else \v{e}\fi{}mec}}, \bibinfo
  {author} {\bibfnamefont {J.}~\bibnamefont {\ifmmode~\check{C}\else
  \v{C}\fi{}erne}}, \ and\ \bibinfo {author} {\bibfnamefont {K.}~\bibnamefont
  {V\'yborn\'y}},\ }\bibfield  {title} {\enquote {\bibinfo {title} {Systematic
  study of magnetic linear dichroism and birefringence in (ga,mn)as},}\ }\href
  {\doibase 10.1103/PhysRevB.89.085203} {\bibfield  {journal} {\bibinfo
  {journal} {Phys. Rev. B}\ }\textbf {\bibinfo {volume} {89}},\ \bibinfo
  {pages} {085203} (\bibinfo {year} {2014})}\BibitemShut {NoStop}%
\bibitem [{\citenamefont {Sang-Wook~Cheong}\ and\ \citenamefont
  {Saxena}(2018)}]{Cheong2018}%
  \BibitemOpen
  \bibfield  {author} {\bibinfo {author} {\bibfnamefont {Valery~Kiryukhin}\
  \bibnamefont {Sang-Wook~Cheong}, \bibfnamefont {Diyar~Talbayev}}\ and\
  \bibinfo {author} {\bibfnamefont {Avadh}\ \bibnamefont {Saxena}},\ }\bibfield
   {title} {\enquote {\bibinfo {title} {Broken symmetries, non-reciprocity, and
  multiferroicity},}\ }\href {\doibase 10.1038/s41535-018-0092-5} {\bibfield
  {journal} {\bibinfo  {journal} {npj Quantum Materials}\ }\textbf {\bibinfo
  {volume} {3}},\ \bibinfo {pages} {19} (\bibinfo {year} {2018})}\BibitemShut
  {NoStop}%
\bibitem [{\citenamefont {Lubashevsky}\ \emph {et~al.}(2014)\citenamefont
  {Lubashevsky}, \citenamefont {Pan}, \citenamefont {Kirzhner}, \citenamefont
  {Koren},\ and\ \citenamefont {Armitage}}]{Superconductor}%
  \BibitemOpen
  \bibfield  {author} {\bibinfo {author} {\bibfnamefont {Y.}~\bibnamefont
  {Lubashevsky}}, \bibinfo {author} {\bibfnamefont {LiDong}\ \bibnamefont
  {Pan}}, \bibinfo {author} {\bibfnamefont {T.}~\bibnamefont {Kirzhner}},
  \bibinfo {author} {\bibfnamefont {G.}~\bibnamefont {Koren}}, \ and\ \bibinfo
  {author} {\bibfnamefont {N.~P.}\ \bibnamefont {Armitage}},\ }\bibfield
  {title} {\enquote {\bibinfo {title} {Optical birefringence and dichroism of
  cuprate superconductors in the thz regime},}\ }\href {\doibase
  10.1103/PhysRevLett.112.147001} {\bibfield  {journal} {\bibinfo  {journal}
  {Phys. Rev. Lett.}\ }\textbf {\bibinfo {volume} {112}},\ \bibinfo {pages}
  {147001} (\bibinfo {year} {2014})}\BibitemShut {NoStop}%
\bibitem [{\citenamefont {Nandkishore}\ and\ \citenamefont
  {Levitov}(2011)}]{PhysRevLett.107.097402}%
  \BibitemOpen
  \bibfield  {author} {\bibinfo {author} {\bibfnamefont {Rahul}\ \bibnamefont
  {Nandkishore}}\ and\ \bibinfo {author} {\bibfnamefont {Leonid}\ \bibnamefont
  {Levitov}},\ }\bibfield  {title} {\enquote {\bibinfo {title} {{Polar Kerr
  Effect and Time Reversal Symmetry Breaking in Bilayer Graphene}},}\ }\href
  {\doibase 10.1103/PhysRevLett.107.097402} {\bibfield  {journal} {\bibinfo
  {journal} {Phys. Rev. Lett.}\ }\textbf {\bibinfo {volume} {107}},\ \bibinfo
  {pages} {097402} (\bibinfo {year} {2011})}\BibitemShut {NoStop}%
\bibitem [{\citenamefont {Tse}\ and\ \citenamefont
  {MacDonald}(2010)}]{PhysRevLett.105.057401}%
  \BibitemOpen
  \bibfield  {author} {\bibinfo {author} {\bibfnamefont {Wang-Kong}\
  \bibnamefont {Tse}}\ and\ \bibinfo {author} {\bibfnamefont {A.~H.}\
  \bibnamefont {MacDonald}},\ }\bibfield  {title} {\enquote {\bibinfo {title}
  {{Giant Magneto-Optical Kerr Effect and Universal Faraday Effect in Thin-Film
  Topological Insulators}},}\ }\href {\doibase 10.1103/PhysRevLett.105.057401}
  {\bibfield  {journal} {\bibinfo  {journal} {Phys. Rev. Lett.}\ }\textbf
  {\bibinfo {volume} {105}},\ \bibinfo {pages} {057401} (\bibinfo {year}
  {2010})}\BibitemShut {NoStop}%
\bibitem [{\citenamefont {Singh}\ \emph {et~al.}(2017)\citenamefont {Singh},
  \citenamefont {Bolotin}, \citenamefont {Ghosh},\ and\ \citenamefont
  {Agarwal}}]{Singh}%
  \BibitemOpen
  \bibfield  {author} {\bibinfo {author} {\bibfnamefont {Ashutosh}\
  \bibnamefont {Singh}}, \bibinfo {author} {\bibfnamefont {Kirill~I.}\
  \bibnamefont {Bolotin}}, \bibinfo {author} {\bibfnamefont {Saikat}\
  \bibnamefont {Ghosh}}, \ and\ \bibinfo {author} {\bibfnamefont {Amit}\
  \bibnamefont {Agarwal}},\ }\bibfield  {title} {\enquote {\bibinfo {title}
  {Nonlinear optical conductivity of a generic two-band system with application
  to doped and gapped graphene},}\ }\href {\doibase 10.1103/PhysRevB.95.155421}
  {\bibfield  {journal} {\bibinfo  {journal} {Phys. Rev. B}\ }\textbf {\bibinfo
  {volume} {95}},\ \bibinfo {pages} {155421} (\bibinfo {year}
  {2017})}\BibitemShut {NoStop}%
\bibitem [{\citenamefont {Singh}\ \emph {et~al.}(2018)\citenamefont {Singh},
  \citenamefont {Ghosh},\ and\ \citenamefont {Agarwal}}]{Singh2018}%
  \BibitemOpen
  \bibfield  {author} {\bibinfo {author} {\bibfnamefont {Ashutosh}\
  \bibnamefont {Singh}}, \bibinfo {author} {\bibfnamefont {Saikat}\
  \bibnamefont {Ghosh}}, \ and\ \bibinfo {author} {\bibfnamefont {Amit}\
  \bibnamefont {Agarwal}},\ }\bibfield  {title} {\enquote {\bibinfo {title}
  {Nonlinear and anisotropic polarization rotation in two-dimensional dirac
  materials},}\ }\href {\doibase 10.1103/PhysRevB.97.205420} {\bibfield
  {journal} {\bibinfo  {journal} {Phys. Rev. B}\ }\textbf {\bibinfo {volume}
  {97}},\ \bibinfo {pages} {205420} (\bibinfo {year} {2018})}\BibitemShut
  {NoStop}%
\bibitem [{\citenamefont {Zyuzin}\ and\ \citenamefont
  {Tiwari}(2016)}]{Zyuzin2016}%
  \BibitemOpen
  \bibfield  {author} {\bibinfo {author} {\bibfnamefont {A.~A.}\ \bibnamefont
  {Zyuzin}}\ and\ \bibinfo {author} {\bibfnamefont {R.~P.}\ \bibnamefont
  {Tiwari}},\ }\bibfield  {title} {\enquote {\bibinfo {title} {{Intrinsic
  anomalous Hall effect in type-II Weyl semimetals}},}\ }\href {\doibase
  10.1134/S002136401611014X} {\bibfield  {journal} {\bibinfo  {journal} {JETP
  Letters}\ }\textbf {\bibinfo {volume} {103}},\ \bibinfo {pages} {717--722}
  (\bibinfo {year} {2016})}\BibitemShut {NoStop}%
\bibitem [{\citenamefont {Mukherjee}\ and\ \citenamefont
  {Carbotte}(2017)}]{PhysRevB.96.085114}%
  \BibitemOpen
  \bibfield  {author} {\bibinfo {author} {\bibfnamefont {S.~P.}\ \bibnamefont
  {Mukherjee}}\ and\ \bibinfo {author} {\bibfnamefont {J.~P.}\ \bibnamefont
  {Carbotte}},\ }\bibfield  {title} {\enquote {\bibinfo {title} {{Absorption of
  circular polarized light in tilted type-I and type-II Weyl semimetals}},}\
  }\href {\doibase 10.1103/PhysRevB.96.085114} {\bibfield  {journal} {\bibinfo
  {journal} {Phys. Rev. B}\ }\textbf {\bibinfo {volume} {96}},\ \bibinfo
  {pages} {085114} (\bibinfo {year} {2017})}\BibitemShut {NoStop}%
\bibitem [{\citenamefont {Mukherjee}\ and\ \citenamefont
  {Carbotte}(2018{\natexlab{a}})}]{PhysRevB.97.035144}%
  \BibitemOpen
  \bibfield  {author} {\bibinfo {author} {\bibfnamefont {S.~P.}\ \bibnamefont
  {Mukherjee}}\ and\ \bibinfo {author} {\bibfnamefont {J.~P.}\ \bibnamefont
  {Carbotte}},\ }\bibfield  {title} {\enquote {\bibinfo {title} {{Imaginary
  part of Hall conductivity in a tilted doped Weyl semimetal with both broken
  time-reversal and inversion symmetry}},}\ }\href {\doibase
  10.1103/PhysRevB.97.035144} {\bibfield  {journal} {\bibinfo  {journal} {Phys.
  Rev. B}\ }\textbf {\bibinfo {volume} {97}},\ \bibinfo {pages} {035144}
  (\bibinfo {year} {2018}{\natexlab{a}})}\BibitemShut {NoStop}%
\bibitem [{\citenamefont {Mukherjee}\ and\ \citenamefont
  {Carbotte}(2018{\natexlab{b}})}]{PhysRevB.97.045150}%
  \BibitemOpen
  \bibfield  {author} {\bibinfo {author} {\bibfnamefont {S.~P.}\ \bibnamefont
  {Mukherjee}}\ and\ \bibinfo {author} {\bibfnamefont {J.~P.}\ \bibnamefont
  {Carbotte}},\ }\bibfield  {title} {\enquote {\bibinfo {title} {{Doping and
  tilting on optics in noncentrosymmetric multi-Weyl semimetals}},}\ }\href
  {\doibase 10.1103/PhysRevB.97.045150} {\bibfield  {journal} {\bibinfo
  {journal} {Phys. Rev. B}\ }\textbf {\bibinfo {volume} {97}},\ \bibinfo
  {pages} {045150} (\bibinfo {year} {2018}{\natexlab{b}})}\BibitemShut
  {NoStop}%
\bibitem [{\citenamefont {Yoshino}(2013)}]{Yoshino:13}%
  \BibitemOpen
  \bibfield  {author} {\bibinfo {author} {\bibfnamefont {Toshihiko}\
  \bibnamefont {Yoshino}},\ }\bibfield  {title} {\enquote {\bibinfo {title}
  {{Theory for oblique-incidence magneto-optical Faraday and Kerr effects in
  interfaced monolayer graphene and their characteristic features}},}\ }\href
  {\doibase 10.1364/JOSAB.30.001085} {\bibfield  {journal} {\bibinfo  {journal}
  {J. Opt. Soc. Am. B}\ }\textbf {\bibinfo {volume} {30}},\ \bibinfo {pages}
  {1085--1091} (\bibinfo {year} {2013})}\BibitemShut {NoStop}%
\bibitem [{\citenamefont {Tesarova}\ \emph {et~al.}(2012)\citenamefont
  {Tesarova}, \citenamefont {Nemec}, \citenamefont {Rozkotova}, \citenamefont
  {Subrt}, \citenamefont {Reichlova}, \citenamefont {Butkovicova},
  \citenamefont {Trojanek}, \citenamefont {Maly}, \citenamefont {Novak},\ and\
  \citenamefont {Jungwirth}}]{MLD}%
  \BibitemOpen
  \bibfield  {author} {\bibinfo {author} {\bibfnamefont {N.}~\bibnamefont
  {Tesarova}}, \bibinfo {author} {\bibfnamefont {P.}~\bibnamefont {Nemec}},
  \bibinfo {author} {\bibfnamefont {E.}~\bibnamefont {Rozkotova}}, \bibinfo
  {author} {\bibfnamefont {J.}~\bibnamefont {Subrt}}, \bibinfo {author}
  {\bibfnamefont {H.}~\bibnamefont {Reichlova}}, \bibinfo {author}
  {\bibfnamefont {D.}~\bibnamefont {Butkovicova}}, \bibinfo {author}
  {\bibfnamefont {F.}~\bibnamefont {Trojanek}}, \bibinfo {author}
  {\bibfnamefont {P.}~\bibnamefont {Maly}}, \bibinfo {author} {\bibfnamefont
  {V.}~\bibnamefont {Novak}}, \ and\ \bibinfo {author} {\bibfnamefont
  {T.}~\bibnamefont {Jungwirth}},\ }\bibfield  {title} {\enquote {\bibinfo
  {title} {Direct measurement of the three-dimensional magnetization vector
  trajectory in gamnas by a magneto-optical pump-and-probe method},}\ }\href
  {\doibase 10.1063/1.3692599} {\bibfield  {journal} {\bibinfo  {journal}
  {Applied Physics Letters}\ }\textbf {\bibinfo {volume} {100}},\ \bibinfo
  {pages} {102403} (\bibinfo {year} {2012})}\BibitemShut {NoStop}%
\bibitem [{\citenamefont {Carbotte}(2016)}]{PhysRevB.94.165111}%
  \BibitemOpen
  \bibfield  {author} {\bibinfo {author} {\bibfnamefont {J.~P.}\ \bibnamefont
  {Carbotte}},\ }\bibfield  {title} {\enquote {\bibinfo {title} {{Dirac cone
  tilt on interband optical background of type-I and type-II Weyl
  semimetals}},}\ }\href {\doibase 10.1103/PhysRevB.94.165111} {\bibfield
  {journal} {\bibinfo  {journal} {Phys. Rev. B}\ }\textbf {\bibinfo {volume}
  {94}},\ \bibinfo {pages} {165111} (\bibinfo {year} {2016})}\BibitemShut
  {NoStop}%
\bibitem [{\citenamefont {Gabor~Szechenyi}\ and\ \citenamefont
  {Cserti}(2016)}]{Cserti}%
  \BibitemOpen
  \bibfield  {author} {\bibinfo {author} {\bibfnamefont {Andor~Kormanyos}\
  \bibnamefont {Gabor~Szechenyi}, \bibfnamefont {Mate~Vigh}}\ and\ \bibinfo
  {author} {\bibfnamefont {Jozsef}\ \bibnamefont {Cserti}},\ }\bibfield
  {title} {\enquote {\bibinfo {title} {Transfer matrix approach for the kerr
  and faraday rotation in layered nanostructures},}\ }\href
  {http://stacks.iop.org/0953-8984/28/i=37/a=375802} {\bibfield  {journal}
  {\bibinfo  {journal} {Journal of Physics: Condensed Matter}\ }\textbf
  {\bibinfo {volume} {28}},\ \bibinfo {pages} {375802} (\bibinfo {year}
  {2016})}\BibitemShut {NoStop}%
\bibitem [{\citenamefont {Li}\ \emph {et~al.}(2017)\citenamefont {Li},
  \citenamefont {Wen}, \citenamefont {He}, \citenamefont {Zhang}, \citenamefont
  {Xia}, \citenamefont {Yu}, \citenamefont {Yang}, \citenamefont {Zhu},
  \citenamefont {Alshareef},\ and\ \citenamefont {Zhang}}]{Peng}%
  \BibitemOpen
  \bibfield  {author} {\bibinfo {author} {\bibfnamefont {Peng}\ \bibnamefont
  {Li}}, \bibinfo {author} {\bibfnamefont {Yan}\ \bibnamefont {Wen}}, \bibinfo
  {author} {\bibfnamefont {Xin}\ \bibnamefont {He}}, \bibinfo {author}
  {\bibfnamefont {Qiang}\ \bibnamefont {Zhang}}, \bibinfo {author}
  {\bibfnamefont {Chuan}\ \bibnamefont {Xia}}, \bibinfo {author} {\bibfnamefont
  {Zhi-Ming}\ \bibnamefont {Yu}}, \bibinfo {author} {\bibfnamefont
  {Shengyuan~A.}\ \bibnamefont {Yang}}, \bibinfo {author} {\bibfnamefont
  {Zhiyong}\ \bibnamefont {Zhu}}, \bibinfo {author} {\bibfnamefont {Husam~N.}\
  \bibnamefont {Alshareef}}, \ and\ \bibinfo {author} {\bibfnamefont
  {Xi-Xiang}\ \bibnamefont {Zhang}},\ }\bibfield  {title} {\enquote {\bibinfo
  {title} {Evidence for topological type-ii weyl semimetal wte2},}\ }\href
  {\doibase 10.1038/s41467-017-02237-1} {\bibfield  {journal} {\bibinfo
  {journal} {Nature Communications}\ }\textbf {\bibinfo {volume} {8}},\
  \bibinfo {pages} {2150} (\bibinfo {year} {2017})}\BibitemShut {NoStop}%
\bibitem [{\citenamefont {Qi}\ \emph {et~al.}(2008)\citenamefont {Qi},
  \citenamefont {Hughes},\ and\ \citenamefont {Zhang}}]{Shou-cheng}%
  \BibitemOpen
  \bibfield  {author} {\bibinfo {author} {\bibfnamefont {Xiao-Liang}\
  \bibnamefont {Qi}}, \bibinfo {author} {\bibfnamefont {Taylor~L.}\
  \bibnamefont {Hughes}}, \ and\ \bibinfo {author} {\bibfnamefont {Shou-Cheng}\
  \bibnamefont {Zhang}},\ }\bibfield  {title} {\enquote {\bibinfo {title}
  {Topological field theory of time-reversal invariant insulators},}\ }\href
  {\doibase 10.1103/PhysRevB.78.195424} {\bibfield  {journal} {\bibinfo
  {journal} {Phys. Rev. B}\ }\textbf {\bibinfo {volume} {78}},\ \bibinfo
  {pages} {195424} (\bibinfo {year} {2008})}\BibitemShut {NoStop}%
\bibitem [{\citenamefont {Visnovsky}(2006)}]{visnovsky2006optics}%
  \BibitemOpen
  \bibfield  {author} {\bibinfo {author} {\bibfnamefont {S.}~\bibnamefont
  {Visnovsky}},\ }\href {https://books.google.co.in/books?id=d27LBQAAQBAJ}
  {\emph {\bibinfo {title} {Optics in Magnetic Multilayers and
  Nanostructures}}},\ Optical Science and Engineering\ (\bibinfo  {publisher}
  {CRC Press},\ \bibinfo {year} {2006})\BibitemShut {NoStop}%
\bibitem [{\citenamefont {Shibata}\ \emph {et~al.}(2018)\citenamefont
  {Shibata}, \citenamefont {Takeuchi}, \citenamefont {Kohno},\ and\
  \citenamefont {Tatara}}]{Shibata}%
  \BibitemOpen
  \bibfield  {author} {\bibinfo {author} {\bibfnamefont {Junya}\ \bibnamefont
  {Shibata}}, \bibinfo {author} {\bibfnamefont {Akihito}\ \bibnamefont
  {Takeuchi}}, \bibinfo {author} {\bibfnamefont {Hiroshi}\ \bibnamefont
  {Kohno}}, \ and\ \bibinfo {author} {\bibfnamefont {Gen}\ \bibnamefont
  {Tatara}},\ }\bibfield  {title} {\enquote {\bibinfo {title} {Theory of
  electromagnetic wave propagation in ferromagnetic rashba conductor},}\ }\href
  {\doibase 10.1063/1.5011130} {\bibfield  {journal} {\bibinfo  {journal}
  {Journal of Applied Physics}\ }\textbf {\bibinfo {volume} {123}},\ \bibinfo
  {pages} {063902} (\bibinfo {year} {2018})}\BibitemShut {NoStop}%
\end{thebibliography}%

\end{document}